\documentclass[aps,prb,twocolumn,superscriptaddress]{revtex4-2}
\usepackage{amsmath,amssymb}
\usepackage{graphicx}
\usepackage{subfigure}
\usepackage{verbatim}
\usepackage{amsfonts}
\usepackage{dcolumn}
\usepackage{bm}
\usepackage{ulem}
\usepackage{color}
\usepackage[colorlinks,citecolor=blue]{hyperref}
\usepackage{dcolumn}
\usepackage{bm}
\usepackage{mathrsfs}
\usepackage{cancel}
\usepackage{multirow}
\usepackage{CJK}
\usepackage{verbatim}

\usepackage{xcolor}
\definecolor{orange}{rgb}{1,0.5,0}
\definecolor{darkblue}{rgb}{0.,0.,0.4}
\definecolor{darkred}{rgb}{0.5,0.,0.}

\newcommand{\wei}[1]{ { \color{darkred} \footnotesize (\textsf{WZ}) \textsf{\textsl{#1}} }}

\newcommand{\hu}[1]{ { \color{orange} \footnotesize (\textsf{HU}) \textsf{\textsl{#1}} }}

\begin{document}

\title{
Symmetry-breaking line defects embedded to a 3D $O(N)$ critical bulk
}

\author{Shuai Yang}	
\affiliation{Department of Physics and State Key Laboratory of Surface Physics, Fudan University, Shanghai 200433, P.R. China}
\affiliation{Department of Physics, Hong Kong University of Science and Technology, Clear Water Bay, Hong Kong, China}
\author{Liang-dong Hu}
\affiliation{School of Physics and Electronic, Hunan University, Changsha 410082, China}
\author{Yan Chen}
\email{yanchen99@fudan.edu.cn}
\affiliation{Department of Physics and State Key Laboratory of Surface Physics, Fudan University, Shanghai 200433, P.R. China}
\affiliation{Shanghai Branch, Hefei National Laboratory, Shanghai 201315, P.R. China}
\author{W. Zhu}
\email{zhuwei@westlake.edu.cn}
\affiliation{Institute of Natural Sciences, Westlake Institute for Advanced Study, Hangzhou 310024, China}
\affiliation{Department of Physics, School of Science, Westlake University, Hangzhou 310030, China }

\date{\today}

\begin{abstract}
While spontaneous breaking of a discrete symmetry in one-dimensional classical systems with short-range interactions is absent, it is expected that a line defect embedded in a bulk criticality exhibits a stable discrete symmetry spontaneous breaking. 
Here, we investigate the behavior of a pinning-field line defect immersed in a 3D bulk that remains tuned to the $O(N)$ Wilson-Fisher critical point.  
    Employing the fuzzy sphere technique, we provide convincing evidence of the existence of stable defect conformal fixed points, and we demonstrate their renormalization group stability by showing no relevant operator and less effective degrees of freedom than that at bulk fixed point via $g$-function. 
    Moreover, we investigate the defect domain wall operator for various $N$, and we identify that it becomes irrelevance for $N\gtrsim 3$ but it is relevant for $N<3$.  
    These evidence indicate that a one-dimensional defect coupled to a critical bulk cannot support a stable symmetry spontaneously broken defect fixed point due to domain wall proliferation for $N<3$ Wilson-Fisher universality, while in the case of $N \gtrsim 3$ a symmetry broken defect is possible.  
\end{abstract}
\maketitle
\section{Introduction}
Spontaneous symmetry breaking (SSB) is generally forbidden in one-dimensional systems with short-range interactions\cite{Mermin_Wagner_PhysRevLett.17.1133,Hohenberg_PhysRev.158.383}. A familiar example is the one-dimensional Ising model, which lacks long-range order at any finite temperature\cite{Peierls_1936On}. The underlying mechanism is the proliferation of domain walls separating different symmetry-breaking sectors: the associated entropic gain always outweighs their finite energy cost, destabilizing any ordered phase. This conclusion can, however, be circumvented in the presence of sufficiently long-range interactions\cite{Dyson:1968up,Anderson_PhysRevB.1.4464}. In such cases, the energy required to separate a pair of domain walls grows with their distance, suppressing their proliferation and allowing long-range order to emerge.

In this context, a natural question is whether an analogous mechanism can arise when a defect is coupled to a higher-dimensional critical bulk\cite{A.J.Bray_1977,Diehl_1997}. Since critical fluctuations may induce effective long-range interactions along the defect, it is therefore conceivable that a defect embedded in a critical environment may support an ordered phase or symmetry broken patterns even when an isolated one-dimensional system cannot.

A systematic investigation of this question requires understanding the infrared behavior of the defect itself. When the bulk is tuned to a conformal fixed point, and the defect flows to a scale-invariant fixed point, the resulting system is naturally described within the framework of defect conformal field theory (dCFT)\cite{Bill__2016,LACES21_CFTdefects}.  The study of dCFT can help us understand the extended objects in quantum field theory. Prominent examples include Wilson\cite{Wilson_OP_PhysRevD.10.2445} and t$^\prime$Hooft\cite{THOOFT19781} line operators in the study of confinement phenomena. 
DCFTs also have numerous manifestations in condensed matter and statistical physics, such as spin impurities in the Kondo problem\cite{Kondo_paper10.1143/PTP.32.37,Kondo_wilson_RevModPhys.47.773,affleck1995conformalfieldtheoryapproach}, boundary theories of gapless symmetry-protected topological phases\cite{gSPT_Ruben_PhysRevX.11.041059,gSPTs_PhysRevX.7.041048,Prembabu_2024} and surface critical phenomena\cite{Diehl:1981jgg,Cardy:1984bb,BINDER199017}. More recently, it has been recognized that dCFTs naturally emerge in the description of quantum states under certain quantum measurements or decoherence\cite{Garratt_2023,JYLee_PRXQuantum.4.030317}.



\begin{figure}[t] 
    \centering
    \includegraphics[width=0.48\textwidth]{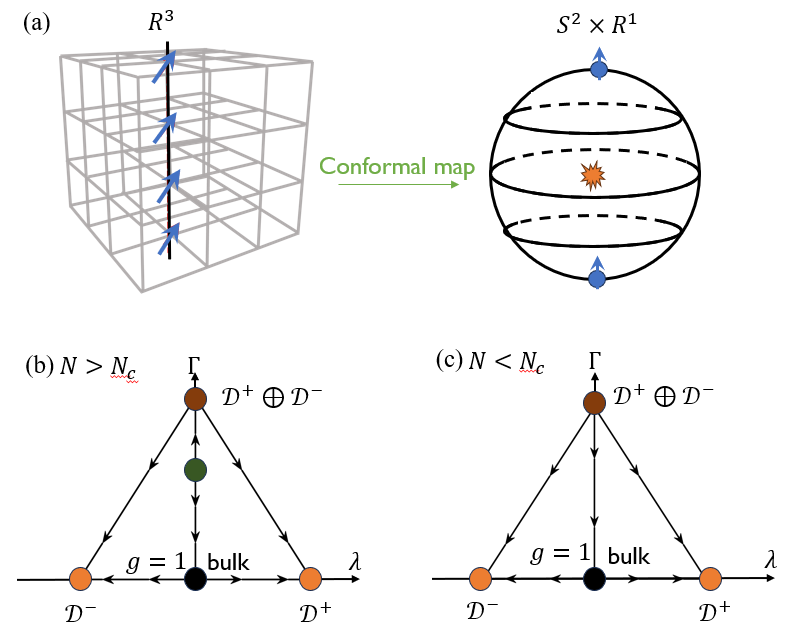} 
    \caption{(a) Schematic plot of line defect living on 3D critical bulk(left panel) and radial quantization view of line defect on $S^{d-1}$(right panel). (b-c) RG flows between the bulk fixed point(black dot) and the fixed point of magnetic line defect(orange dot), as well as SSB defect(brown dot). Depending on the symmetry of the bulk $O(N)$ Wilson–Fisher theory in which the defect is embedded, two distinct RG flow scenarios can arise: (b) corresponds to the case where a particular spontaneous-symmetry-breaking fixed point is stable ($N>N_c$), while (c) corresponds to the case where this SSB fixed point is unstable ($N<N_c$). In this work, we assume the SSB defect is the superposition of two symmetry-related pinning-field defects with opposite orientations. 
    }
    \label{fig:sch_plot}
\end{figure}

Recent progress in dCFT has made this question more concrete. A symmetry-breaking line defect can be explicitly realized by turning on a relevant defect perturbation that selects one symmetry-breaking sector, while a putative spontaneously symmetry-breaking defect is naturally described as a non-simple conformal defect containing several superselection sectors\cite{lanzetta2025beginningendpointbootstrapconformal,komargodski2025defectanomaliesspinfluxduality}. The stability of such an ordered defect is then controlled by symmetry-allowed defect-changing operators, which create domain walls between different sectors on the defect line. In the three-dimensional Ising CFT, the pinning-field defect provides the simplest example: positive and negative local fields flow to two conjugate defects\cite{Hu_2024}, usually denoted by $\mathcal{D}^+$ and $\mathcal{D}^-$, and the candidate SSB defect is $\mathcal D^+\oplus \mathcal D^-$ (see Fig. \ref{fig:sch_plot} (b-c)). Recent fuzzy-sphere calculations, overlap methods, cusp analyses, and defect bootstrap studies have found that the leading domain-wall operator connecting $\mathcal{D}^+$ and $\mathcal{D}^-$ has scaling dimension $\Delta_{\rm DW}\simeq 0.84<1$\cite{Zhou_2024,Cuomo:2024psk,lanzetta2025beginningendpointbootstrapconformal}. This implies that the domain-wall perturbation is relevant, so the minimal Ising SSB defect is unstable against domain-wall proliferation.
This negative result naturally raises the central question addressed in this work: in the three-dimensional $O(N)$ Wilson--Fisher CFT, can increasing $N$ make the corresponding defect-changing operator irrelevant, thereby allowing a stable SSB line defect(see the RG diagram in Fig.\ref{fig:sch_plot}(b-c))? 

Recently, fuzzy-sphere regularization \cite{PhysRevX.13.021009,Fuzzysphere_review2026}offers a non-perturbative approach to explore CFTs, and it enables direct access to a broad range of conformal data, including complete operator information\cite{PhysRevX.13.021009}, correlation functions\cite{four_point_corr_PhysRevB.108.235123}, OPE coefficients\cite{ope_PhysRevLett.131.031601}, conformal generators\cite{generators_10.21468/SciPostPhys.18.3.086,fan2024noteexplicitconstructionconformal}, and renormalization group monotonic functions\cite{F_func_PhysRevB.111.155151,Zhou_2024}. 
Importantly, fuzzy-sphere regularization provides a complementary approach to defect CFTs \cite{Hu_2024,Zhou_2024}.
For the problem considered in this work, this approach allows us to study line defects in the $O(N)$ Wilson–Fisher CFTs for arbitrary integer values of $N$, thereby filling a gap left by existing methods. 
We first explain how to construct such defects for general $O(N)$ models on the fuzzy sphere.
Then, we examine the existence of a defect induced conformal fixed point. The structure of the defect operator conformal multiplets indicates the existence of an attractive defect fixed point. Finally, 
we demonstrate that this line defect can undergo spontaneous symmetry breaking when $N\gtrsim 3$. This conclusion is supported by both the scaling dimensions of the domain wall operators($\Delta_{\phi^{+-}}>1$) and the values of the defect $g$-function($g<0.5$).

\section{Theoretical Background and Stability Criterion}
\label{sec:theoretical_background}

In this section, we formulate the line defects studied in this work and explain the diagnostics used to test their stability. We first review the pinning-field defect in the three-dimensional $O(N)$ Wilson--Fisher CFT. We then introduce the $\mathbb{Z}_2$ symmetric defect\cite{lanzetta2025beginningendpointbootstrapconformal}, whose stability is controlled by defect-changing operators. Finally, we describe a microscopic impurity realization, the radial-quantization interpretation of the relevant Hilbert spaces, the complementary constraint from the defect $g$-function, and the prior landscape that motivates the finite-$N$ stability test.

This problem was introduced and studied in early literature\cite{Allais_2014}. Various perturbative techniques($4-\epsilon$ expansion, large $N$ expansion) have been utilized extensively to tackle this problem\cite{Nishioka_2023,girault2026consequencessymmetrybreakingconformaldefect}. It has also been realized in lattice models and investigated via Monte Carlo simulations\cite{allais2014magneticdefectlinecritical,PhysRevB.95.014401}. Although the conformal bootstrap approach that incorporates defects remains highly challenging, partial operator spectrum have been obtained\cite{Gimenez_Grau_2022,Belton_2025}, with recent progress summarized in the Ref. \cite{lanzetta2025beginningendpointbootstrapconformal,wenliang_li_Hu:2025yrs}.

\subsection{Pinning-field line defects}
\label{subsec:pinning_defect}

We consider the three-dimensional $O(N)$ Wilson--Fisher CFT with fundamental order parameter $\phi^a$, $a=1,\ldots,N$. A magnetic line defect is introduced by coupling the bulk order parameter to a localized source along the defect worldline,
\begin{equation}
    S_{\rm def}
    =
    S_{\rm WF}
    +
    h\int d\tau\,
    \delta_{ab}n^a(\tau)\phi^b(\tau,\mathbf 0).
    \label{eq:pinning_defect_action}
\end{equation}
Here the integral is taken along the defect worldline. We take $h>0$ and regard $n^a(\tau)$ as a dimensionless vector source profile, so that $h n^a(\tau)$ is the physical source. Taking $n^a(\tau)=\hat n^a$, with $\hat n\in S^{N-1}$, gives the ordinary pinning-field defect. More general $\tau$-dependent source profiles describe other defect configurations. For example, a profile that changes from $n^a=0$ to $n^a=\hat n^a$ across a point such as $\tau=0$ creates a pinning defect, while one that changes from $n^a=\hat n_1^a$ to $n^a=\hat n_2^a$ describes a defect-changing process. In the dCFT language, such a localized change of defect condition is represented by the insertion of a local defect operator: a defect-creation operator in the first case and a defect-changing operator in the second\cite{lanzetta2025beginningendpointbootstrapconformal}.

Since the perturbation is supported on a one-dimensional defect, the RG eigenvalue of $h$ near the trivial defect is
$    y_h=1-\Delta_{\vec\phi} $.
Since the three-dimensional $O(N)$ Wilson--Fisher CFT, $\Delta_{\vec\phi}<1$, the pinning field is a relevant defect perturbation, and an infinitesimal $h$ is expected to drive the line to a nontrivial infrared conformal defect\cite{Hanke2000,Allais_2014,Bill__2016,Nishioka_2023,girault2026consequencessymmetrybreakingconformaldefect,allais2014magneticdefectlinecritical,PhysRevB.95.014401,Hu_2024}. We denote the resulting pinning-field defect by $\mathcal{D}^{\hat n}$ for a constant orientation $\hat n$.
A fixed orientation preserves the stabilizer subgroup $O(N-1)\subset O(N)$, and the family of symmetry-related pinning defects forms the orbit
\begin{equation}
    \mathcal D^{\hat n},
    \qquad
    \hat n\in O(N)/O(N-1)\simeq S^{N-1}. \label{eq:phase-slip-defect}
\end{equation}
Defects with different orientations are related by global $O(N)$ transformations. In particular, for $N=1$, the orbit reduces to two points,
$S^0=\{+,-\}$,
corresponding to the two Ising pinning defects $\mathcal D^+$ and $\mathcal D^-$. This two-point orbit is the prototype for the antipodal direct-sum construction introduced below.

Relative to the trivial defect, the pinning defect breaks both transverse translations and the part of the internal symmetry that does not preserve $\hat n$. The defect line still preserves translations and conformal transformations along the defect, while the Ward identity for broken transverse translations gives the displacement operator $\hat D$ with protected scaling dimension
\begin{equation} \label{eq:D_op}
    \Delta_{\hat D}=2.
\end{equation}
Similarly, the Ward identity for the broken $O(N)$ currents gives internal tilt operators with protected dimension. In a frame where $\hat n=\hat e_1$, we denote them by $\hat t_a$ with
\begin{equation} \label{eq:t_op}
    \Delta_{\hat t}=1,
    \qquad
    a=2,\ldots,N .
\end{equation}
In the numerical spectra below, the levels at $\Delta_{\hat t}=1$ and $\Delta_{\hat D}=2$ provide benchmarks for identifying the emergent pinning-field defect fixed point.

\subsection{$\mathbb{Z}_2$ symmetric defect and defect-changing operators}
\label{subsec:antipodal_defect}

To formulate a candidate ordered defect, we choose an unoriented easy axis in the internal space and denote it by $\hat e$.
This axis is part of the defect data: the possible long-range order discussed below is therefore an Ising-like order associated with the residual $\mathbb{Z}_2$ symmetry exchanging the two antipodal directions.
The two antipodal pinning defects are
\begin{equation}
    \mathcal{D}^+\equiv \mathcal D^{+\hat e},
    \qquad
    \mathcal{D}^-\equiv \mathcal{D}^{-\hat e}.
\end{equation}
The putative symmetry-breaking defect considered in this work is the non-simple, exchange-symmetric direct-sum defect \cite{lanzetta2025beginningendpointbootstrapconformal}
\begin{equation}
    \mathcal{D}^{\ell}
    =
    \mathcal{D}^+\oplus \mathcal{D}^- .
    \label{eq:anti_direct_sum}
\end{equation}
$\mathcal{D}^{\ell}$ is the candidate SSB phase on the line.
It has two simple components, corresponding to the two antipodal pinning directions.
The residual $\mathbb{Z}_2$ symmetry maps $\mathcal{D}^+$ to $\mathcal{D}^-$ and exchanges the projectors $P_+$ and $P_-$; hence $\mathcal{D}^{\ell}$ is invariant as a non-simple defect, while a pure component selects one of the two pinning directions.

Because $\mathcal{D}^{\ell}=\mathcal{D}^+\oplus\mathcal{D}^-$ is a direct sum, its defect-operator algebra contains two orthogonal topological component projectors, $P_\alpha P_\beta=\delta_{\alpha\beta}P_\alpha$ and $P_++P_-=\mathbf{1}$. The intrinsic order parameter is therefore not the bulk vector field $\phi^a$ itself, but the exchange-odd component label
\begin{equation}
    s^z=P_+-P_-,
    \qquad
    \Delta_{s^z}=0,
    \qquad
    (s^z)^2=\mathbf{1},
    \label{eq:defect_order_parameter}
\end{equation}
Accordingly, $\langle s^z\rangle_{\mathcal{D}^{\pm}}=\pm1$, whereas exchange symmetry gives $\langle s^z\rangle_{\mathcal{D}^{\ell}}=0$. For a defect primary, conformal invariance fixes the two-point function to
\begin{equation}
    \langle s^z(\tau)s^z(0)\rangle_{\mathcal{D}^{\ell}}
    =\frac{C_s}{|\tau|^{2\Delta_{s^z}}}.
    \label{eq:defect_lro}
\end{equation}
Here the projector algebra fixes $C_s=1$ in the above normalization, so $\Delta_{s^z}=0$ makes the correlator independent of separation. Because the one-point function vanishes in the exchange-symmetric description, this is also a nondecaying connected correlator and hence establishes defect long-range order (LRO). Conversely, a nonzero constant long-distance limit for a nontrivial exchange-odd defect primary is compatible with scale invariance only if its dimension is zero. Thus, for this component-label operator, defect LRO is equivalent to $\Delta_{s^z}=0$ \cite{lanzetta2025beginningendpointbootstrapconformal}.

The stability of $\mathcal{D}^{\ell}$ is controlled in part by operators that change the defect condition from one component to the other.
We denote the leading defect-changing operator, equivalently the domain-wall operator on the line, by $\hat\phi^{+-}:\mathcal{D}^+\rightarrow \mathcal{D}^-$, with conjugate operator $\hat\phi^{-+}:\mathcal{D}^-\rightarrow \mathcal{D}^+$.
The exchange-even Hermitian combination generates the domain-wall perturbation
\begin{equation}
    \delta S_{\rm DW}
    =
    \gamma\int d\tau\,
    \left(
        \hat\phi^{+-}
        +
        \hat\phi^{-+}
    \right).
    \label{eq:domain_wall_perturbation}
\end{equation}
The exchange-odd combination is not included when the residual $\mathbb{Z}_2$ symmetry is imposed.
Since the defect is one-dimensional, the leading RG equation for the domain-wall fugacity is
\begin{equation}
    \frac{d\gamma}{d\ell}
    =
    \left(1-\Delta_{\hat\phi^{+-}}\right)\gamma+\cdots ,
    \label{eq:domain_wall_beta}
\end{equation}
where $\Delta_{\hat\phi^{+-}}$ is the scaling dimension of $\hat\phi^{+-}$. Therefore, $\Delta_{\hat\phi^{+-}}<1$ makes the domain-wall perturbation relevant and destabilizes $\mathcal{D}^\ell$ by domain-wall proliferation, while $\Delta_{\hat\phi^{+-}}>1$ makes this perturbation irrelevant.
This statement assumes the exchange symmetry and the preserved $O(N-1)$ symmetry defining the easy-axis problem, so perturbations that explicitly select one component or rotate the chosen axis are not included.
Full stability also requires that the lowest nontrivial defect operator living within a single component $\mathcal{D}^\pm$, among operators allowed by these same symmetries, is irrelevant.

For $N=1$, $\mathcal{D}^\ell$ is precisely the Ising construction $\mathcal{D}^+\oplus \mathcal{D}^-$. The operator $\hat\phi^{+-}$ is then the ordinary Ising domain-wall operator on the line.
For $N>1$, Eq.~\eqref{eq:anti_direct_sum} should be understood as the easy-axis construction described above.
This should be distinguished from the Bose--Kondo impurity defects of Ref.~\cite{Sarma_2026_BoseKondoImpurity}.
There, dynamical spin-$S$ impurities couple isotropically to the full $O(3)$ order parameter through the $SO(3)$-invariant vector interaction $\mathbf S\mathbin{\cdot}\boldsymbol\phi$; in the fuzzy-sphere radial-quantization setup, the corresponding impurity spins are placed at the north and south poles and couple with opposite signs, as required by CPT.
In this work, by contrast, the coupling is Ising-like, $J s^z\phi^1$, selecting a single easy axis and leaving $O(N-1)$ together with the residual diagonal $\mathbb Z_2$.
Accordingly, the relevant diagnostic is the stability of the two-component easy-axis defect $\mathcal D^\ell$ against the residual-$\mathbb Z_2$ domain-wall perturbation.

Let $\Delta_{\rm min}^{\pm,\pm}$ denote the scaling dimension of the lowest nontrivial defect operator inside a single component $D^\pm$ that is allowed by the symmetries defining the easy-axis problem.
The stability criteria studied in this work are therefore
\begin{equation}
    \Delta_{\hat\phi^{+-}}>1,
    \qquad
    \Delta_{\rm min}^{\pm,\pm}>1
    \label{eq:ssb_stability_criterion}
\end{equation}
The first condition tests the irrelevance of the domain-wall perturbation, and the second excludes relevant perturbations internal to each component.
In radial quantization, $\Delta_{\hat\phi^{+-}}$ is extracted from the lowest primary in the defect-changing Hilbert space $\mathcal H^{+-}$, as described below.

\subsection{Microscopic impurity realization}
\label{subsec:impurity_model}

$\mathcal{D}^\ell$ can be realized microscopically by coupling the critical bulk to an Ising impurity. A minimal Hamiltonian is\cite{lanzetta2025beginningendpointbootstrapconformal,komargodski2025defectanomaliesspinfluxduality}
\begin{equation}
    H_{\rm imp}
    =
    H_{O(N)}^\ast
    +
    J s^z \phi^{1}(\mathbf 0)
    +
    \Gamma s^x
    +
    \lambda s^z ,
    \label{eq:impurity_hamiltonian}
\end{equation}
where $s^{x,z}$ are Pauli matrices acting on the impurity Hilbert space, and $\phi^1$ is the component of the order parameter along the selected easy axis $\hat e_1$. We take $J>0$ for definiteness; changing the sign of $J$ only relabels the two components $\mathcal{D}^+$ and $\mathcal{D}^-$.

At $\lambda=0$, this Hamiltonian preserves an $O(N-1)$ symmetry rotating the components transverse to $\hat e_1$, together with a diagonal $\mathbb{Z}_2$ symmetry combining the internal reflection of the selected order-parameter component with an impurity-spin flip,
\begin{equation}
    \phi^1\mapsto-\phi^1,
    \qquad
    s^z\mapsto -s^z,
    \qquad
    s^x\mapsto s^x .
    \label{eq:diag_z2}
\end{equation}
At $\Gamma=\lambda=0$, the impurity spin component $s^z$ is conserved. In each conserved impurity-spin sector, the coupling $J s^z\phi^1(\mathbf 0)$ is an ordinary pinning field with opposite orientation. The two impurity-spin sectors then flow separately to the two pinning components:
\begin{equation}
    s^z=+1:
    \quad
    \mathcal{D}^+,
    \qquad
    s^z=-1:
    \quad
    \mathcal{D}^- .
\end{equation}
Thus the infrared defect at this fine-tuned point is the non-simple SSB defect with LRO
\begin{equation} \label{eq:SSBfixedpoint}
    \mathcal{D}^{\ell} = \mathcal{D}^+\oplus \mathcal{D}^- .
\end{equation}

The transverse field $\Gamma s^x$ flips the impurity spin. Near $\mathcal{D}^\ell$, it therefore changes the defect condition from $\mathcal{D}^+$ to $\mathcal{D}^-$ or from $\mathcal{D}^-$ to $\mathcal{D}^+$. In the dCFT language, the microscopic flip operator has a nonzero projection onto the leading exchange-even domain-wall operator,
\begin{equation}
    s^x
    \quad\longrightarrow\quad
    c_{\rm DW}
    \left(
        \hat\phi^{+-}
        +
        \hat\phi^{-+}
    \right)
    +\cdots ,
\end{equation}
where $c_{\rm DW}$ is nonzero and the ellipsis denotes descendants and less relevant defect-changing operators. Hence $\Gamma$ is the microscopic coupling to the domain-wall perturbation in Eq.~\eqref{eq:domain_wall_perturbation}. Its leading RG flow near $\mathcal{D}^\ell$ is
\begin{equation}
    \frac{d\Gamma}{d\ell}
    =
    \left(1-\Delta_{\hat\phi^{+-}}\right)\Gamma+\cdots .
\end{equation}
Thus a nonzero but small $\Gamma$ destabilizes $\mathcal D^\ell$ when $\Delta_{\hat\phi^{+-}}<1$, while it is irrelevant when $\Delta_{\hat\phi^{+-}}>1$.

The longitudinal field $\lambda s^z$ explicitly breaks the $\mathbb{Z}_2$ symmetry. It couples directly to the component $s^z=1$ and $s^z=-1$. Therefore any nonzero $\lambda$ selects one of the two pinning components and drives the defect to either $\mathcal D^+$ or $\mathcal D^-$, rather than to $\mathcal D^\ell$. In other words,
\begin{equation}
    \lambda\neq0:
    \qquad
    \mathcal{D}^\ell
    \longrightarrow
    \mathcal{D}^+
    \ \text{or}\
    \mathcal{D}^- .
\end{equation}
The local phase diagram near $\mathcal{D}^\ell$ is therefore organized by two distinct perturbations. The longitudinal field $\lambda$ is an explicitly symmetry-breaking direction and is excluded when testing the exchange-symmetric ordered defect. The transverse field $\Gamma$ preserves the $\mathbb{Z}_2$ and is the physical domain-wall perturbation. Stability of the SSB defect is therefore the statement that this symmetry-preserving perturbation is irrelevant near $\mathcal{D}^\ell$, rather than merely the statement that the fine-tuned Hamiltonian with $\Gamma=0$ has two decoupled impurity-spin sectors.

In the numerical calculation below, the impurity model is used to identify the allowed perturbations, while the scaling dimension that controls $\Gamma$ is extracted more directly from the defect-changing Hilbert space $\mathcal H^{+-}$.

\subsection{Radial quantization and defect-changing Hilbert spaces}
\label{subsec:radial_quantization}

The fuzzy-sphere calculation naturally implements radial quantization. A line defect in flat space is mapped to a pair of pointlike defects inserted at the north and south poles of $S^2$, extending along the Euclidean time direction(see Fig.\ref{fig:sch_plot} (a) for illustration). This gives a direct spectral realization of the defect Hilbert spaces needed above. We denote by
\begin{equation} \label{eq:defectHilbert}
    \mathcal H^{\alpha\beta}
\end{equation}
the Hilbert space obtained by imposing defect condition $\alpha$ at the north pole and $\beta$ at the south pole, with $\alpha,\beta\in\{+, -,0\}$.
Here $0$ denotes the absence of a pinning defect. The three sectors used below have distinct physical roles. 
The same-component sector $\mathcal H^{++}$ contains ordinary local defect operators living on a single pinning component $\mathcal{D}^+$; by exchange symmetry, $\mathcal H^{--}$ gives the corresponding spectrum on $\mathcal{D}^-$.
The exchange symmetry relates $\mathcal H^{+-}$ and $\mathcal H^{-+}$, so the
ordering convention only fixes the direction of the operator
$\hat\phi^{+-}:\mathcal{D}^+\rightarrow \mathcal{D}^-$.
Finally, the mixed sector $\mathcal H^{+0}$
contains endpoint operators connecting the trivial defect and the pinning defect $\mathcal{D}^+$. Depending on the orientation, these operators may be viewed as creating or ending the pinning defect.
Thus $\mathcal H^{++}$ diagnoses perturbations internal to a component, $\mathcal H^{+-}$ diagnoses the domain-wall perturbation that can destroy $\mathcal D^\ell$, and $\mathcal H^{+0}$ measures the cost of ending or creating a pinning defect.

In the numerical implementation, scaling dimensions are extracted from the finite-size energy spectrum on the fuzzy sphere. Let $E_n^{\alpha\beta}$ denote the $n$th energy level in $\mathcal H^{\alpha\beta}$, and let $E_0^{00}$ and $E_{T_{\mu\nu}}^{00}$ be the defect-free ground-state energy and stress-tensor energy at the same finite system size. We use the bulk stress tensor to fix the velocity normalization by setting
$    \Delta_{T_{\mu\nu}}=3 .$
With this common normalization, ordinary defect operators in $\mathcal H^{++}$ are extracted from excitation gaps above the same-component ground state\cite{Hu_2024,Zhou_2024}:
\begin{equation}
    \Delta_{\hat{\mathcal O}}
    =
    3
    \frac{
        E_{\hat{\mathcal O}}^{++}-E_0^{++}
    }{
        E_{T_{\mu\nu}}^{00}-E_0^{00}
    },
    \label{eq:defect_dimension_extraction}
\end{equation}
The domain-wall operator is different: it is the ground state of the mixed sector $\mathcal H^{+-}$, compared against the same-pinning reference sector\cite{Zhou_2024}:
\begin{equation}
    \Delta_{\hat\phi^{+-}}
    =
    3
    \frac{
        E_0^{+-}-E_0^{++}
    }{
        E_{T_{\mu\nu}}^{00}-E_0^{00}
    } .
    \label{eq:defect_changing_dimension_extraction}
\end{equation}
Equivalently, one may use $E_0^{--}$ as the reference energy; exchange symmetry makes $E_0^{++}=E_0^{--}$ up to finite-size numerical accuracy.
The endpoint dimension in $\mathcal H^{+0}$ uses the analogous mixed-sector subtraction, with the reference energy taken to be the average of the pinning and trivial sectors\cite{Zhou_2024}:
\begin{equation}
    \Delta_{\hat\phi^{+0}}
    =
    3
    \frac{
        E_0^{+0}-\frac{1}{2}\left(E_0^{++}+E_0^{00}\right)
    }{
        E_{T_{\mu\nu}}^{00}-E_0^{00}
    } .
    \label{eq:defect_creation_dimension_extraction}
\end{equation}
This construction allows the stability-relevant domain-wall dimension to be measured directly from the pinning-field Hamiltonian, without explicitly retaining the impurity spin in Eq.~\eqref{eq:impurity_hamiltonian}. The same spectra also provide same-component data for the stability criteria and endpoint data for the numerical characterization of creating or ending a pinning defect.

\subsection{Defect $g$-function}
\label{subsec:g_function_theory}

The defect $g$-function characterizes the universal contribution of a line defect to the partition function. We normalize it so that the trivial line has $g_{\rm trivial}=1$. For unitary RG flows between conformal line defects in a fixed bulk CFT, the line-defect $g$-theorem implies $g_{\rm UV}\geq g_{\rm IR}$\cite{Cuomo_g_theorem__PhysRevLett.128.021603}; this result generalizes the boundary $g$-theorem of Ref.~\cite{Affleck_Ludwig_g_theorem_PhysRevLett.67.161}. For the exchange-symmetric direct sum $\mathcal{D}^\ell=\mathcal{D}^+\oplus\mathcal{D}^-$, additivity and exchange symmetry give $g_{\mathcal{D}^\ell}=g_{\mathcal{D}^+}+g_{\mathcal{D}^-}=2g_{\mathcal{D}}$, where $g_{\mathcal{D}}$ denotes the value for either pinning component.

It follows that $g_{\mathcal{D}}<1/2$ excludes the trivial line as a direct infrared endpoint of $\mathcal{D}^\ell$, because $g_{\mathcal{D}^\ell}=2g_{\mathcal{D}}<1=g_{\rm trivial}$. Nontrivial exchange-symmetric endpoints with $g_{\rm IR}\leq 2g_{\mathcal{D}}$ remain compatible with monotonicity, so this constraint complements the operator-spectrum stability tests.

\subsection{Stability criteria and prior constraints}
\label{subsec:ssb_stability_constraints}

The ordered phase under consideration is the easy-axis direct-sum defect $\mathcal{D}^\ell$ in Eq.~\eqref{eq:anti_direct_sum}, whose two components $\mathcal{D}^\pm$ are the antipodal pinning-field defects along the chosen easy axis. Its residual Ising-like exchange symmetry $\mathcal{D}^+\leftrightarrow\mathcal{D}^-$ is encoded by the nontrivial dimension-zero component-label operator $s^z=P_+-P_-$, which diagnoses defect long-range order.

Stability within this symmetry class requires two operator-spectrum conditions. First, the exchange-even domain-wall perturbation $\hat\phi^{+-}+\hat\phi^{-+}$ must be irrelevant, which requires $\Delta_{\hat\phi^{+-}}>1$. Second, each component must contain no relevant operator allowed by the easy-axis, residual exchange, and preserved $O(N-1)$ symmetries. The $g$-function supplies the separate RG-endpoint constraint derived in Sec.~\ref{subsec:g_function_theory}.

Existing results provide useful benchmarks. In two-dimensional Ising boundary CFT, Oshikawa and Affleck found an exactly marginal boundary domain-wall operator with $\Delta_{\hat\phi^{+-}}=1$\cite{OSHIKAWA1997533}. For the magnetic pinning line in the three-dimensional Ising CFT, fuzzy-sphere calculations give $\Delta_{\hat\phi^{+-}}\simeq0.84<1$ and $g_{\mathcal{D}}\simeq0.60$\cite{Zhou_2024}. Cusp and endpoint-bootstrap analyses are consistent with the resulting domain-wall instability\cite{Cuomo:2024psk,lanzetta2025beginningendpointbootstrapconformal}. Thus the relevant domain-wall perturbation destabilizes the Ising direct sum, and $g_{\mathcal{D}}>1/2$ makes the trivial line an allowed direct RG endpoint under $g$-monotonicity.

Two considerations motivate extending the analysis to higher $N$. In the free-scalar limit, integrating out the bulk field produces an effective interaction along the pinning line with power $d-2$\cite{pirsa_PIRSA_23110068}. This connects the defect problem to one-dimensional systems with slowly decaying interactions, which can support an ordered phase\cite{Dyson:1968up,Anderson_PhysRevB.1.4464}. Large-$N$ and $\epsilon$-expansion analyses further indicate that $\Delta_{\phi^{+-}}$ can increase with $N$, while leaving the finite-$N$ threshold undetermined\cite{Cuomo:2024psk,Cuomo_2022}. The finite-$N$ question is therefore whether $\Delta_{\hat\phi^{+-}}$ exceeds $1$ along the three-dimensional $O(N)$ Wilson--Fisher sequence.

The numerical analysis implements these diagnostics directly. The mixed Hilbert space $\mathcal H^{+-}$ determines $\Delta_{\hat\phi^{+-}}$, the same-component spectrum in $\mathcal H^{++}$ tests for additional symmetry-allowed relevant perturbations, and wavefunction overlaps determine $g_{\mathcal{D}}$\cite{Zhou_2024}. These quantities probe, respectively, the domain-wall stability, same-component stability, and allowed direct RG endpoints of $\mathcal{D}^\ell$.


\section{Numerical Results}

\subsection{Fuzzy Sphere Model}
Following the general construction of the $O(N)$ Wilson-Fisher models on the fuzzy sphere outlined in Ref. \cite{PhysRevX.13.021009,he2025freerealscalarcft,guo2025onfreescalarwilsonfisherconformal}, we consider $N+1$-flavor fermions $\psi_a$ as the fundamental building blocks. Here, the component with $a=0$  serves as a reference direction, while the remaining $N$ components \(a=1, \dots, N\) span the $O(N)$ vector space. Within this framework, the vector field transforming in the fundamental $O(N)$ representation can be defined as
\begin{equation}
    V_a(x) = \psi_a^\dagger \psi_0 + \psi_0^\dagger \psi_a,
\end{equation}
which generates the rotation from the reference $0$ direction to the $a$-th direction. This operator plays the role of the order parameter when analyzing the phase transition propertities. Without loss of generality, we choose the polarization direction to be $a = N$ in the subsequent parts of this paper. That is, we introduce the following pinning-field perturbation to the bulk critical Hamiltonian:
\begin{equation}
    \begin{small}
    H_d=H_{\text{wf}}+h_N V_N(\theta=0)+ h_SV_N(\theta=\pi)
    \end{small}
\end{equation}
Here, for bulk theory, we adopt the same Hamiltonian $H_{\text{wf}}$ as in Ref.\cite{guo2025onfreescalarwilsonfisherconformal}, to ensure that the system flows to the Wilson-Fisher CFT in the infrared limit.
Regarding the defect part, $h_{S(N)}$ is the pinning field strength at the south (north) pole on the fuzzy sphere. 
From the perspective of radial quantization, the insertion of a straight line defect in flat Euclidean space $\mathbb{R}^d$ maps to a pair of antipodal line defects on $S^{d-1} \times \mathbb{R}$(see Fig.\ref{fig:sch_plot}(a) for illustration), extending along the time direction  \cite{Hu_2024}. From this viewpoint, 
we thus add $(0+1)$-D point like defects at the north pole and south pole of the quantum fuzzy sphere. This construction is the fuzzy-sphere realization of a straight line defect in flat space. 
In the following, we will show numerical evidence to examine that the pinning field $h_{S(N)}$ introduces a relevant perturbation, which drives the system toward a nontrivial defect conformal fixed point. 
Say, $h_S=h_N$ corresponds to $\mathcal H^{\pm \pm}$ in Eq. \eqref{eq:defectHilbert} while $h_S=-h_N$ relates to $\mathcal H^{+-(-+)}$.
Next we denote the strength of pinning field as $h_d=|h_S|=|h_N|$. 

\subsection{Defect conformal fixed points}

\begin{figure}[t] 
    \centering
    \includegraphics[width=0.45\textwidth]{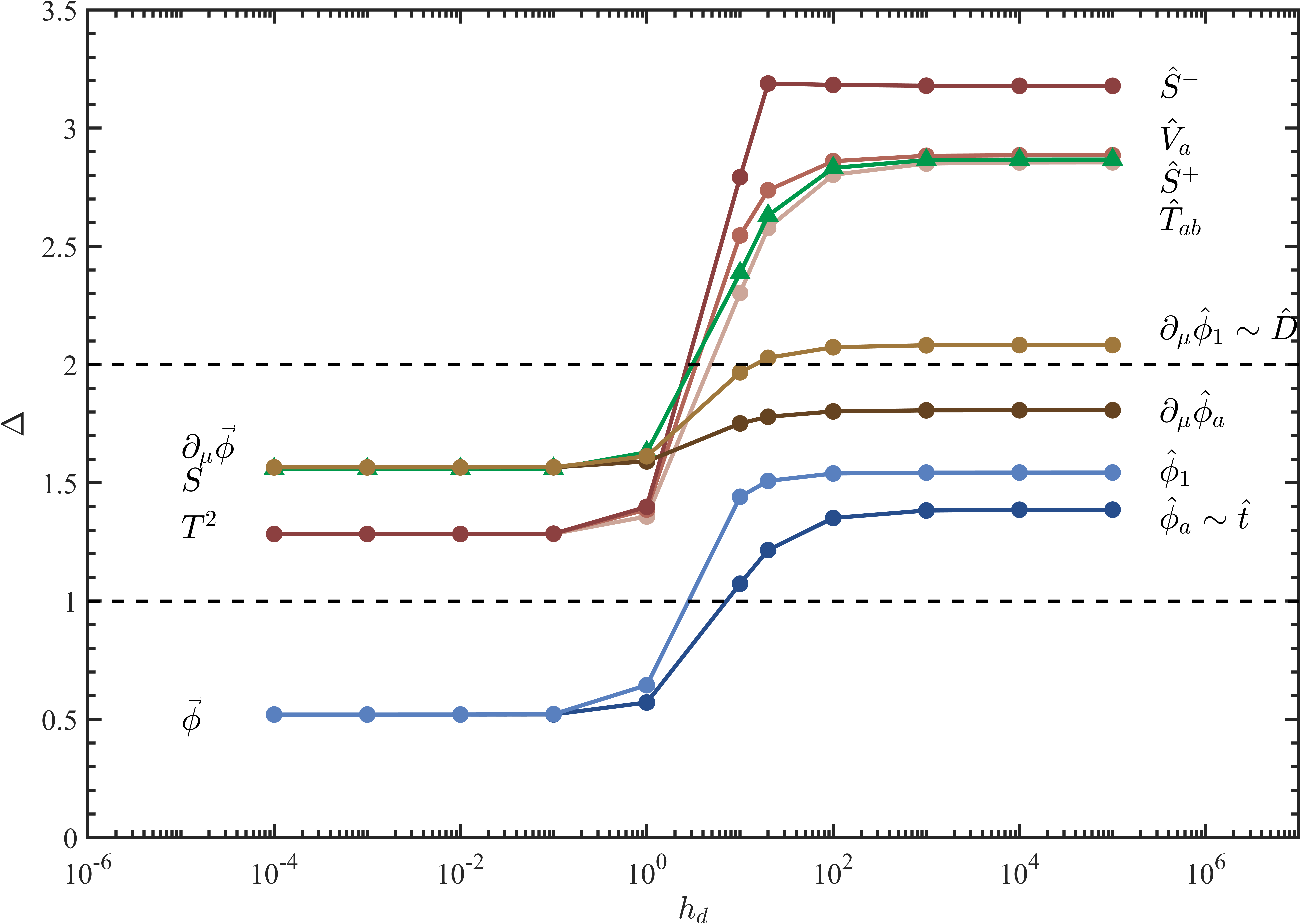} 
    \caption{Evolution of several low-lying excitation states along the RG flow from the bulk fixed point$(h_
d=0)$ to the defect fixed point$(h_
d\to \infty)$. For $h_d\ge100$, the extracted scaling dimensions approach stable values, indicating that the system is already close to the defect conformal fixed point. The results shown here are obtained for the O(3) Wilson–Fisher bulk fixed point on a system with size $N_o=6$.
}
    \label{fig:O3_wf_op_flow}
\end{figure}

We begin with the operator-level evidence for defect-driven line defect conformal fixed point. 
We extract scaling dimensions of defect operators according to Eq. \ref{eq:defect_dimension_extraction}.
First, we take the line defect immersed in the critical $O(3)$ model as an example to demonstrate how several low-energy operators in the bulk evolve when approaching the defect fixed point within the finite-size energy spectrum. 
In Fig. \ref{fig:O3_wf_op_flow}, by increasing the defect strength $h_d$, the scaling dimensions of operators evolve into a set of values different from the bulk values. The operator spectra
become independent of the strength of defect $h_d$ when  $h_d\gtrsim 1$, showing the fixed point at large $h_d$ limit is attractive.

Another observation in Fig. \ref{fig:O3_wf_op_flow} is the bulk operators split into different defect operators. 
In specific, the evolution of operators in the Fig.\ref{fig:O3_wf_op_flow} can be essentially understood as a consequence of explicit symmetry breaking: originally degenerate states within different $L_z$ sectors split according to the branching rule $SO(3) \supset SO(2)$, and degenerate states within different $S_z$ subspaces split according to the branching rule of $SO(N)\supset SO(N-1)$. We briefly summarize as follows (see Supple. Mat. \cite{sm} for details):
\begin{enumerate}
    \item The $O(N)$ vector $\boldsymbol{\vec{\phi}}$ in the bulk spectrum splits into the $O(N-1)$ vector operator $\hat{\phi}_a(a=2,\cdots,N)$ and the singlet operator $\hat{\phi}_1$. Both remain in the scalar representation under the reduced rotational symmetry group. Here, $\hat{\phi}_1$ is associated with the defect one-point function, while $\hat{\phi}_a$ corresponds to the tilt operator $\hat t$   in the dCFT(see below).
    
\item The bulk operator $\partial_\mu \boldsymbol{\vec{\phi}}$ in $L_z = 1$ sector splits into $\partial_\mu \hat{\phi}_1$ and $\partial_\mu \hat{\phi}_a$. $\partial_\mu \hat{\phi}_a$ operator corresponds to the displacement operator $\hat D$  in the dCFT (see below).

\item  After symmetry breaking, the original rank-2 traceless symmetric  tensor operator $T$ in the $O(N)$ bulk spectrum splits into: an $O(N-1)$ vector operator $\hat{V}_a$, a new traceless symmetric rank-2 tensor operator $\hat{T}_{ab}$, and a singlet operator $\hat{S}^+$.

\item The original $O(N)$ singlet operator $S$ corresponds to the $\hat{S}^-$ singlet in the defect spectrum.
\end{enumerate}

Second, a characteristic feature of a defect fixed point is the appearance of protected operators associated with broken symmetries. The line defect explicitly breaks both spatial $SO(3)$ rotation symmetry and the global $O(N$) symmetry. Correspondingly, two universal defect operators are expected to appear:
(1) the displacement operator $\hat{D}$, which describes local deformation transverse to the defect and (2) the tilt operator $\hat{t}$ which generates local rotations of $O(N-1)$ vector in the internal symmetry space. Since both operators descend from conserved currents, their scaling dimensions are protected and fixed entirely by the defect dimension $p=1$: $\Delta_{\hat{t}} = p$ (Eq. \eqref{eq:t_op}) and $\Delta_{\hat{D}} = p+1$ (Eq. \eqref{eq:D_op}). The appearance of these protected operators with the expected scaling dimensions provides a nontrivial check of the emergent infrared defect fixed point.

\begin{figure}[b] 
    \centering
    \includegraphics[width=0.37\textwidth]{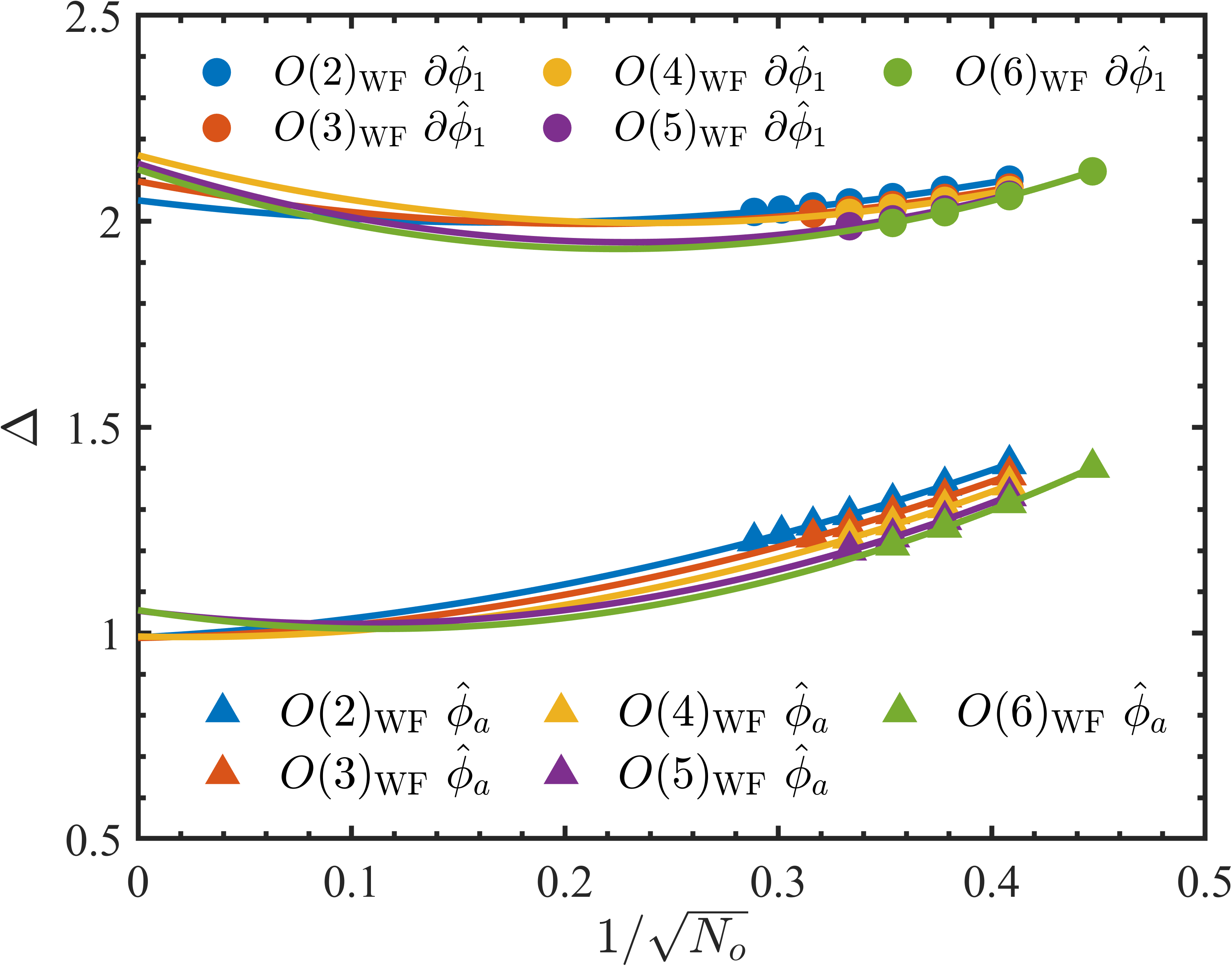} 
    \includegraphics[width=0.37\textwidth]{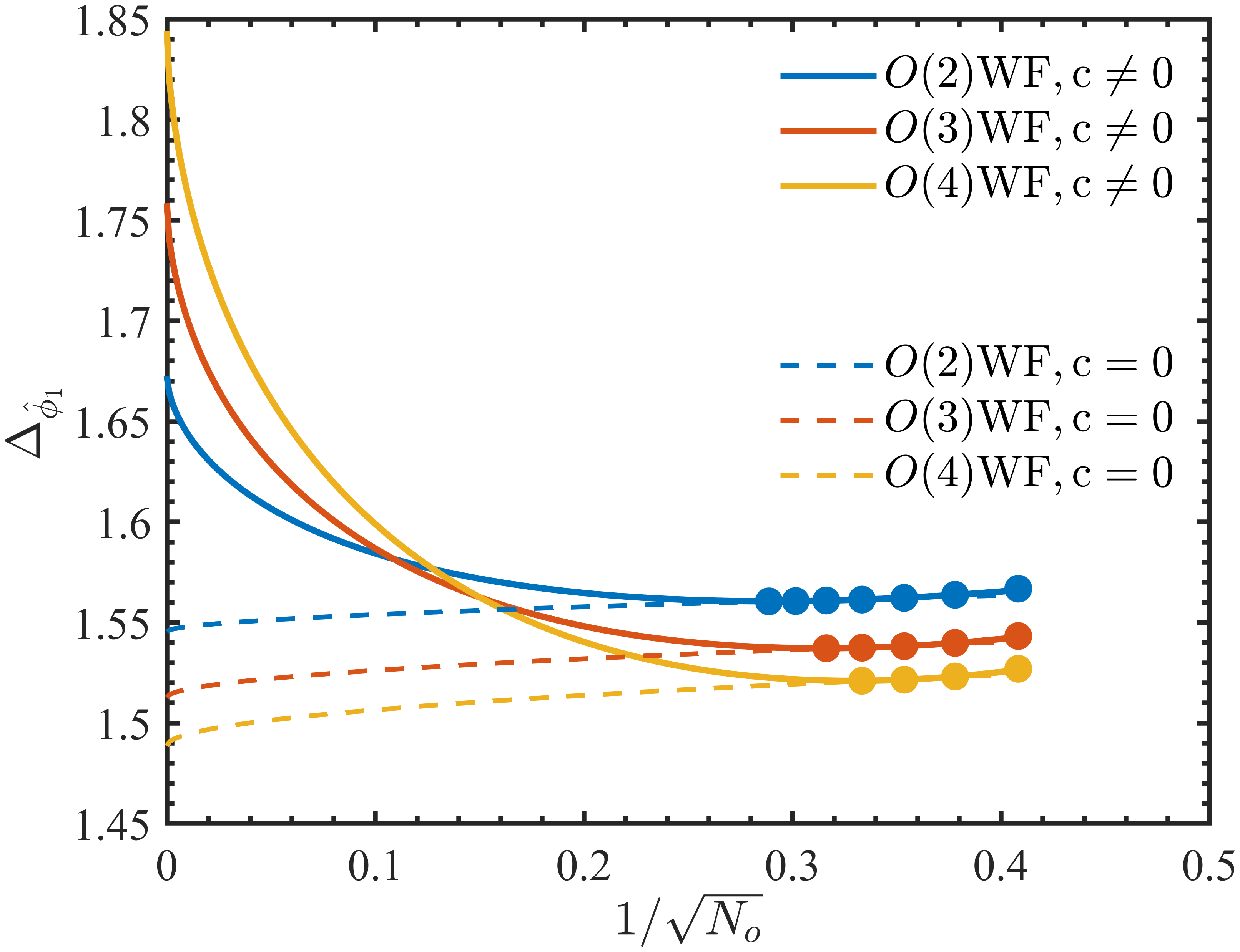}
    \caption{The extrapolation for (a) tilt operator $\hat{\phi}_a$, displacement operator ${\partial\hat{\phi}_1}$ at different bulk $O(N=2\sim 6)$ Wilson Fisher conformal field theories, and (b) $\hat{\phi}_1$ at different bulk $O(N=2\sim 4)$ Wilson Fisher conformal field theories. We use Eq. (\ref{eq:finite_size_consistent_scaling}) to obtain the extrapolated values. Using the same data, we also present the fitting results(dashed lines) obtained by including only the correction from the lowest-lying singlet operator(i.e. $c=0$). 
    } 
    \label{fig:disp_tilt_phi1_scaling}
\end{figure}

In Fig. \ref{fig:disp_tilt_phi1_scaling}(a), we demonstrate the identified tilt operator $\hat t\sim \hat\phi_a$ and displacement operator $\hat D\sim \partial \hat\phi_1$, for various system sizes.
Their dimensions on given system sizes are largely close to the expected values. To cue the finite-size effect, we perform a finite-size extrapolation to get their thermodynamic-limit values. We take into account that the defect spectrum mainly corrected by the two symmetric irrelevant scalar operators $\hat{\phi}_1,\ \hat{S}^-$ in the $L_z=0$ sector. Accordingly, for the scaling dimension of an operator $\hat{O}$, we consider the following finite-size scaling form
\begin{equation}
    \Delta_{\hat{O}}(N)=
\Delta_{\hat{O}}(\infty)
+
\frac{b}{R^{\Delta_{\hat{\phi}_1}-1}}
+
\frac{c}{R^{\Delta_{\hat{S}^-}-1}}.
\label{eq:finite_size_consistent_scaling}
\end{equation}
After performing a finite-size extrapolation, the extrapolated values are closer to the theoretical expectations: $\Delta_{\hat{\phi}_a}=1, \Delta_{\partial \hat{\phi}_1}=2$, for all universality classes $N=2-6$ (see Fig. \ref{fig:disp_tilt_phi1_scaling}(a)). 
The appearance of tilt and displacement operator confirms the emergence of a defect fixed point driven by the line defect.

Third, a stable attractive fixed point requires that no relevant operator exists at this fixed point. The lowest operator we identified is $\hat \phi_1$ living in $L_z=0$. We show the finite-size data of $\hat \phi_1$ in Fig. \ref{fig:disp_tilt_phi1_scaling}(b), and the extrapolated data in Tab. \ref{tab:delta_phi1_ON}. We ensure this operator is irrelevant.
As a comparison, the large-N calculation gives the leading nontrivial $O(N-1)$-singlet defect operator with the dimension around
$\Delta_{\hat{\phi}_1}\approx 1.542$ in $d=3$ \cite{Cuomo_2022}. Our results confirm that the lowest primary at the defect fixed point is irrelevant. 

\begin{figure*}[t] 
    \centering
\includegraphics[width=0.16\textwidth]{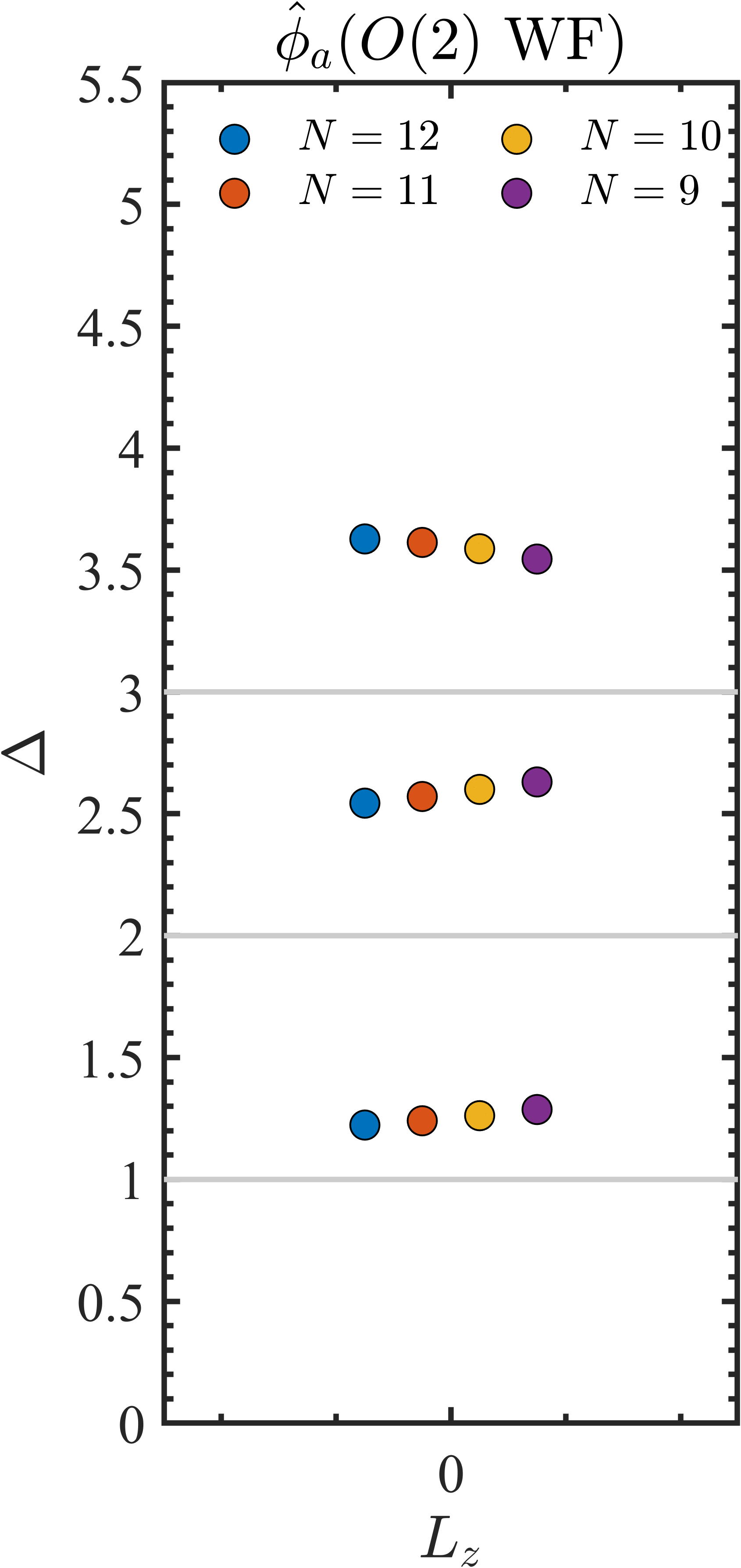} 
\includegraphics[width=0.16\textwidth]{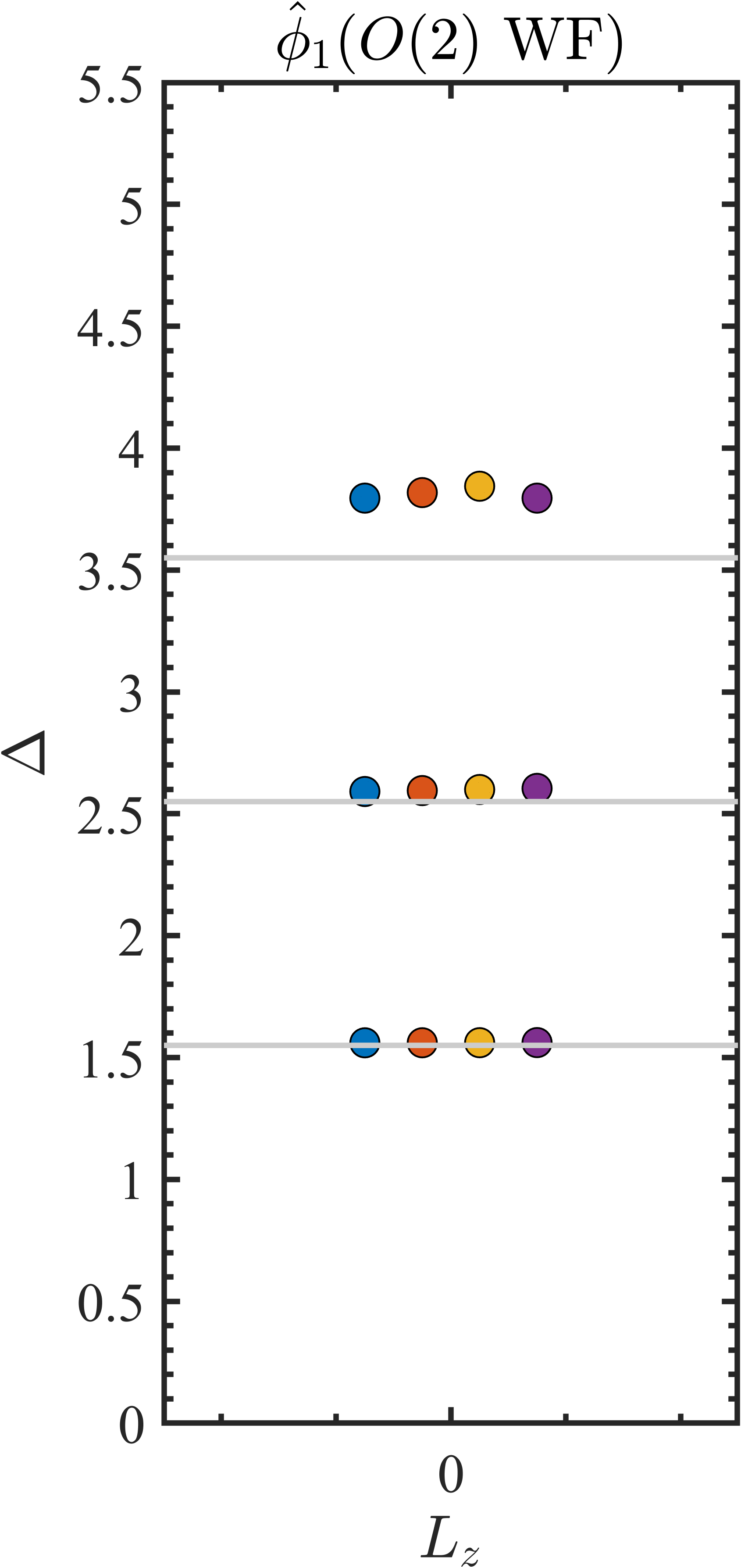}
\includegraphics[width=0.16\textwidth]{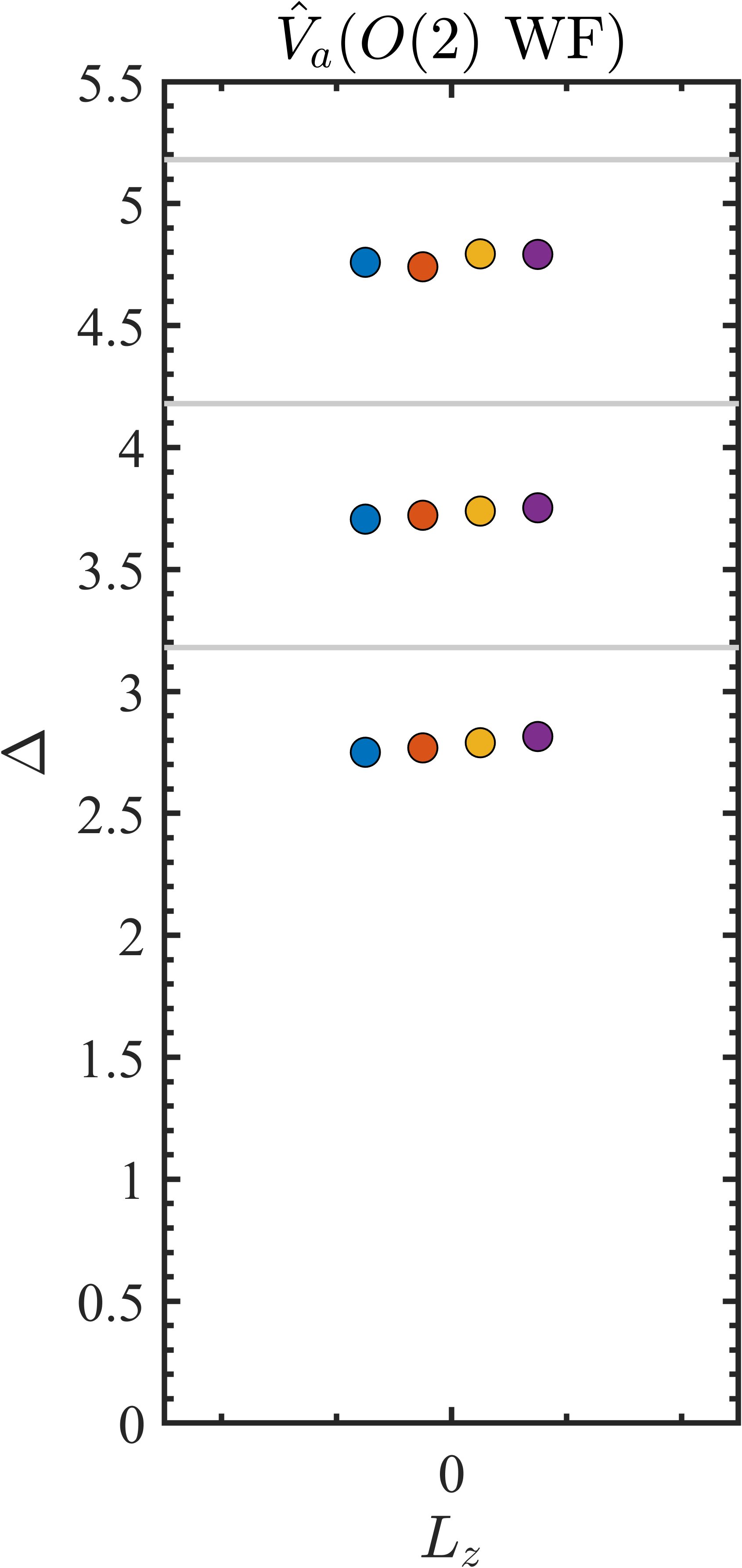}
\includegraphics[width=0.16\textwidth]{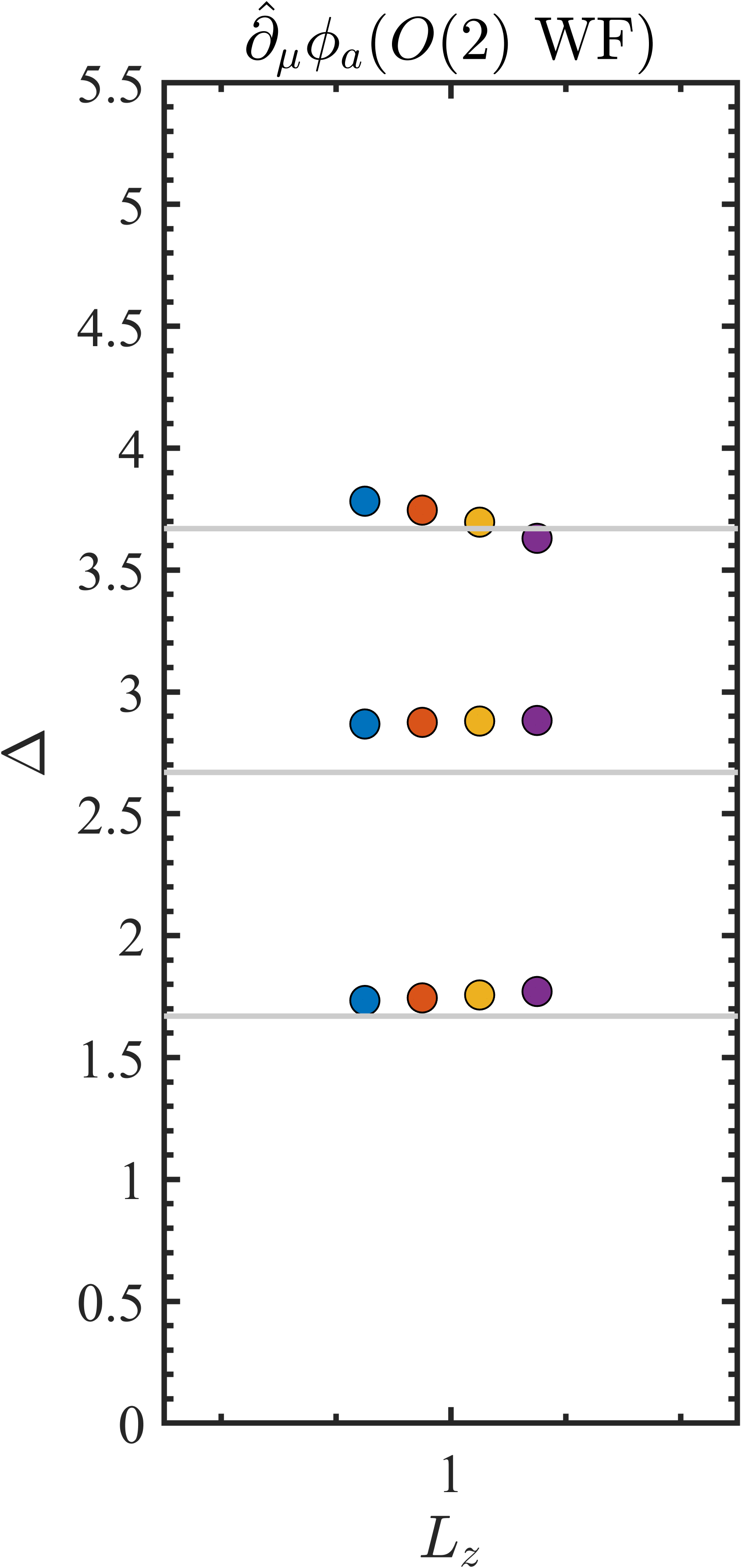}\includegraphics[width=0.16\textwidth]{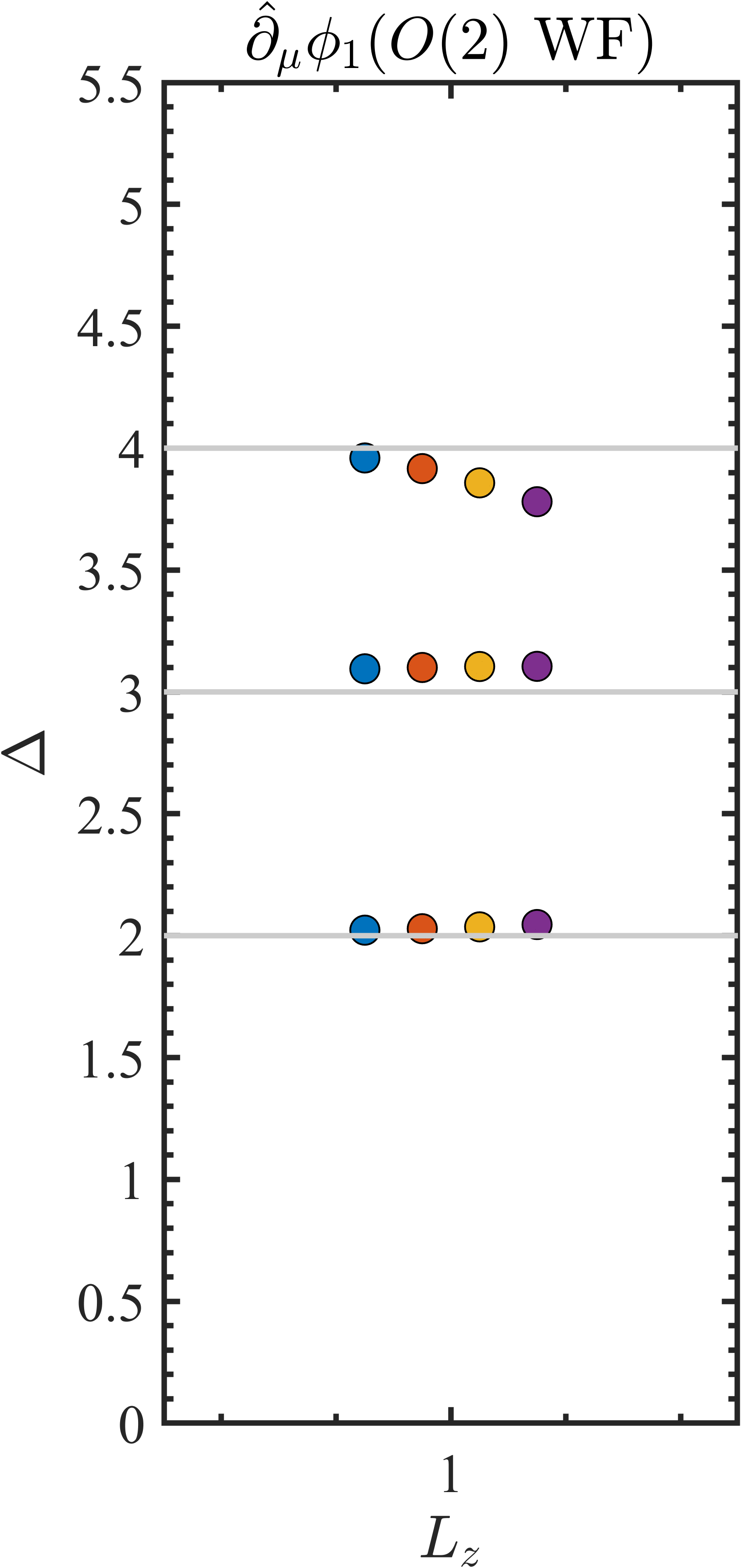}\\
\includegraphics[width=0.16\textwidth]{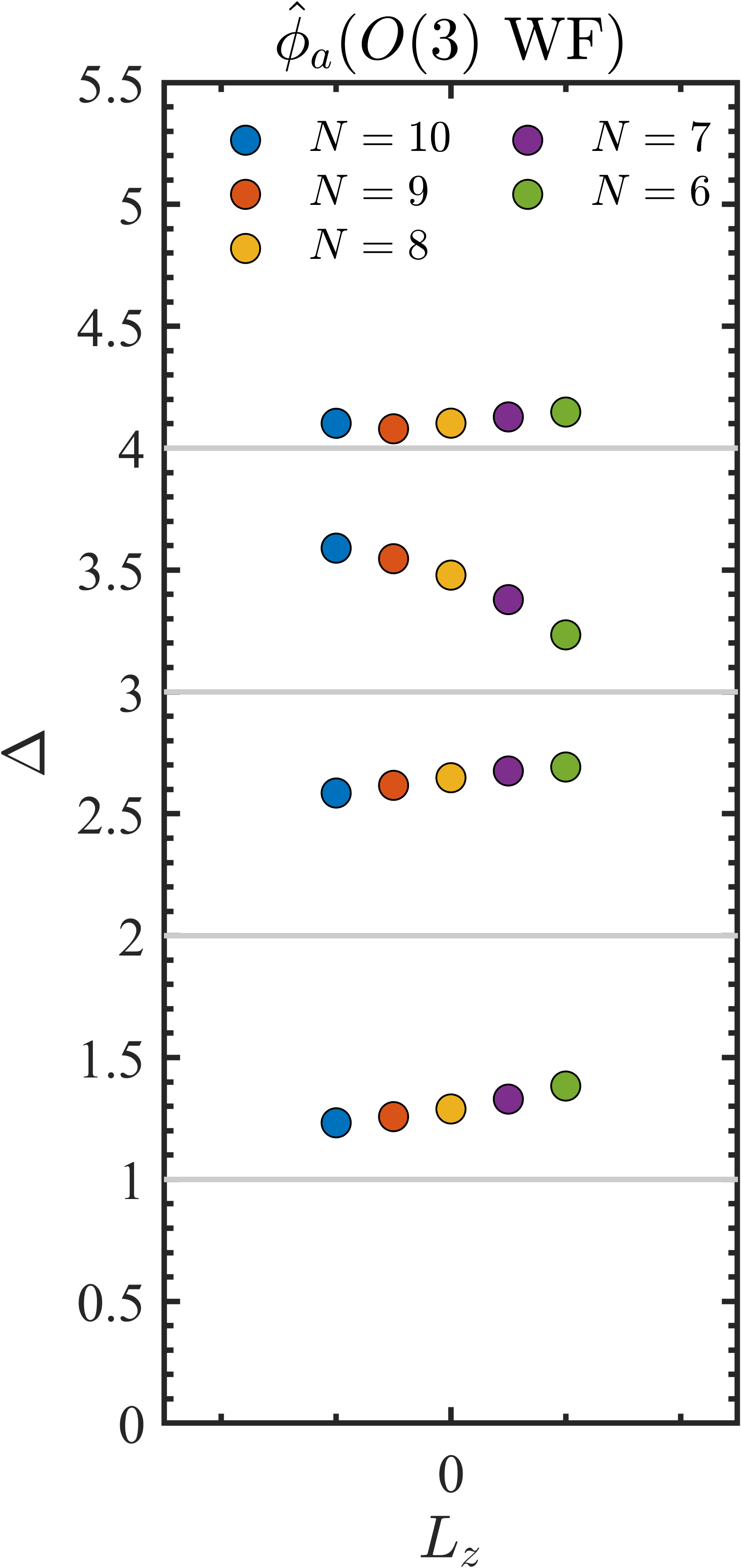} 
\includegraphics[width=0.16\textwidth]{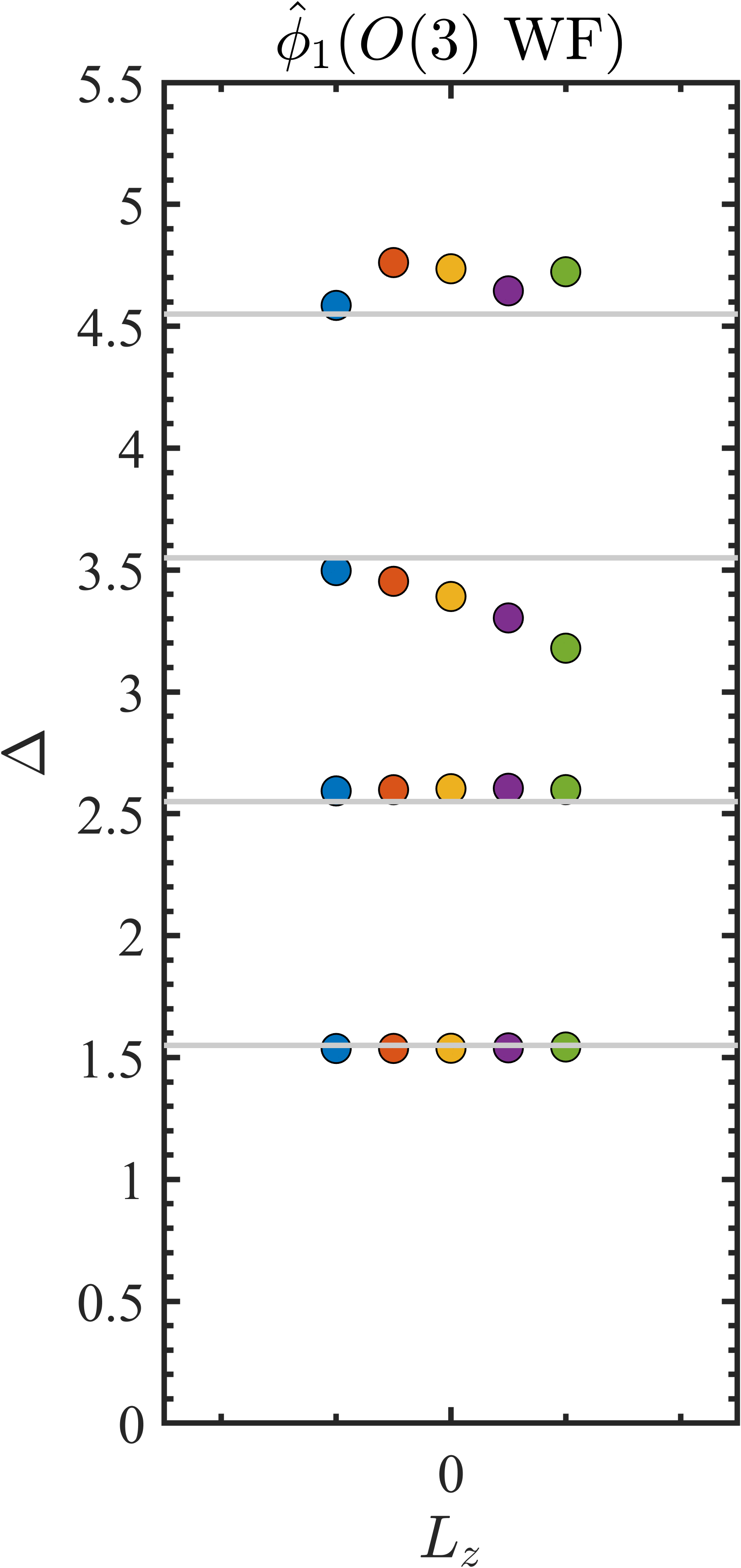}
\includegraphics[width=0.16\textwidth]{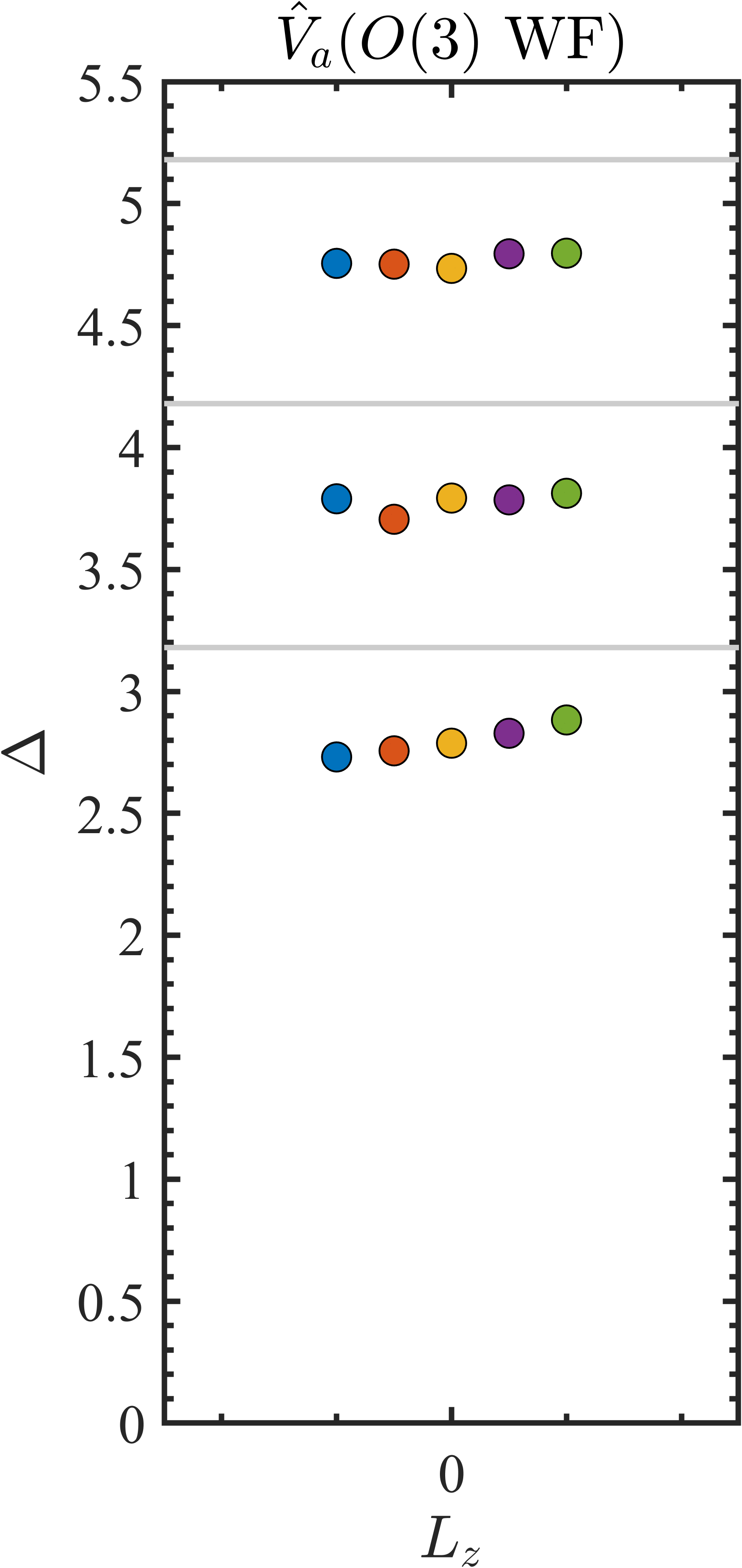}
\includegraphics[width=0.16\textwidth]{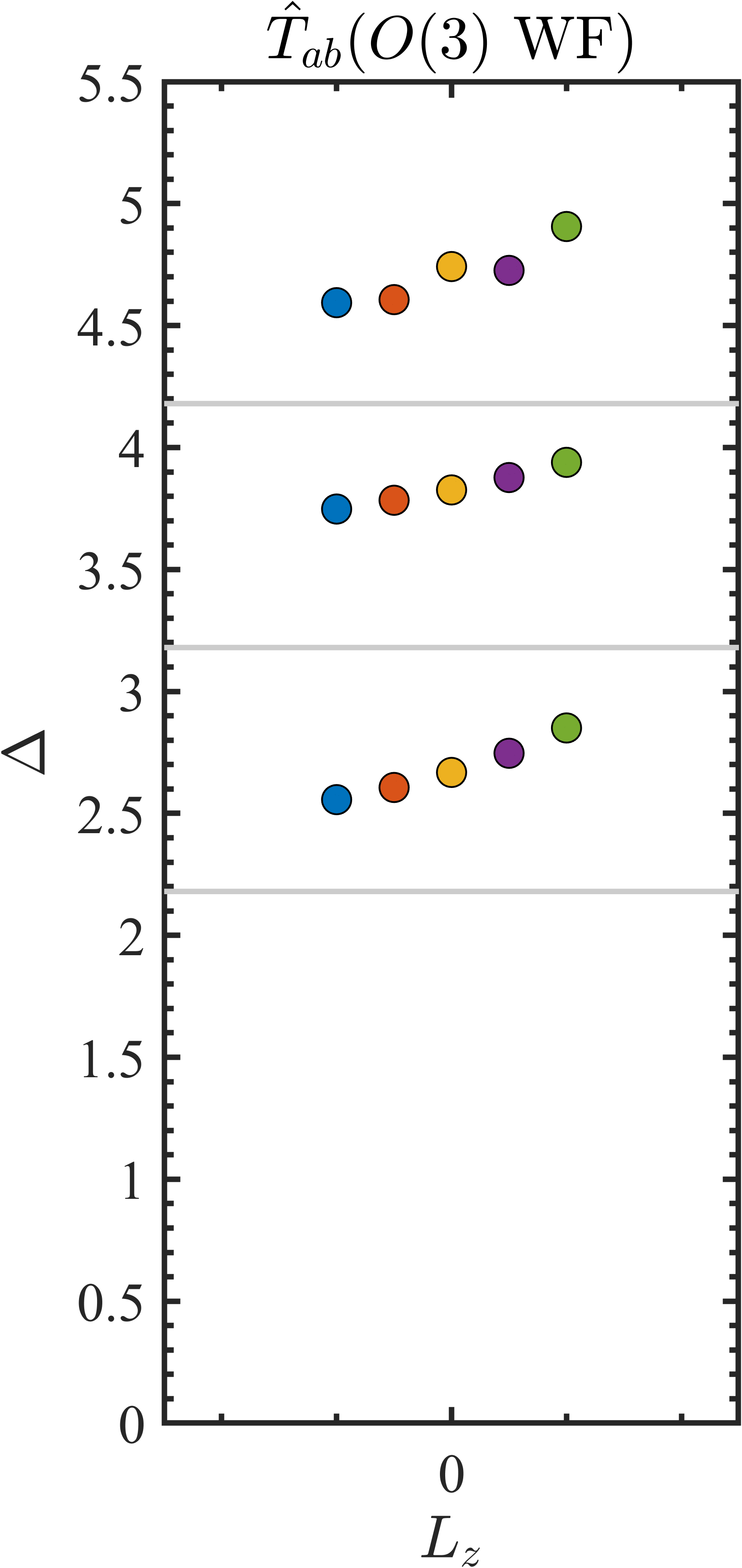}
\includegraphics[width=0.16\textwidth]{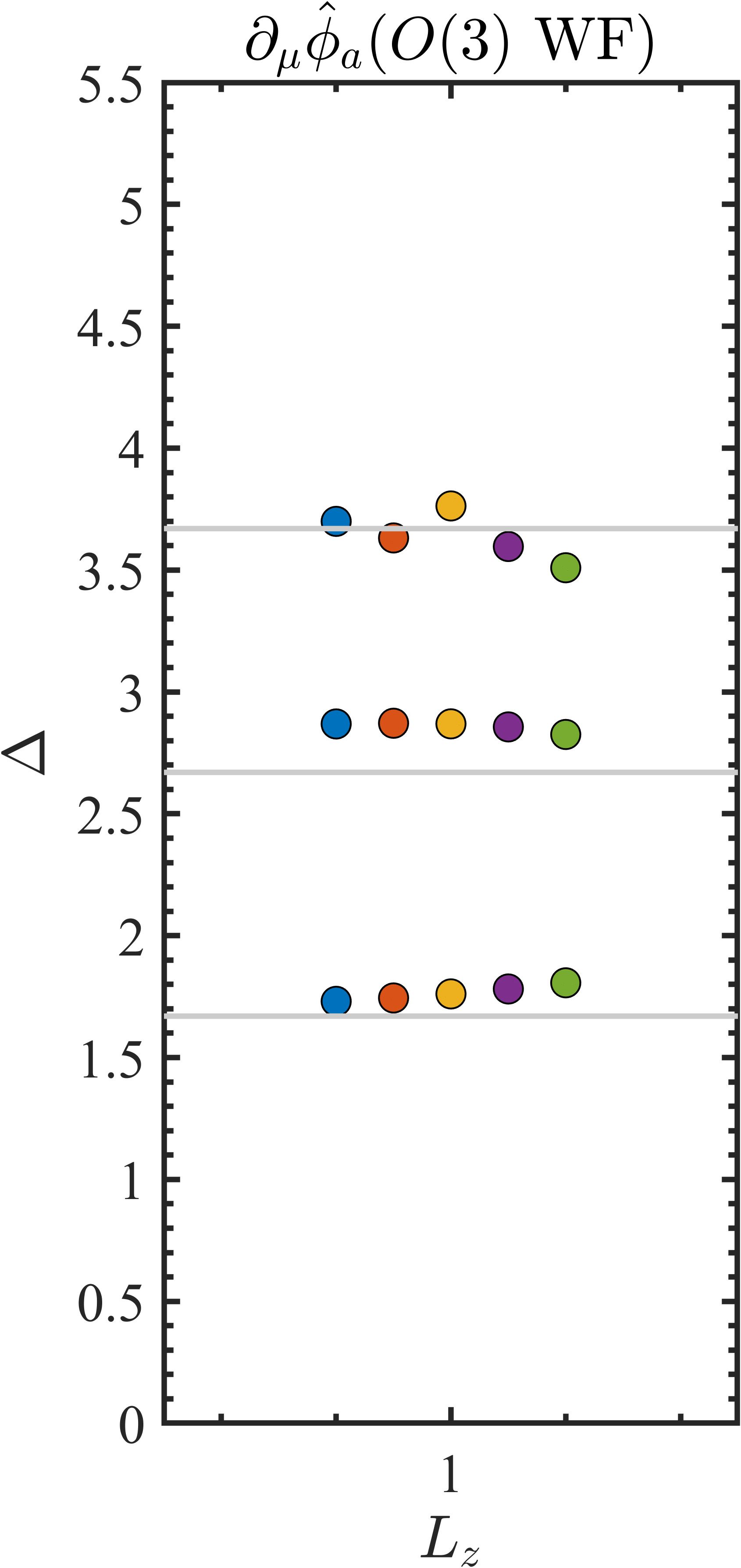}\includegraphics[width=0.16\textwidth]{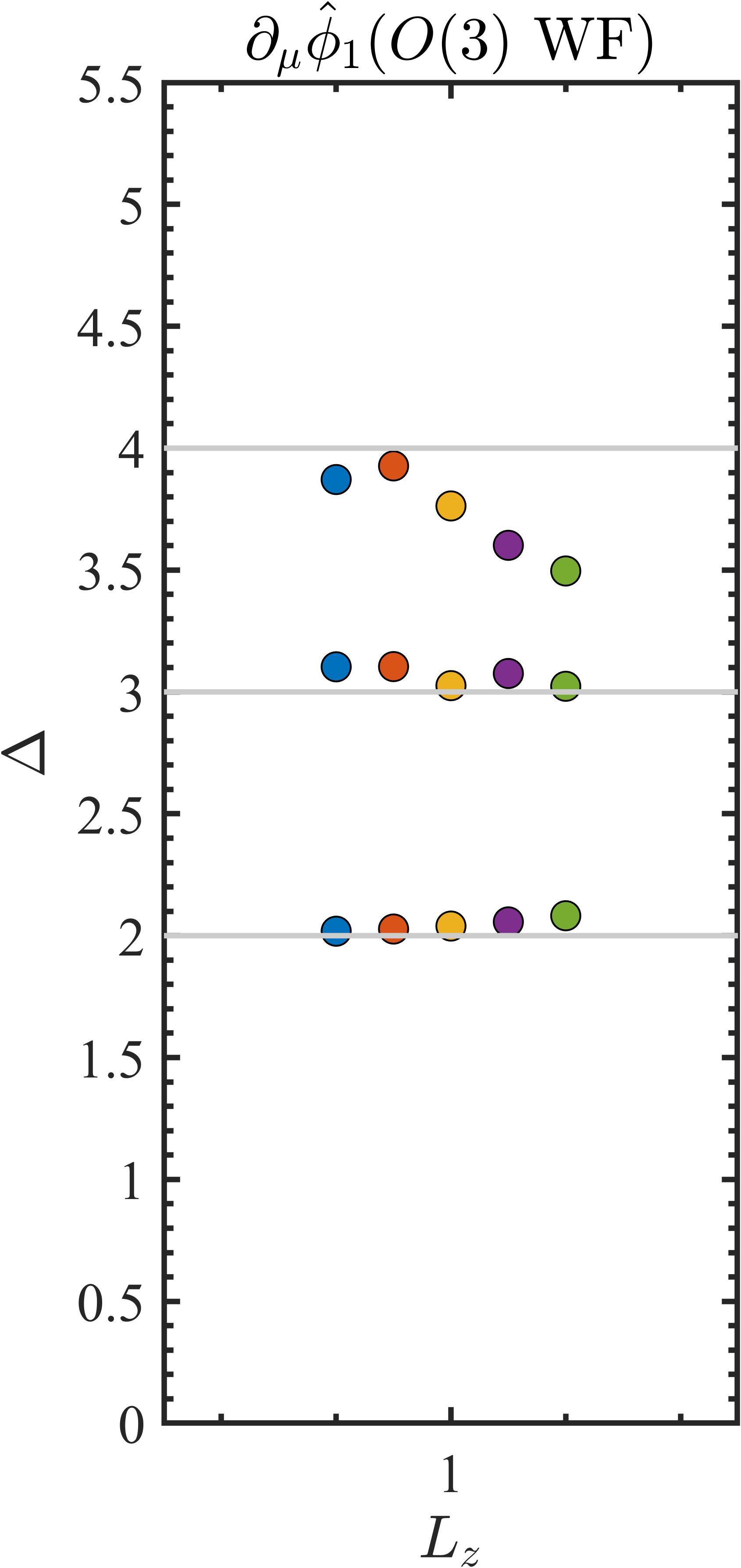}\\
    \caption{Line defect conformal multiplets, we present several lowest $O(N-1)$ singlets $\hat{\phi}_1,\ \partial_\mu\hat\phi_1$, vectors $\hat{\phi}_a,\ \hat{V}_a,\ \partial_\mu\hat\phi_a$ and rank-2 tensor $\hat{T}_{ab}$. The gray horizontal line is taken from the result of the $\epsilon$ expansion \cite{Cuomo_2022,Gimenez_Grau_2022,girault2026consequencessymmetrybreakingconformaldefect}.  These results are produced at $h_d=1000$ with subject to critical $O(N=2,3)$ bulk theories.
    }
    \label{fig:O3_wf_multi}
\end{figure*}

\begin{table}[t]
\centering
\caption{Scaling dimension $\Delta_{\hat{\phi}_1}(\infty)$ estimated by the Fuzzy Sphere model  for different bulk $O(N)$ symmetry. The finite-size extrapolation is performed based on Eq. \eqref{eq:finite_size_consistent_scaling}, and the error bar is estimated by using $b,c\neq 0$ and $b\neq0,c=0$. The analytic results are taken from Ref.\cite{Cuomo_2022}.
}
\label{tab:delta_phi1_ON}

\begin{tabular}{cccc}
\hline
\hline
  &$N=2$ & $N=3$ & $N=4$  \\
 \hline
$\text{Fuzzy Sphere }\Delta_{\hat{\phi}_1}(\infty)$  & $1.609(11)$ & $1.635(58)$ & $1.666(38)$ \\
$\text{large-N}$ &1.542 &1.542 &1.542 \\
$\epsilon-\text{expansion}$ & $1.55$ & $1.55$ & $1.55$ \\
\hline
\hline
\end{tabular}
\end{table}

Fourth, in Fig. \ref{fig:O3_wf_multi}, we show the low-energy operator spectrum, which implies the emergent conformal symmetry from the multiplets of primary operators. Most of them exhibit integer spacing in good agreement with the predictions of conformal symmetry. However, some states still suffer from significant finite-size effects. 
For comparison, we also show the $\epsilon$-expansion results (grey horizontal lines) \cite{Cuomo_2022,Gimenez_Grau_2022,girault2026consequencessymmetrybreakingconformaldefect}. We observe that the lowest conformal multiplets $\hat \phi_1$ and displacement operator fit $\epsilon$-expansion results quite well. For the vector and tensor fields, such as $\hat V_a, \hat T_{ab}$, an obvious discrepancy is found. 


\subsection{Defect changing operator}

Next we turn to explore the question whether or not SSB fixed point dubbed as $\mathcal{D}^+\oplus \mathcal{D}^- $ (see Eq. \eqref{eq:SSBfixedpoint}) on a one-dimensional defect is stable. The underlying mechanism is that the infinite correlation length of the bulk may effectively mediate long-range interactions on the defect. Such long-range interactions can potentially stabilize SSB even in one-dimension.  
Recently, for the line defect in 3D Ising CFT, studies using fuzzy sphere \cite{Zhou_2024} and conformal bootstrap \cite{lanzetta2025beginningendpointbootstrapconformal} methods  have estimated domain wall operator $\Delta_{\hat\phi^{+-}} \approx 0.8$. These values lie below the SSB threshold, indicating that SSB is not favored for $N=1$ case.
Large-$N$ calculation predicts that the scaling dimension domain wall operator $\Delta_{\hat\phi^{+-}}$ grows linearly with $N$ \cite{Cuomo:2024psk}.  This suggests that SSB could be allowed for sufficiently large $N$, but the critical value of $N_c$ has not been studied in microscopic study before.

Given this landscape, a central open question is to estimate the critical value $N_c$, where the SSB is allowed when $N>N_c$. 
To test it,  we probe the SSB possibility by testing a putative spontaneous symmetry breaking fixed point against the domain wall fluctuations, which would disrupt the potential long-range order. 
We construct a defect changing Hamiltonian $\mathcal{H}^{+-}$ by inserting two opposite pinning field defects $\mathcal{D}^+$ and $\mathcal{D}^-$ at the South and North Pole of fuzzy sphere. The leading primary of $\mathcal{H}^{+-}$ determines the dimension of defect changing operator $\hat\phi^{+-}$. The scaling dimension $\Delta_{\hat\phi^{+-}}$ of this domain wall operator will help to elucidate the possibility of SSB.

\begin{figure}[b] 
    \centering
    \includegraphics[width=0.65\linewidth]{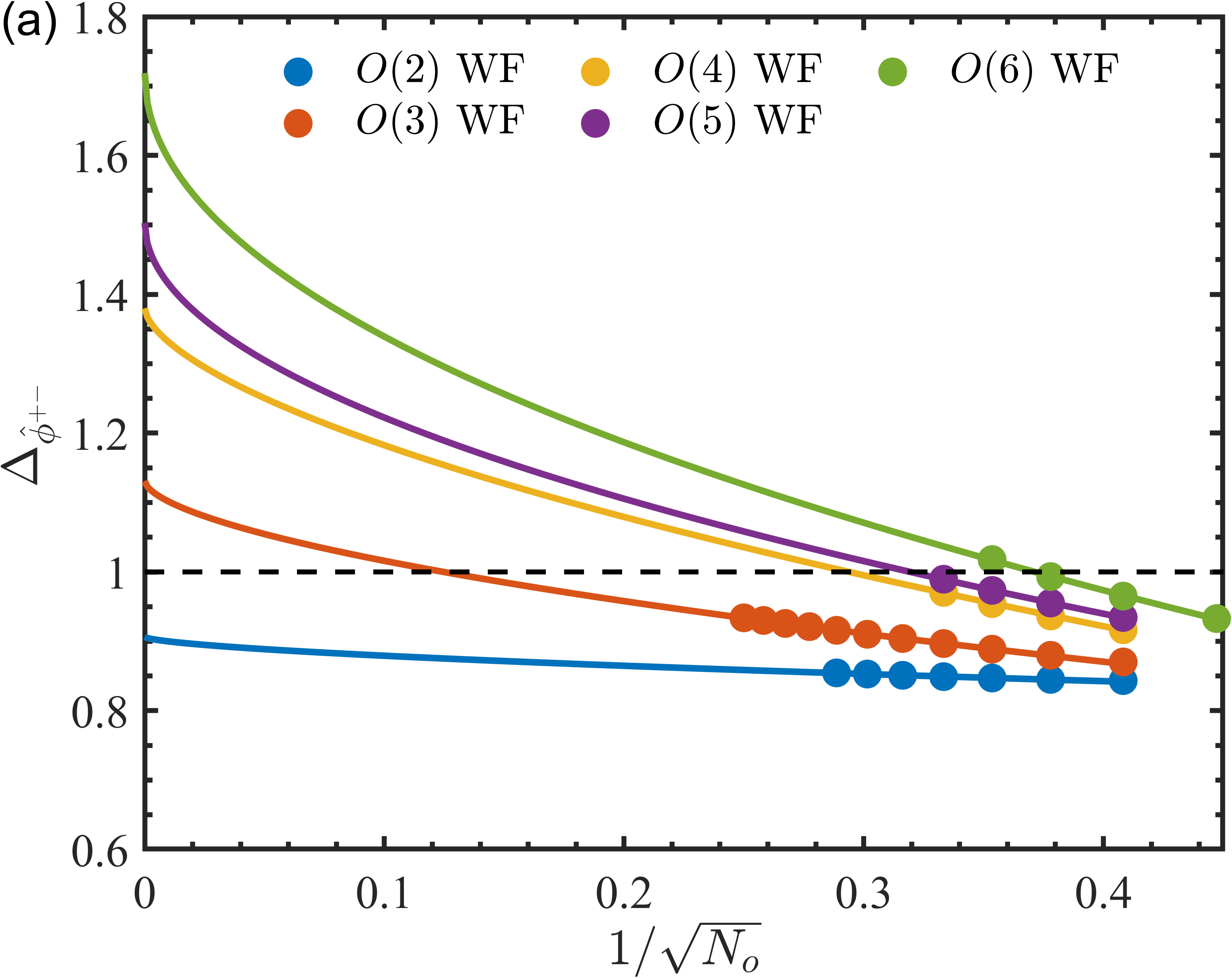} 
\includegraphics[width=0.65\linewidth]{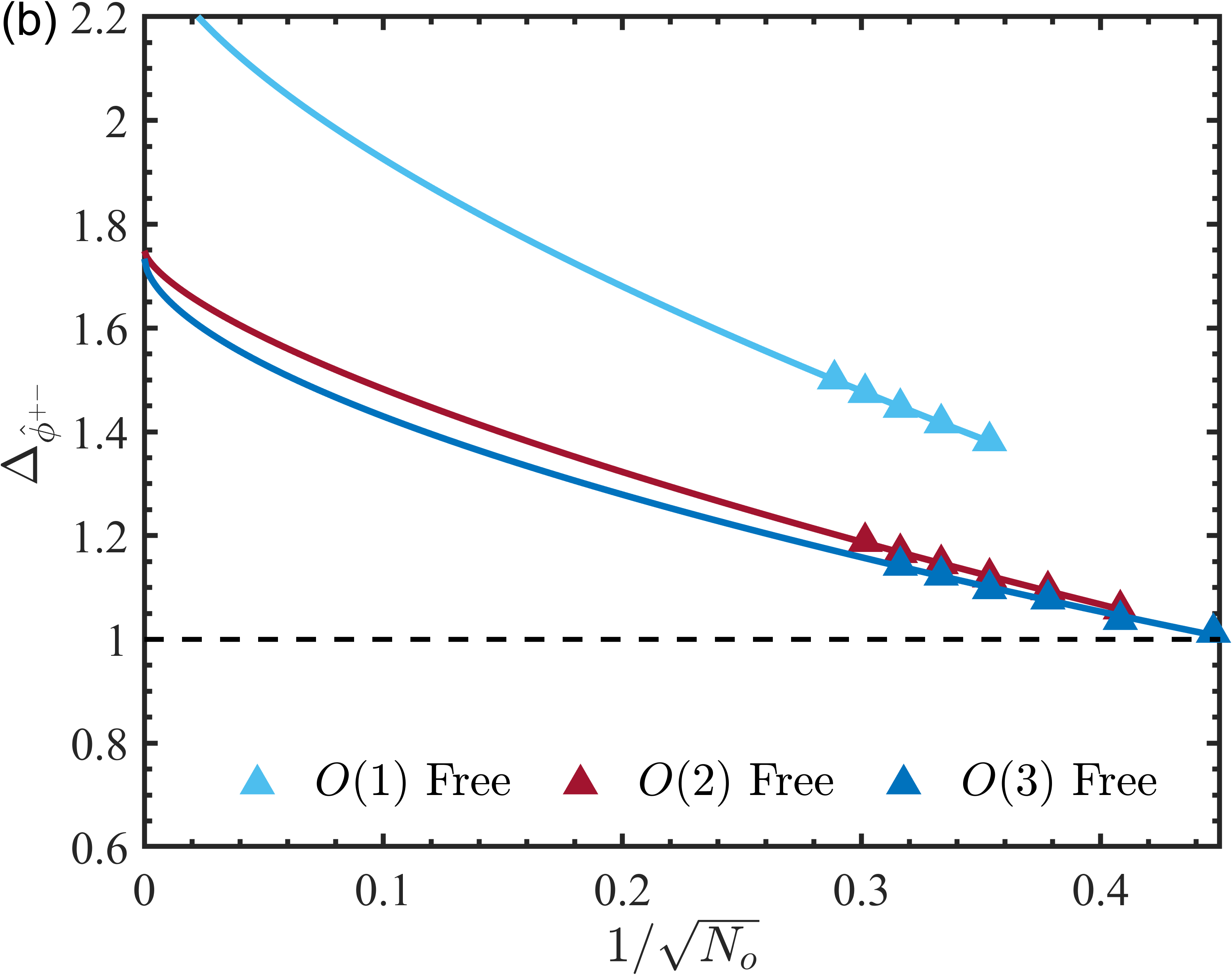}
    \caption{The extrapolation for scaling dimension of the defect changing operator at different bulk (a)$O(N=2\sim6)$ Wilson Fisher conformal field theories and (b) $O(N=1\sim3)$ free scalar theories. The scaling form is listed in Eq.\ref{eq:finite_size_consistent_scaling}.
    } 
\label{fig:defect_chaing_op_dim_extrapolate}
\end{figure}

In Fig.\ref{fig:defect_chaing_op_dim_extrapolate}, we present the extracted values of $\Delta_{\hat\phi^{+-}}$ from the energy spectrum of $\mathcal{H}^{+-}$. 
As a comparison, we calculate the case where the bulk is a massless free scalar theory (i.e., the Gaussian fixed point) with the same pinning-field defect configuration.
The obtained values of $\Delta_{\hat\phi^{+-}}$ clearly larger than 1 on all system sizes and drift upwards, demonstrating that domain wall operator is irrelevant. Thus, when domain wall fluctuation is irrelevant, the SSB is stable for the free scalar theory. This fully agrees well with the free scalar theory: For a pinning field defect in a free scalar CFT, integrating out the $d$-dimensional bulk theory induces long-range interactions with the power of $d-2$ on the line defect. This allows for the possibility of SSB on the defect when $d<4$\cite{pirsa_PIRSA_23110068}. 

For the Wilson-Fisher $O(N)$ theory, we observe  important features as we change the flavor number $N$. One the one hand, the dimension of domain wall operator monotonically increases as $N$. On the other hand,  
the domain wall operator evolves from relevance towards irrelevance as the system size increases. 
Using the extrapolation following Eq. \ref{eq:finite_size_consistent_scaling}, we determine the domain wall operator is relevant for $O(2)$ case. For $N\ge 4$ cases, we ensure finite-size extrapolated values become irrelevant. For $N=3$ case, although the finite-size data all below $1$, the extrapolated value of domain wall operator is scaled to the value larger than $1$. Based on these observations, we  estimates 
\begin{equation}
\begin{aligned}\label{eq:changingopNc}
        \Delta_{\hat\phi^{+-}}(N\le 2) \lesssim 1\lesssim \Delta_{\hat\phi^{+-}}(N\ge 3)
\end{aligned}
\end{equation}
Th relevance of domain wall operator gives an estimation of $2<N_c<3$.

\subsection{Defect $g$-function}
\begin{figure}[!htbp] 
    \centering
    \includegraphics[width=0.7\linewidth]{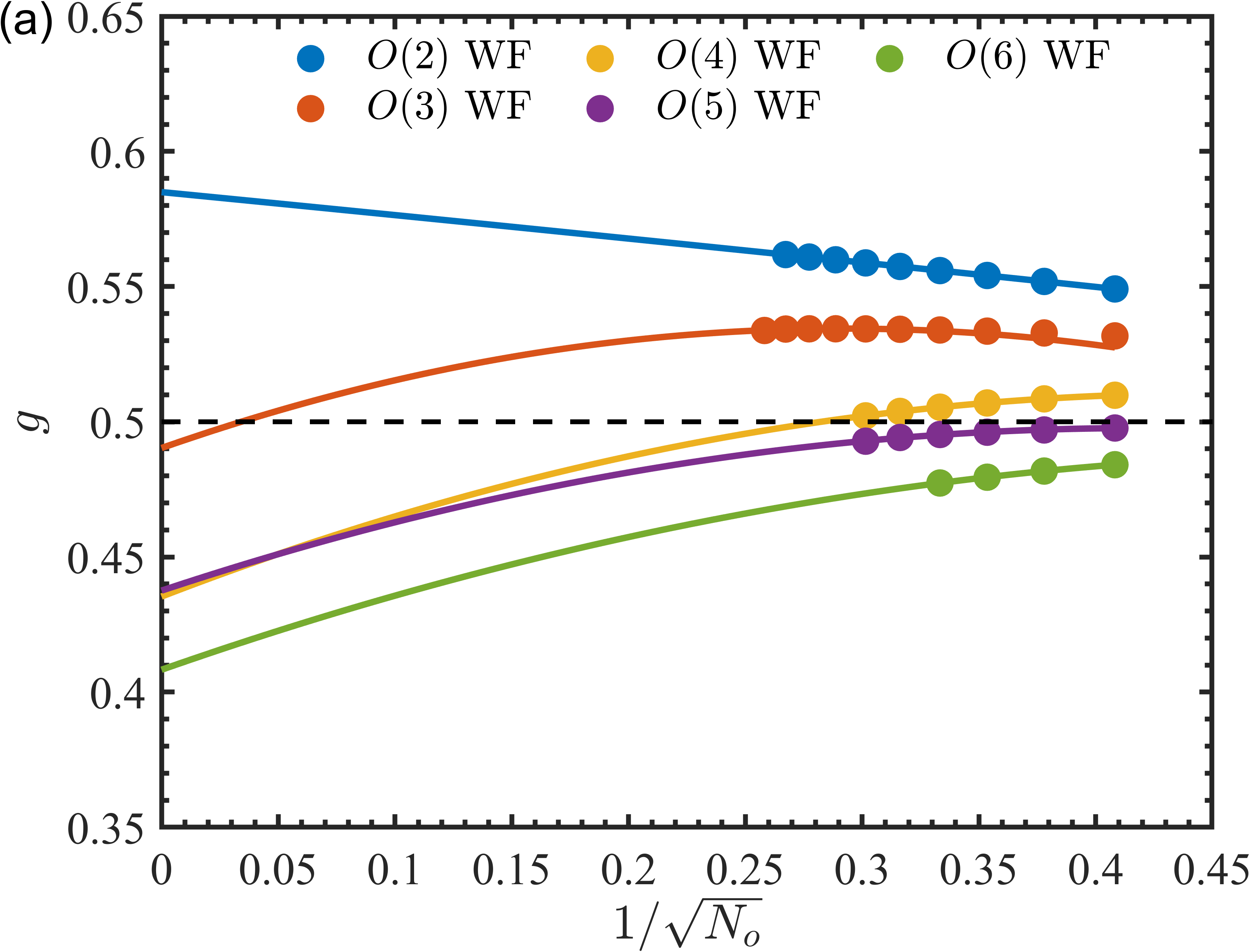}
    \includegraphics[width=0.7\linewidth]{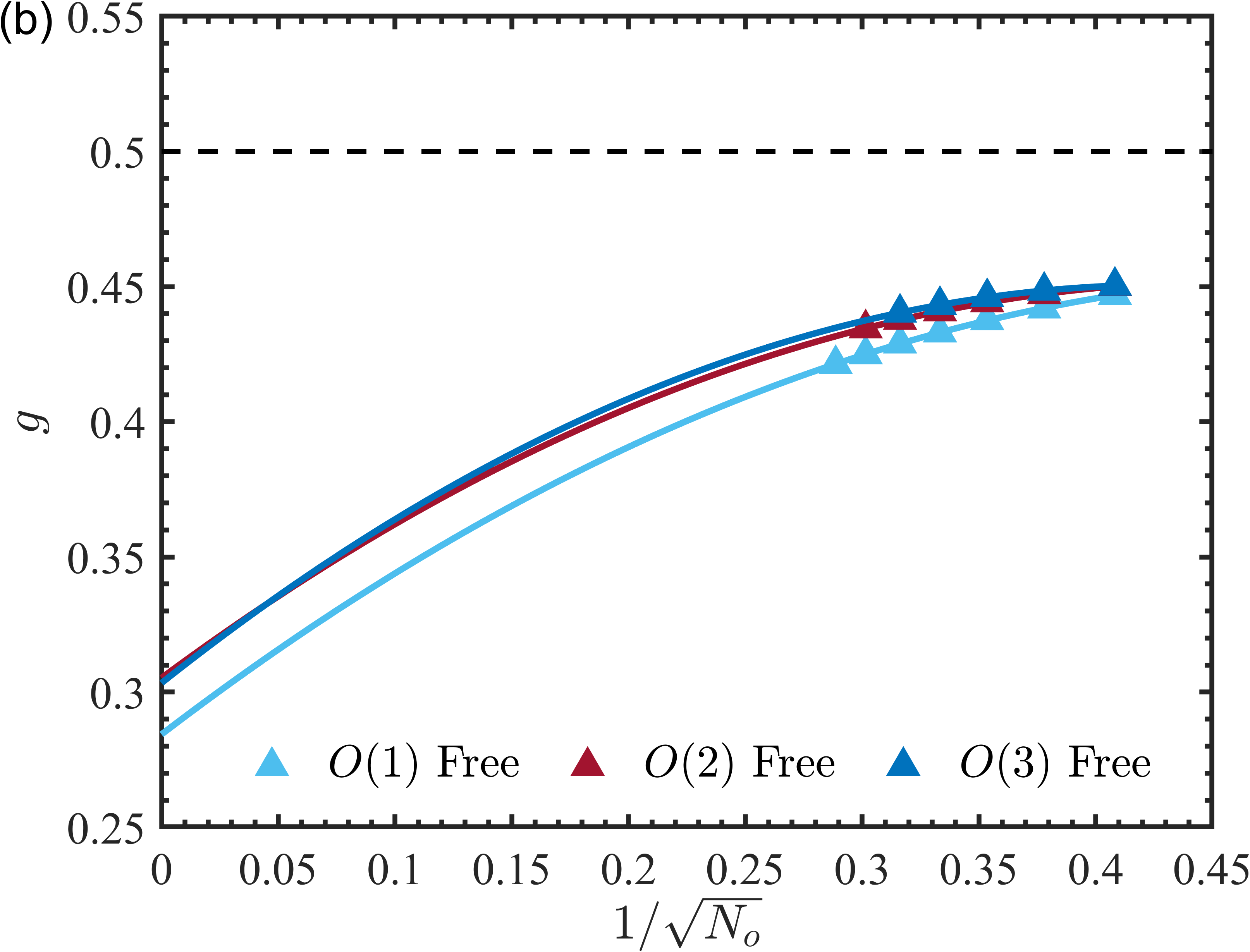}
    \caption{The finite size extrapolation of defect $g-$function at different bulk (a)$O(N=2\sim6)$ Wilson Fisher conformal field theories and (b) $O(N=1\sim3)$ free scalar theories. We extrapolate to the thermodynamic limit through scaling form of $g(N_o)=g(\infty)+aN_o^{-1/2}+bN_o^{-1}$. 
    }
    \label{fig:g_func_extrapolate_diff_ON_wf}
\end{figure}

We can also measure the defect $g$-function to investigate the effective degrees of freedom induced by the defect. This fundamental quantity is defined as the ratio between the partition function in the presence of a defect and that of the defect-free theory on $S^d$\cite{Affleck_Ludwig_g_theorem_PhysRevLett.67.161}. On the fuzzy sphere, this quantity can be extracted from the overlap of wavefunctions\cite{Zhou_2024}:
\begin{equation}
    g=\frac{|\langle+0|00\rangle|^2}{|\langle+0|++\rangle|^2}
    \label{eq:g_func_ovlap}
\end{equation}
where $|ab\rangle$ represents the lowest-energy state of the Hamiltonian by setting different pinning field parameters $h_N$ and $h_S$. 
The evolution of the defect $g$-function with system size is shown in the Fig.\ref{fig:g_func_extrapolate_diff_ON_wf}. These results correspond to a defect subjected to a series of interacting $O(N=2\sim 6)$ Wilson-Fisher CFTs and the non-interacting free scalar CFT. These results demonstrate two key points: 
1) defect $g$-function $g<g_{\text{bulk}}=1$ holds for all defect fixed points considered. According to the $g$-theorem \cite{Casini_2016_gTheorem,Cuomo_g_theorem__PhysRevLett.128.021603}, i.e. the value of the $g$-function decreases monotonically along the renormalization-group flow, it endorses a renormalization group flow from the bulk fixed point to the defect fixed point, supporting the phase diagram in Fig. \ref{fig:sch_plot}. 
2) The finite-size extrapolation of the $g$-function results in 
\begin{equation}
    g_{\mathcal{D}}(N=3)\approx 0.490,\ g_{\mathcal{D}}(N\ge 3)< 0.5.
\end{equation} 
Physically, this value is relevant to the discussion of the stability of $0-\pi$ domain wall, i.e. if $g_{\mathcal{D}^\ell}=g_{\mathcal{D}^+}+g_{\mathcal{D}^-}=2g_{\mathcal{D}}<1$  implies the spontaneous discrete symmetry broken on the defect line is possible\footnote{In this work, we only consider this particular scenario of spontaneous symmetry breaking. The condition $2g_{\mathcal{D}}<1$ indicates that the corresponding symmetry-broken fixed point $\mathcal{D}^+\oplus\mathcal{D}^-$ is stable. Other possible spontaneous symmetry-breaking patterns may lead to different bounds on the value of $g-$function.}. Our $g$-function calculations  point to a critical value of $2< N_c <3$, or  $g_{\mathcal{D}}(N<N_c)>0.5$ and $g_{\mathcal{D}}(N>N_c)<0.5$. This estimation is consistent with the scaling dimensions of the domain wall operators (Eq. \eqref{eq:changingopNc}), but differs from the existing large-N or $\epsilon$-expansion predictions (see Tab. \ref{tab:gfunc}).
Our results imply that spontaneous $\mathbb{Z}_2$ symmetry breaking on the defect line can occur for the line defect in the $O(N\ge 3)$ Wilson-Fisher bulk CFT.

\begin{table}[!htbp]
\centering
\caption{ The defect $g$-function obtained in the Fuzzy Sphere model for various $N=2,3,4$. The $\epsilon$-expansion and large-$N$ results are from Ref. \cite{Cuomo_2022}.
 }
\begin{tabular}{cccc}
\hline
\hline
 $g$ &$N=2$ & $N=3$ & $N=4$  \\
 \hline
Fuzzy Sphere  & $0.585$ & $0.490$ & $0.439$ \\
\hline
$\epsilon$-exp  & $0.535$ & $0.503$ & $0.472$ \\
\hline
large-N  & $0.735$ & $0.630$ & $0.541$ \\
\hline
\hline
\end{tabular} \label{tab:gfunc}
\end{table}


\section{summary and outlook}

With the help of fuzzy sphere simulation, we investigated the behavior of one-dimensional line defects coupled to three-dimensional bulk $O(N)$ Wilson–Fisher conformal field theory. 
We identify that a symmetry explicitly broken line defect always induces a stable defect conformal fixed point, with no relevant operator and $g$-function less than 1.
Moreover, we study the defect changing (domain wall) operator, and we confirm its scaling dimension monotonically increases with $N$. We estimate the critical value between $2<N_c<3$, where defect changing operator envolves from relevant  for $N<N_c$ to irrelevant for $N>N_c$. It is consistent with the $g$-function calculation. This implies that a discrete symmetry spontaneously broken could occur on a line defect embedded in a bulk Wilson-Fisher theory with $O(N\ge 3)$. To further test this possibility, it would be interesting to introduce a \(Z_2\)-symmetric line-defect perturbation in the \(O(N)\) Wilson–Fisher CFT and directly search for an infrared defect fixed point with spontaneous symmetry breaking.

The current work relates to the single pinning component $\mathcal D^{\pm}$ and $\pi$ domain wall $\mathcal D^+\oplus \mathcal{D}^- $ (see Eq.\eqref{eq:SSBfixedpoint}). Many related problems could be studied using the same method shown in this paper. For example, we further calculate the defect creation operator $\hat\phi^{0+}$ (see Eq. \ref{eq:defect_creation_dimension_extraction} Supple. Mat. \cite{sm}). More general $\theta$-domain wall operator can be defined, which describes the interface made of two pinning components with an angle difference $\theta$ (see Eq. \eqref{eq:phase-slip-defect}). For example, the information of $2\pi/3$ domain wall operator  $\Delta^{2\pi/3}$ could elucidate the stability of $\mathbb Z_3$ symmetry spontaneously broken in the line defect setup.

A natural direction for future work is to investigate a closely related question: can a surface defect in a three-dimensional conformal field theory spontaneously break a continuous internal symmetry\cite{Max_boundary_10.21468/SciPostPhys.12.4.131,Krishnan_2023,cuomo2024spontaneoussymmetrybreakingsurface}? This question is directly connected to the emergence of the extraordinary-log universality class observed in the $O(N)$ model\cite{Hu_2021,Parisen_Toldin_2022,Sun_2023,Parisen_Toldin_2025}. A recent study based on the bilayer fuzzy sphere $O(3)$ model\cite{chaohanPhysRevB.110.115113} suggests that the corresponding critical value satisfies \(N_c>3\)\cite{feng2026studying3donsurface}. Following a similar strategy, it may be possible to determine the critical value \(N_c\) at which the extraordinary-log surface universality class terminates\cite{Padayasi_2022,sun2026analyticbootstraponboundary}.

Furthermore, even richer defect universality classes may emerge when both bulk topology and criticality are taken into account simultaneously. Systems whose bulk realizes a deconfined quantum critical point or a critical point separating topologically ordered phases provide ideal platforms for exploring novel defect universality class\cite{long_zhang_Fa_wang_PhysRevLett.118.087201,ding_chengxiang_PhysRevLett.120.235701,StefanWessel__PhysRevB.98.140403,shangliu_PhysRevB.104.104201,zhu_yanzhang_tkx5-kzhh,toldin2025extraordinarytransitionedgecorrelated,liu2025edgemodestopologicalmott,Ma_2022}. Given that several such bulk universality classes have already been successfully realized within the fuzzy sphere framework\cite{Voinea_2025,Zhou_2025,zhou2025chernsimonsmatterconformalfieldtheory,Zhou_2026,Voinea_2026,tang2025emergence3dsuperconformalising}, extending fuzzy sphere techniques to investigate their defect conformal field theories appears particularly promising. It's also interesting to study the defect conformal field theories in CFTs that containing fermionic operators\cite{BCFT_GNY_MC,Shen_2025_BCFT_TI_TSC,ge2025boundarycriticalitytwodimensionalcorrelated}.




\begin{acknowledgements}
    We are grateful for Zheng Zhou and Jiechao Feng for explaining the conformal perturbation theory. We also thank Yin-Chen He and Shang Liu for the fruitful discussion. We use FuzzifiED\cite{zhou2025fuzzifiedjuliapackage} and ITensors\cite{Fishman_2022} to perform numerical calculations. L.D.H. and W.Z. are supported by NSFC under No. 12474144. L.D.H. is supported by the Fundamental Research Funds for the Central Universities. S. Y. and Y. C. are supported by the National Key Research and Development Program of China Grant No. 2022YFA1402204 and the National Natural Science Foundation of China Grant No. 12274086. 
\end{acknowledgements}

\bibliography{CFT_Sphere.bib}

\begin{thebibliography}{85}%
\makeatletter
\providecommand \@ifxundefined [1]{%
 \@ifx{#1\undefined}
}%
\providecommand \@ifnum [1]{%
 \ifnum #1\expandafter \@firstoftwo
 \else \expandafter \@secondoftwo
 \fi
}%
\providecommand \@ifx [1]{%
 \ifx #1\expandafter \@firstoftwo
 \else \expandafter \@secondoftwo
 \fi
}%
\providecommand \natexlab [1]{#1}%
\providecommand \enquote  [1]{``#1''}%
\providecommand \bibnamefont  [1]{#1}%
\providecommand \bibfnamefont [1]{#1}%
\providecommand \citenamefont [1]{#1}%
\providecommand \href@noop [0]{\@secondoftwo}%
\providecommand \href [0]{\begingroup \@sanitize@url \@href}%
\providecommand \@href[1]{\@@startlink{#1}\@@href}%
\providecommand \@@href[1]{\endgroup#1\@@endlink}%
\providecommand \@sanitize@url [0]{\catcode `\\12\catcode `\$12\catcode `\&12\catcode `\#12\catcode `\^12\catcode `\_12\catcode `\%12\relax}%
\providecommand \@@startlink[1]{}%
\providecommand \@@endlink[0]{}%
\providecommand \url  [0]{\begingroup\@sanitize@url \@url }%
\providecommand \@url [1]{\endgroup\@href {#1}{\urlprefix }}%
\providecommand \urlprefix  [0]{URL }%
\providecommand \Eprint [0]{\href }%
\providecommand \doibase [0]{https://doi.org/}%
\providecommand \selectlanguage [0]{\@gobble}%
\providecommand \bibinfo  [0]{\@secondoftwo}%
\providecommand \bibfield  [0]{\@secondoftwo}%
\providecommand \translation [1]{[#1]}%
\providecommand \BibitemOpen [0]{}%
\providecommand \bibitemStop [0]{}%
\providecommand \bibitemNoStop [0]{.\EOS\space}%
\providecommand \EOS [0]{\spacefactor3000\relax}%
\providecommand \BibitemShut  [1]{\csname bibitem#1\endcsname}%
\let\auto@bib@innerbib\@empty
\bibitem [{\citenamefont {Mermin}\ and\ \citenamefont {Wagner}(1966)}]{Mermin_Wagner_PhysRevLett.17.1133}%
  \BibitemOpen
  \bibfield  {author} {\bibinfo {author} {\bibfnamefont {N.~D.}\ \bibnamefont {Mermin}}\ and\ \bibinfo {author} {\bibfnamefont {H.}~\bibnamefont {Wagner}},\ }\bibfield  {title} {\bibinfo {title} {Absence of ferromagnetism or antiferromagnetism in one- or two-dimensional isotropic heisenberg models},\ }\href {https://doi.org/10.1103/PhysRevLett.17.1133} {\bibfield  {journal} {\bibinfo  {journal} {Phys. Rev. Lett.}\ }\textbf {\bibinfo {volume} {17}},\ \bibinfo {pages} {1133} (\bibinfo {year} {1966})}\BibitemShut {NoStop}%
\bibitem [{\citenamefont {Hohenberg}(1967)}]{Hohenberg_PhysRev.158.383}%
  \BibitemOpen
  \bibfield  {author} {\bibinfo {author} {\bibfnamefont {P.~C.}\ \bibnamefont {Hohenberg}},\ }\bibfield  {title} {\bibinfo {title} {Existence of long-range order in one and two dimensions},\ }\href {https://doi.org/10.1103/PhysRev.158.383} {\bibfield  {journal} {\bibinfo  {journal} {Phys. Rev.}\ }\textbf {\bibinfo {volume} {158}},\ \bibinfo {pages} {383} (\bibinfo {year} {1967})}\BibitemShut {NoStop}%
\bibitem [{\citenamefont {Peierls}(1936)}]{Peierls_1936On}%
  \BibitemOpen
  \bibfield  {author} {\bibinfo {author} {\bibfnamefont {R.~E.}\ \bibnamefont {Peierls}},\ }\bibfield  {title} {\bibinfo {title} {On ising's model of ferromagnetism},\ }\href@noop {} {\bibfield  {journal} {\bibinfo  {journal} {Mathematical Proceedings of the Cambridge Philosophical Society}\ }\textbf {\bibinfo {volume} {32}},\ \bibinfo {pages} {477} (\bibinfo {year} {1936})}\BibitemShut {NoStop}%
\bibitem [{\citenamefont {Dyson}(1969)}]{Dyson:1968up}%
  \BibitemOpen
  \bibfield  {author} {\bibinfo {author} {\bibfnamefont {F.~J.}\ \bibnamefont {Dyson}},\ }\bibfield  {title} {\bibinfo {title} {{Existence of a phase transition in a one-dimensional Ising ferromagnet}},\ }\href {https://doi.org/10.1007/BF01645907} {\bibfield  {journal} {\bibinfo  {journal} {Commun. Math. Phys.}\ }\textbf {\bibinfo {volume} {12}},\ \bibinfo {pages} {91} (\bibinfo {year} {1969})}\BibitemShut {NoStop}%
\bibitem [{\citenamefont {Anderson}\ \emph {et~al.}(1970)\citenamefont {Anderson}, \citenamefont {Yuval},\ and\ \citenamefont {Hamann}}]{Anderson_PhysRevB.1.4464}%
  \BibitemOpen
  \bibfield  {author} {\bibinfo {author} {\bibfnamefont {P.~W.}\ \bibnamefont {Anderson}}, \bibinfo {author} {\bibfnamefont {G.}~\bibnamefont {Yuval}},\ and\ \bibinfo {author} {\bibfnamefont {D.~R.}\ \bibnamefont {Hamann}},\ }\bibfield  {title} {\bibinfo {title} {Exact results in the kondo problem. ii. scaling theory, qualitatively correct solution, and some new results on one-dimensional classical statistical models},\ }\href {https://doi.org/10.1103/PhysRevB.1.4464} {\bibfield  {journal} {\bibinfo  {journal} {Phys. Rev. B}\ }\textbf {\bibinfo {volume} {1}},\ \bibinfo {pages} {4464} (\bibinfo {year} {1970})}\BibitemShut {NoStop}%
\bibitem [{\citenamefont {Bray}\ and\ \citenamefont {Moore}(1977)}]{A.J.Bray_1977}%
  \BibitemOpen
  \bibfield  {author} {\bibinfo {author} {\bibfnamefont {A.~J.}\ \bibnamefont {Bray}}\ and\ \bibinfo {author} {\bibfnamefont {M.~A.}\ \bibnamefont {Moore}},\ }\bibfield  {title} {\bibinfo {title} {Critical behaviour of semi-infinite systems},\ }\href {https://doi.org/10.1088/0305-4470/10/11/021} {\bibfield  {journal} {\bibinfo  {journal} {Journal of Physics A: Mathematical and General}\ }\textbf {\bibinfo {volume} {10}},\ \bibinfo {pages} {1927} (\bibinfo {year} {1977})}\BibitemShut {NoStop}%
\bibitem [{\citenamefont {Diehl}(1997)}]{Diehl_1997}%
  \BibitemOpen
  \bibfield  {author} {\bibinfo {author} {\bibfnamefont {H.~W.}\ \bibnamefont {Diehl}},\ }\bibfield  {title} {\bibinfo {title} {The theory of boundary critical phenomena},\ }\href {https://doi.org/10.1142/s0217979297001751} {\bibfield  {journal} {\bibinfo  {journal} {International Journal of Modern Physics B}\ }\textbf {\bibinfo {volume} {11}},\ \bibinfo {pages} {3503–3523} (\bibinfo {year} {1997})}\BibitemShut {NoStop}%
\bibitem [{\citenamefont {Billò}\ \emph {et~al.}(2016)\citenamefont {Billò}, \citenamefont {Gonçalves}, \citenamefont {Lauria},\ and\ \citenamefont {Meineri}}]{Bill__2016}%
  \BibitemOpen
  \bibfield  {author} {\bibinfo {author} {\bibfnamefont {M.}~\bibnamefont {Billò}}, \bibinfo {author} {\bibfnamefont {V.}~\bibnamefont {Gonçalves}}, \bibinfo {author} {\bibfnamefont {E.}~\bibnamefont {Lauria}},\ and\ \bibinfo {author} {\bibfnamefont {M.}~\bibnamefont {Meineri}},\ }\bibfield  {title} {\bibinfo {title} {Defects in conformal field theory},\ }\href {https://doi.org/10.1007/jhep04(2016)091} {\bibfield  {journal} {\bibinfo  {journal} {Journal of High Energy Physics}\ }\textbf {\bibinfo {volume} {2016}},\ \bibinfo {pages} {1–56} (\bibinfo {year} {2016})}\BibitemShut {NoStop}%
\bibitem [{\citenamefont {C.P.Herzog}(2021)}]{LACES21_CFTdefects}%
  \BibitemOpen
  \bibfield  {author} {\bibinfo {author} {\bibnamefont {C.P.Herzog}},\ }\href {https://www.ggi.infn.it/laces/LACES21/CFTdefects21_CFTnotes.pdf} {\bibinfo {title} {Conformal field theory with boundaries and defects}} (\bibinfo {year} {2021}),\ \bibinfo {note} {lACES 2021 School on CFT and Defects}\BibitemShut {NoStop}%
\bibitem [{\citenamefont {Wilson}(1974)}]{Wilson_OP_PhysRevD.10.2445}%
  \BibitemOpen
  \bibfield  {author} {\bibinfo {author} {\bibfnamefont {K.~G.}\ \bibnamefont {Wilson}},\ }\bibfield  {title} {\bibinfo {title} {Confinement of quarks},\ }\href {https://doi.org/10.1103/PhysRevD.10.2445} {\bibfield  {journal} {\bibinfo  {journal} {Phys. Rev. D}\ }\textbf {\bibinfo {volume} {10}},\ \bibinfo {pages} {2445} (\bibinfo {year} {1974})}\BibitemShut {NoStop}%
\bibitem [{\citenamefont {{'t Hooft}}(1978)}]{THOOFT19781}%
  \BibitemOpen
  \bibfield  {author} {\bibinfo {author} {\bibfnamefont {G.}~\bibnamefont {{'t Hooft}}},\ }\bibfield  {title} {\bibinfo {title} {On the phase transition towards permanent quark confinement},\ }\href {https://doi.org/https://doi.org/10.1016/0550-3213(78)90153-0} {\bibfield  {journal} {\bibinfo  {journal} {Nuclear Physics B}\ }\textbf {\bibinfo {volume} {138}},\ \bibinfo {pages} {1} (\bibinfo {year} {1978})}\BibitemShut {NoStop}%
\bibitem [{\citenamefont {Kondo}(1964)}]{Kondo_paper10.1143/PTP.32.37}%
  \BibitemOpen
  \bibfield  {author} {\bibinfo {author} {\bibfnamefont {J.}~\bibnamefont {Kondo}},\ }\bibfield  {title} {\bibinfo {title} {Resistance minimum in dilute magnetic alloys},\ }\href {https://doi.org/10.1143/PTP.32.37} {\bibfield  {journal} {\bibinfo  {journal} {Progress of Theoretical Physics}\ }\textbf {\bibinfo {volume} {32}},\ \bibinfo {pages} {37} (\bibinfo {year} {1964})},\ \Eprint {https://arxiv.org/abs/https://academic.oup.com/ptp/article-pdf/32/1/37/5193092/32-1-37.pdf} {https://academic.oup.com/ptp/article-pdf/32/1/37/5193092/32-1-37.pdf} \BibitemShut {NoStop}%
\bibitem [{\citenamefont {Wilson}(1975)}]{Kondo_wilson_RevModPhys.47.773}%
  \BibitemOpen
  \bibfield  {author} {\bibinfo {author} {\bibfnamefont {K.~G.}\ \bibnamefont {Wilson}},\ }\bibfield  {title} {\bibinfo {title} {The renormalization group: Critical phenomena and the kondo problem},\ }\href {https://doi.org/10.1103/RevModPhys.47.773} {\bibfield  {journal} {\bibinfo  {journal} {Rev. Mod. Phys.}\ }\textbf {\bibinfo {volume} {47}},\ \bibinfo {pages} {773} (\bibinfo {year} {1975})}\BibitemShut {NoStop}%
\bibitem [{\citenamefont {Affleck}(1995)}]{affleck1995conformalfieldtheoryapproach}%
  \BibitemOpen
  \bibfield  {author} {\bibinfo {author} {\bibfnamefont {I.}~\bibnamefont {Affleck}},\ }\href {https://arxiv.org/abs/cond-mat/9512099} {\bibinfo {title} {Conformal field theory approach to the kondo effect}} (\bibinfo {year} {1995}),\ \Eprint {https://arxiv.org/abs/cond-mat/9512099} {arXiv:cond-mat/9512099 [cond-mat]} \BibitemShut {NoStop}%
\bibitem [{\citenamefont {Verresen}\ \emph {et~al.}(2021)\citenamefont {Verresen}, \citenamefont {Thorngren}, \citenamefont {Jones},\ and\ \citenamefont {Pollmann}}]{gSPT_Ruben_PhysRevX.11.041059}%
  \BibitemOpen
  \bibfield  {author} {\bibinfo {author} {\bibfnamefont {R.}~\bibnamefont {Verresen}}, \bibinfo {author} {\bibfnamefont {R.}~\bibnamefont {Thorngren}}, \bibinfo {author} {\bibfnamefont {N.~G.}\ \bibnamefont {Jones}},\ and\ \bibinfo {author} {\bibfnamefont {F.}~\bibnamefont {Pollmann}},\ }\bibfield  {title} {\bibinfo {title} {Gapless topological phases and symmetry-enriched quantum criticality},\ }\href {https://doi.org/10.1103/PhysRevX.11.041059} {\bibfield  {journal} {\bibinfo  {journal} {Phys. Rev. X}\ }\textbf {\bibinfo {volume} {11}},\ \bibinfo {pages} {041059} (\bibinfo {year} {2021})}\BibitemShut {NoStop}%
\bibitem [{\citenamefont {Scaffidi}\ \emph {et~al.}(2017)\citenamefont {Scaffidi}, \citenamefont {Parker},\ and\ \citenamefont {Vasseur}}]{gSPTs_PhysRevX.7.041048}%
  \BibitemOpen
  \bibfield  {author} {\bibinfo {author} {\bibfnamefont {T.}~\bibnamefont {Scaffidi}}, \bibinfo {author} {\bibfnamefont {D.~E.}\ \bibnamefont {Parker}},\ and\ \bibinfo {author} {\bibfnamefont {R.}~\bibnamefont {Vasseur}},\ }\bibfield  {title} {\bibinfo {title} {Gapless symmetry-protected topological order},\ }\href {https://doi.org/10.1103/PhysRevX.7.041048} {\bibfield  {journal} {\bibinfo  {journal} {Phys. Rev. X}\ }\textbf {\bibinfo {volume} {7}},\ \bibinfo {pages} {041048} (\bibinfo {year} {2017})}\BibitemShut {NoStop}%
\bibitem [{\citenamefont {Prembabu}\ \emph {et~al.}(2024)\citenamefont {Prembabu}, \citenamefont {Thorngren},\ and\ \citenamefont {Verresen}}]{Prembabu_2024}%
  \BibitemOpen
  \bibfield  {author} {\bibinfo {author} {\bibfnamefont {S.}~\bibnamefont {Prembabu}}, \bibinfo {author} {\bibfnamefont {R.}~\bibnamefont {Thorngren}},\ and\ \bibinfo {author} {\bibfnamefont {R.}~\bibnamefont {Verresen}},\ }\bibfield  {title} {\bibinfo {title} {Boundary-deconfined quantum criticality at transitions between symmetry-protected topological chains},\ }\bibfield  {journal} {\bibinfo  {journal} {Physical Review B}\ }\textbf {\bibinfo {volume} {109}},\ \href {https://doi.org/10.1103/physrevb.109.l201112} {10.1103/physrevb.109.l201112} (\bibinfo {year} {2024})\BibitemShut {NoStop}%
\bibitem [{\citenamefont {Diehl}\ and\ \citenamefont {Dietrich}(1981)}]{Diehl:1981jgg}%
  \BibitemOpen
  \bibfield  {author} {\bibinfo {author} {\bibfnamefont {H.~W.}\ \bibnamefont {Diehl}}\ and\ \bibinfo {author} {\bibfnamefont {S.}~\bibnamefont {Dietrich}},\ }\bibfield  {title} {\bibinfo {title} {{Field-theoretical approach to static critical phenomena in semi-infinite systems}},\ }\href {https://doi.org/10.1007/BF01298293} {\bibfield  {journal} {\bibinfo  {journal} {Z. Phys. B}\ }\textbf {\bibinfo {volume} {42}},\ \bibinfo {pages} {65} (\bibinfo {year} {1981})}\BibitemShut {NoStop}%
\bibitem [{\citenamefont {Cardy}(1984)}]{Cardy:1984bb}%
  \BibitemOpen
  \bibfield  {author} {\bibinfo {author} {\bibfnamefont {J.~L.}\ \bibnamefont {Cardy}},\ }\bibfield  {title} {\bibinfo {title} {{Conformal Invariance and Surface Critical Behavior}},\ }\href {https://doi.org/10.1016/0550-3213(84)90241-4} {\bibfield  {journal} {\bibinfo  {journal} {Nucl. Phys. B}\ }\textbf {\bibinfo {volume} {240}},\ \bibinfo {pages} {514} (\bibinfo {year} {1984})}\BibitemShut {NoStop}%
\bibitem [{\citenamefont {Binder}\ and\ \citenamefont {Landau}(1990)}]{BINDER199017}%
  \BibitemOpen
  \bibfield  {author} {\bibinfo {author} {\bibfnamefont {K.}~\bibnamefont {Binder}}\ and\ \bibinfo {author} {\bibfnamefont {D.}~\bibnamefont {Landau}},\ }\bibfield  {title} {\bibinfo {title} {Critical phenomena at surfaces},\ }\href {https://doi.org/https://doi.org/10.1016/0378-4371(90)90311-F} {\bibfield  {journal} {\bibinfo  {journal} {Physica A: Statistical Mechanics and its Applications}\ }\textbf {\bibinfo {volume} {163}},\ \bibinfo {pages} {17} (\bibinfo {year} {1990})}\BibitemShut {NoStop}%
\bibitem [{\citenamefont {Garratt}\ \emph {et~al.}(2023)\citenamefont {Garratt}, \citenamefont {Weinstein},\ and\ \citenamefont {Altman}}]{Garratt_2023}%
  \BibitemOpen
  \bibfield  {author} {\bibinfo {author} {\bibfnamefont {S.~J.}\ \bibnamefont {Garratt}}, \bibinfo {author} {\bibfnamefont {Z.}~\bibnamefont {Weinstein}},\ and\ \bibinfo {author} {\bibfnamefont {E.}~\bibnamefont {Altman}},\ }\bibfield  {title} {\bibinfo {title} {Measurements conspire nonlocally to restructure critical quantum states},\ }\bibfield  {journal} {\bibinfo  {journal} {Physical Review X}\ }\textbf {\bibinfo {volume} {13}},\ \href {https://doi.org/10.1103/physrevx.13.021026} {10.1103/physrevx.13.021026} (\bibinfo {year} {2023})\BibitemShut {NoStop}%
\bibitem [{\citenamefont {Lee}\ \emph {et~al.}(2023)\citenamefont {Lee}, \citenamefont {Jian},\ and\ \citenamefont {Xu}}]{JYLee_PRXQuantum.4.030317}%
  \BibitemOpen
  \bibfield  {author} {\bibinfo {author} {\bibfnamefont {J.~Y.}\ \bibnamefont {Lee}}, \bibinfo {author} {\bibfnamefont {C.-M.}\ \bibnamefont {Jian}},\ and\ \bibinfo {author} {\bibfnamefont {C.}~\bibnamefont {Xu}},\ }\bibfield  {title} {\bibinfo {title} {Quantum criticality under decoherence or weak measurement},\ }\href {https://doi.org/10.1103/PRXQuantum.4.030317} {\bibfield  {journal} {\bibinfo  {journal} {PRX Quantum}\ }\textbf {\bibinfo {volume} {4}},\ \bibinfo {pages} {030317} (\bibinfo {year} {2023})}\BibitemShut {NoStop}%
\bibitem [{\citenamefont {Lanzetta}\ \emph {et~al.}(2025)\citenamefont {Lanzetta}, \citenamefont {Liu},\ and\ \citenamefont {Metlitski}}]{lanzetta2025beginningendpointbootstrapconformal}%
  \BibitemOpen
  \bibfield  {author} {\bibinfo {author} {\bibfnamefont {R.~A.}\ \bibnamefont {Lanzetta}}, \bibinfo {author} {\bibfnamefont {S.}~\bibnamefont {Liu}},\ and\ \bibinfo {author} {\bibfnamefont {M.~A.}\ \bibnamefont {Metlitski}},\ }\href {https://arxiv.org/abs/2508.14964} {\bibinfo {title} {The beginning of the endpoint bootstrap for conformal line defects}} (\bibinfo {year} {2025}),\ \Eprint {https://arxiv.org/abs/2508.14964} {arXiv:2508.14964 [cond-mat.str-el]} \BibitemShut {NoStop}%
\bibitem [{\citenamefont {Komargodski}\ \emph {et~al.}(2025)\citenamefont {Komargodski}, \citenamefont {Popov},\ and\ \citenamefont {Rayhaun}}]{komargodski2025defectanomaliesspinfluxduality}%
  \BibitemOpen
  \bibfield  {author} {\bibinfo {author} {\bibfnamefont {Z.}~\bibnamefont {Komargodski}}, \bibinfo {author} {\bibfnamefont {F.~K.}\ \bibnamefont {Popov}},\ and\ \bibinfo {author} {\bibfnamefont {B.~C.}\ \bibnamefont {Rayhaun}},\ }\href {https://arxiv.org/abs/2508.14963} {\bibinfo {title} {Defect anomalies, a spin-flux duality, and boson-kondo problems}} (\bibinfo {year} {2025}),\ \Eprint {https://arxiv.org/abs/2508.14963} {arXiv:2508.14963 [hep-th]} \BibitemShut {NoStop}%
\bibitem [{\citenamefont {Hu}\ \emph {et~al.}(2024)\citenamefont {Hu}, \citenamefont {He},\ and\ \citenamefont {Zhu}}]{Hu_2024}%
  \BibitemOpen
  \bibfield  {author} {\bibinfo {author} {\bibfnamefont {L.}~\bibnamefont {Hu}}, \bibinfo {author} {\bibfnamefont {Y.-C.}\ \bibnamefont {He}},\ and\ \bibinfo {author} {\bibfnamefont {W.}~\bibnamefont {Zhu}},\ }\bibfield  {title} {\bibinfo {title} {Solving conformal defects in 3d conformal field theory using fuzzy sphere regularization},\ }\bibfield  {journal} {\bibinfo  {journal} {Nature Communications}\ }\textbf {\bibinfo {volume} {15}},\ \href {https://doi.org/10.1038/s41467-024-47978-y} {10.1038/s41467-024-47978-y} (\bibinfo {year} {2024})\BibitemShut {NoStop}%
\bibitem [{\citenamefont {Zhou}\ \emph {et~al.}(2024)\citenamefont {Zhou}, \citenamefont {Gaiotto}, \citenamefont {He},\ and\ \citenamefont {Zou}}]{Zhou_2024}%
  \BibitemOpen
  \bibfield  {author} {\bibinfo {author} {\bibfnamefont {Z.}~\bibnamefont {Zhou}}, \bibinfo {author} {\bibfnamefont {D.}~\bibnamefont {Gaiotto}}, \bibinfo {author} {\bibfnamefont {Y.-C.}\ \bibnamefont {He}},\ and\ \bibinfo {author} {\bibfnamefont {Y.}~\bibnamefont {Zou}},\ }\bibfield  {title} {\bibinfo {title} {The $g$-function and defect changing operators from wavefunction overlap on a fuzzy sphere},\ }\bibfield  {journal} {\bibinfo  {journal} {SciPost Physics}\ }\textbf {\bibinfo {volume} {17}},\ \href {https://doi.org/10.21468/scipostphys.17.1.021} {10.21468/scipostphys.17.1.021} (\bibinfo {year} {2024})\BibitemShut {NoStop}%
\bibitem [{\citenamefont {Cuomo}\ \emph {et~al.}(2024)\citenamefont {Cuomo}, \citenamefont {He},\ and\ \citenamefont {Komargodski}}]{Cuomo:2024psk}%
  \BibitemOpen
  \bibfield  {author} {\bibinfo {author} {\bibfnamefont {G.}~\bibnamefont {Cuomo}}, \bibinfo {author} {\bibfnamefont {Y.-C.}\ \bibnamefont {He}},\ and\ \bibinfo {author} {\bibfnamefont {Z.}~\bibnamefont {Komargodski}},\ }\bibfield  {title} {\bibinfo {title} {{Impurities with a cusp: general theory and 3d Ising}},\ }\href {https://doi.org/10.1007/JHEP11(2024)061} {\bibfield  {journal} {\bibinfo  {journal} {JHEP}\ }\textbf {\bibinfo {volume} {11}},\ \bibinfo {pages} {061}},\ \Eprint {https://arxiv.org/abs/2406.10186} {arXiv:2406.10186 [hep-th]} \BibitemShut {NoStop}%
\bibitem [{\citenamefont {Zhu}\ \emph {et~al.}(2023)\citenamefont {Zhu}, \citenamefont {Han}, \citenamefont {Huffman}, \citenamefont {Hofmann},\ and\ \citenamefont {He}}]{PhysRevX.13.021009}%
  \BibitemOpen
  \bibfield  {author} {\bibinfo {author} {\bibfnamefont {W.}~\bibnamefont {Zhu}}, \bibinfo {author} {\bibfnamefont {C.}~\bibnamefont {Han}}, \bibinfo {author} {\bibfnamefont {E.}~\bibnamefont {Huffman}}, \bibinfo {author} {\bibfnamefont {J.~S.}\ \bibnamefont {Hofmann}},\ and\ \bibinfo {author} {\bibfnamefont {Y.-C.}\ \bibnamefont {He}},\ }\bibfield  {title} {\bibinfo {title} {Uncovering conformal symmetry in the 3d ising transition: State-operator correspondence from a quantum fuzzy sphere regularization},\ }\href {https://doi.org/10.1103/PhysRevX.13.021009} {\bibfield  {journal} {\bibinfo  {journal} {Phys. Rev. X}\ }\textbf {\bibinfo {volume} {13}},\ \bibinfo {pages} {021009} (\bibinfo {year} {2023})}\BibitemShut {NoStop}%
\bibitem [{\citenamefont {He}\ and\ \citenamefont {Zhu}(2026)}]{Fuzzysphere_review2026}%
  \BibitemOpen
  \bibfield  {author} {\bibinfo {author} {\bibfnamefont {Y.-C.}\ \bibnamefont {He}}\ and\ \bibinfo {author} {\bibfnamefont {W.}~\bibnamefont {Zhu}},\ }\bibfield  {title} {\bibinfo {title} {A fuzzy sphere journey in critical phenomena},\ }\href {https://doi.org/10.1146/annurev-conmatphys-031424-020256} {\bibfield  {journal} {\bibinfo  {journal} {Annual Review of Condensed Matter Physics}\ }\textbf {\bibinfo {volume} {17}},\ \bibinfo {pages} {1–25} (\bibinfo {year} {2026})}\BibitemShut {NoStop}%
\bibitem [{\citenamefont {Han}\ \emph {et~al.}(2023)\citenamefont {Han}, \citenamefont {Hu}, \citenamefont {Zhu},\ and\ \citenamefont {He}}]{four_point_corr_PhysRevB.108.235123}%
  \BibitemOpen
  \bibfield  {author} {\bibinfo {author} {\bibfnamefont {C.}~\bibnamefont {Han}}, \bibinfo {author} {\bibfnamefont {L.}~\bibnamefont {Hu}}, \bibinfo {author} {\bibfnamefont {W.}~\bibnamefont {Zhu}},\ and\ \bibinfo {author} {\bibfnamefont {Y.-C.}\ \bibnamefont {He}},\ }\bibfield  {title} {\bibinfo {title} {Conformal four-point correlators of the three-dimensional ising transition via the quantum fuzzy sphere},\ }\href {https://doi.org/10.1103/PhysRevB.108.235123} {\bibfield  {journal} {\bibinfo  {journal} {Phys. Rev. B}\ }\textbf {\bibinfo {volume} {108}},\ \bibinfo {pages} {235123} (\bibinfo {year} {2023})}\BibitemShut {NoStop}%
\bibitem [{\citenamefont {Hu}\ \emph {et~al.}(2023)\citenamefont {Hu}, \citenamefont {He},\ and\ \citenamefont {Zhu}}]{ope_PhysRevLett.131.031601}%
  \BibitemOpen
  \bibfield  {author} {\bibinfo {author} {\bibfnamefont {L.}~\bibnamefont {Hu}}, \bibinfo {author} {\bibfnamefont {Y.-C.}\ \bibnamefont {He}},\ and\ \bibinfo {author} {\bibfnamefont {W.}~\bibnamefont {Zhu}},\ }\bibfield  {title} {\bibinfo {title} {Operator product expansion coefficients of the 3d ising criticality via quantum fuzzy spheres},\ }\href {https://doi.org/10.1103/PhysRevLett.131.031601} {\bibfield  {journal} {\bibinfo  {journal} {Phys. Rev. Lett.}\ }\textbf {\bibinfo {volume} {131}},\ \bibinfo {pages} {031601} (\bibinfo {year} {2023})}\BibitemShut {NoStop}%
\bibitem [{\citenamefont {Fardelli}\ \emph {et~al.}(2025)\citenamefont {Fardelli}, \citenamefont {Fitzpatrick},\ and\ \citenamefont {Katz}}]{generators_10.21468/SciPostPhys.18.3.086}%
  \BibitemOpen
  \bibfield  {author} {\bibinfo {author} {\bibfnamefont {G.}~\bibnamefont {Fardelli}}, \bibinfo {author} {\bibfnamefont {A.~L.}\ \bibnamefont {Fitzpatrick}},\ and\ \bibinfo {author} {\bibfnamefont {E.}~\bibnamefont {Katz}},\ }\bibfield  {title} {\bibinfo {title} {{Constructing the infrared conformal generators on the fuzzy sphere}},\ }\href {https://doi.org/10.21468/SciPostPhys.18.3.086} {\bibfield  {journal} {\bibinfo  {journal} {SciPost Phys.}\ }\textbf {\bibinfo {volume} {18}},\ \bibinfo {pages} {086} (\bibinfo {year} {2025})}\BibitemShut {NoStop}%
\bibitem [{\citenamefont {Fan}(2024)}]{fan2024noteexplicitconstructionconformal}%
  \BibitemOpen
  \bibfield  {author} {\bibinfo {author} {\bibfnamefont {R.}~\bibnamefont {Fan}},\ }\href {https://arxiv.org/abs/2409.08257} {\bibinfo {title} {Note on explicit construction of conformal generators on the fuzzy sphere}} (\bibinfo {year} {2024}),\ \Eprint {https://arxiv.org/abs/2409.08257} {arXiv:2409.08257 [hep-th]} \BibitemShut {NoStop}%
\bibitem [{\citenamefont {Hu}\ \emph {et~al.}(2025)\citenamefont {Hu}, \citenamefont {Zhu},\ and\ \citenamefont {He}}]{F_func_PhysRevB.111.155151}%
  \BibitemOpen
  \bibfield  {author} {\bibinfo {author} {\bibfnamefont {L.}~\bibnamefont {Hu}}, \bibinfo {author} {\bibfnamefont {W.}~\bibnamefont {Zhu}},\ and\ \bibinfo {author} {\bibfnamefont {Y.-C.}\ \bibnamefont {He}},\ }\bibfield  {title} {\bibinfo {title} {Entropic $f$ function of three-dimensional ising conformal field theory via fuzzy sphere regularization},\ }\href {https://doi.org/10.1103/PhysRevB.111.155151} {\bibfield  {journal} {\bibinfo  {journal} {Phys. Rev. B}\ }\textbf {\bibinfo {volume} {111}},\ \bibinfo {pages} {155151} (\bibinfo {year} {2025})}\BibitemShut {NoStop}%
\bibitem [{\citenamefont {Allais}\ and\ \citenamefont {Sachdev}(2014)}]{Allais_2014}%
  \BibitemOpen
  \bibfield  {author} {\bibinfo {author} {\bibfnamefont {A.}~\bibnamefont {Allais}}\ and\ \bibinfo {author} {\bibfnamefont {S.}~\bibnamefont {Sachdev}},\ }\bibfield  {title} {\bibinfo {title} {Spectral function of a localized fermion coupled to the wilson-fisher conformal field theory},\ }\bibfield  {journal} {\bibinfo  {journal} {Physical Review B}\ }\textbf {\bibinfo {volume} {90}},\ \href {https://doi.org/10.1103/physrevb.90.035131} {10.1103/physrevb.90.035131} (\bibinfo {year} {2014})\BibitemShut {NoStop}%
\bibitem [{\citenamefont {Nishioka}\ \emph {et~al.}(2023)\citenamefont {Nishioka}, \citenamefont {Okuyama},\ and\ \citenamefont {Shimamori}}]{Nishioka_2023}%
  \BibitemOpen
  \bibfield  {author} {\bibinfo {author} {\bibfnamefont {T.}~\bibnamefont {Nishioka}}, \bibinfo {author} {\bibfnamefont {Y.}~\bibnamefont {Okuyama}},\ and\ \bibinfo {author} {\bibfnamefont {S.}~\bibnamefont {Shimamori}},\ }\bibfield  {title} {\bibinfo {title} {The epsilon expansion of the o(n) model with line defect from conformal field theory},\ }\bibfield  {journal} {\bibinfo  {journal} {Journal of High Energy Physics}\ }\textbf {\bibinfo {volume} {2023}},\ \href {https://doi.org/10.1007/jhep03(2023)203} {10.1007/jhep03(2023)203} (\bibinfo {year} {2023})\BibitemShut {NoStop}%
\bibitem [{\citenamefont {Girault}\ \emph {et~al.}(2026)\citenamefont {Girault}, \citenamefont {Paulos},\ and\ \citenamefont {van Vliet}}]{girault2026consequencessymmetrybreakingconformaldefect}%
  \BibitemOpen
  \bibfield  {author} {\bibinfo {author} {\bibfnamefont {B.}~\bibnamefont {Girault}}, \bibinfo {author} {\bibfnamefont {M.~F.}\ \bibnamefont {Paulos}},\ and\ \bibinfo {author} {\bibfnamefont {P.}~\bibnamefont {van Vliet}},\ }\href {https://arxiv.org/abs/2509.26561} {\bibinfo {title} {Consequences of symmetry-breaking on conformal defect data}} (\bibinfo {year} {2026}),\ \Eprint {https://arxiv.org/abs/2509.26561} {arXiv:2509.26561 [hep-th]} \BibitemShut {NoStop}%
\bibitem [{\citenamefont {Allais}(2014)}]{allais2014magneticdefectlinecritical}%
  \BibitemOpen
  \bibfield  {author} {\bibinfo {author} {\bibfnamefont {A.}~\bibnamefont {Allais}},\ }\href {https://arxiv.org/abs/1412.3449} {\bibinfo {title} {Magnetic defect line in a critical ising bath}} (\bibinfo {year} {2014}),\ \Eprint {https://arxiv.org/abs/1412.3449} {arXiv:1412.3449 [cond-mat.str-el]} \BibitemShut {NoStop}%
\bibitem [{\citenamefont {Parisen~Toldin}\ \emph {et~al.}(2017)\citenamefont {Parisen~Toldin}, \citenamefont {Assaad},\ and\ \citenamefont {Wessel}}]{PhysRevB.95.014401}%
  \BibitemOpen
  \bibfield  {author} {\bibinfo {author} {\bibfnamefont {F.}~\bibnamefont {Parisen~Toldin}}, \bibinfo {author} {\bibfnamefont {F.~F.}\ \bibnamefont {Assaad}},\ and\ \bibinfo {author} {\bibfnamefont {S.}~\bibnamefont {Wessel}},\ }\bibfield  {title} {\bibinfo {title} {Critical behavior in the presence of an order-parameter pinning field},\ }\href {https://doi.org/10.1103/PhysRevB.95.014401} {\bibfield  {journal} {\bibinfo  {journal} {Phys. Rev. B}\ }\textbf {\bibinfo {volume} {95}},\ \bibinfo {pages} {014401} (\bibinfo {year} {2017})}\BibitemShut {NoStop}%
\bibitem [{\citenamefont {Gimenez-Grau}\ \emph {et~al.}(2022)\citenamefont {Gimenez-Grau}, \citenamefont {Lauria}, \citenamefont {Liendo},\ and\ \citenamefont {van Vliet}}]{Gimenez_Grau_2022}%
  \BibitemOpen
  \bibfield  {author} {\bibinfo {author} {\bibfnamefont {A.}~\bibnamefont {Gimenez-Grau}}, \bibinfo {author} {\bibfnamefont {E.}~\bibnamefont {Lauria}}, \bibinfo {author} {\bibfnamefont {P.}~\bibnamefont {Liendo}},\ and\ \bibinfo {author} {\bibfnamefont {P.}~\bibnamefont {van Vliet}},\ }\bibfield  {title} {\bibinfo {title} {Bootstrapping line defects with o(2) global symmetry},\ }\bibfield  {journal} {\bibinfo  {journal} {Journal of High Energy Physics}\ }\textbf {\bibinfo {volume} {2022}},\ \href {https://doi.org/10.1007/jhep11(2022)018} {10.1007/jhep11(2022)018} (\bibinfo {year} {2022})\BibitemShut {NoStop}%
\bibitem [{\citenamefont {Belton}\ \emph {et~al.}(2025)\citenamefont {Belton}, \citenamefont {Drukker}, \citenamefont {Kong},\ and\ \citenamefont {Stergiou}}]{Belton_2025}%
  \BibitemOpen
  \bibfield  {author} {\bibinfo {author} {\bibfnamefont {J.}~\bibnamefont {Belton}}, \bibinfo {author} {\bibfnamefont {N.}~\bibnamefont {Drukker}}, \bibinfo {author} {\bibfnamefont {Z.}~\bibnamefont {Kong}},\ and\ \bibinfo {author} {\bibfnamefont {A.}~\bibnamefont {Stergiou}},\ }\bibfield  {title} {\bibinfo {title} {Fine spectrum from crude analytic bootstrap},\ }\href {https://doi.org/10.1088/1751-8121/adf925} {\bibfield  {journal} {\bibinfo  {journal} {Journal of Physics A: Mathematical and Theoretical}\ }\textbf {\bibinfo {volume} {58}},\ \bibinfo {pages} {345401} (\bibinfo {year} {2025})}\BibitemShut {NoStop}%
\bibitem [{\citenamefont {Hu}\ and\ \citenamefont {Li}(2026)}]{wenliang_li_Hu:2025yrs}%
  \BibitemOpen
  \bibfield  {author} {\bibinfo {author} {\bibfnamefont {R.}~\bibnamefont {Hu}}\ and\ \bibinfo {author} {\bibfnamefont {W.}~\bibnamefont {Li}},\ }\bibfield  {title} {\bibinfo {title} {{Accurate Boundary Bootstrap for the Three-Dimensional O(N) Normal Universality Class}},\ }\href {https://doi.org/10.1103/hnd3-636j} {\bibfield  {journal} {\bibinfo  {journal} {Phys. Rev. Lett.}\ }\textbf {\bibinfo {volume} {136}},\ \bibinfo {pages} {221601} (\bibinfo {year} {2026})},\ \Eprint {https://arxiv.org/abs/2508.20854} {arXiv:2508.20854 [hep-th]} \BibitemShut {NoStop}%
\bibitem [{\citenamefont {Hanke}(2000)}]{Hanke2000}%
  \BibitemOpen
  \bibfield  {author} {\bibinfo {author} {\bibfnamefont {A.}~\bibnamefont {Hanke}},\ }\bibfield  {title} {\bibinfo {title} {Critical adsorption on defects in ising magnets and binary alloys},\ }\href {https://doi.org/10.1103/PhysRevLett.84.2180} {\bibfield  {journal} {\bibinfo  {journal} {Phys. Rev. Lett.}\ }\textbf {\bibinfo {volume} {84}},\ \bibinfo {pages} {2180} (\bibinfo {year} {2000})}\BibitemShut {NoStop}%
\bibitem [{\citenamefont {Sarma}\ \emph {et~al.}(2026)\citenamefont {Sarma}, \citenamefont {Zhou}, \citenamefont {Lanzetta},\ and\ \citenamefont {He}}]{Sarma_2026_BoseKondoImpurity}%
  \BibitemOpen
  \bibfield  {author} {\bibinfo {author} {\bibfnamefont {A.}~\bibnamefont {Sarma}}, \bibinfo {author} {\bibfnamefont {Z.}~\bibnamefont {Zhou}}, \bibinfo {author} {\bibfnamefont {R.~A.}\ \bibnamefont {Lanzetta}},\ and\ \bibinfo {author} {\bibfnamefont {Y.-C.}\ \bibnamefont {He}},\ }\href {https://arxiv.org/abs/2604.07554} {\bibinfo {title} {Fortuitous universality of bose-kondo impurities}} (\bibinfo {year} {2026}),\ \Eprint {https://arxiv.org/abs/2604.07554} {arXiv:2604.07554 [cond-mat.str-el]} \BibitemShut {NoStop}%
\bibitem [{\citenamefont {Cuomo}\ \emph {et~al.}(2022{\natexlab{a}})\citenamefont {Cuomo}, \citenamefont {Komargodski},\ and\ \citenamefont {Raviv-Moshe}}]{Cuomo_g_theorem__PhysRevLett.128.021603}%
  \BibitemOpen
  \bibfield  {author} {\bibinfo {author} {\bibfnamefont {G.}~\bibnamefont {Cuomo}}, \bibinfo {author} {\bibfnamefont {Z.}~\bibnamefont {Komargodski}},\ and\ \bibinfo {author} {\bibfnamefont {A.}~\bibnamefont {Raviv-Moshe}},\ }\bibfield  {title} {\bibinfo {title} {Renormalization group flows on line defects},\ }\href {https://doi.org/10.1103/PhysRevLett.128.021603} {\bibfield  {journal} {\bibinfo  {journal} {Phys. Rev. Lett.}\ }\textbf {\bibinfo {volume} {128}},\ \bibinfo {pages} {021603} (\bibinfo {year} {2022}{\natexlab{a}})}\BibitemShut {NoStop}%
\bibitem [{\citenamefont {Affleck}\ and\ \citenamefont {Ludwig}(1991)}]{Affleck_Ludwig_g_theorem_PhysRevLett.67.161}%
  \BibitemOpen
  \bibfield  {author} {\bibinfo {author} {\bibfnamefont {I.}~\bibnamefont {Affleck}}\ and\ \bibinfo {author} {\bibfnamefont {A.~W.~W.}\ \bibnamefont {Ludwig}},\ }\bibfield  {title} {\bibinfo {title} {Universal noninteger ``ground-state degeneracy'' in critical quantum systems},\ }\href {https://doi.org/10.1103/PhysRevLett.67.161} {\bibfield  {journal} {\bibinfo  {journal} {Phys. Rev. Lett.}\ }\textbf {\bibinfo {volume} {67}},\ \bibinfo {pages} {161} (\bibinfo {year} {1991})}\BibitemShut {NoStop}%
\bibitem [{\citenamefont {Oshikawa}\ and\ \citenamefont {Affleck}(1997)}]{OSHIKAWA1997533}%
  \BibitemOpen
  \bibfield  {author} {\bibinfo {author} {\bibfnamefont {M.}~\bibnamefont {Oshikawa}}\ and\ \bibinfo {author} {\bibfnamefont {I.}~\bibnamefont {Affleck}},\ }\bibfield  {title} {\bibinfo {title} {Boundary conformal field theory approach to the critical two-dimensional ising model with a defect line},\ }\href {https://doi.org/https://doi.org/10.1016/S0550-3213(97)00219-8} {\bibfield  {journal} {\bibinfo  {journal} {Nuclear Physics B}\ }\textbf {\bibinfo {volume} {495}},\ \bibinfo {pages} {533} (\bibinfo {year} {1997})}\BibitemShut {NoStop}%
\bibitem [{\citenamefont {Lanzetta}(2023)}]{pirsa_PIRSA_23110068}%
  \BibitemOpen
  \bibfield  {author} {\bibinfo {author} {\bibfnamefont {R.}~\bibnamefont {Lanzetta}},\ }\href {https://doi.org/10.48660/23110068} {\bibinfo {title} {Long-range order on line defects in ising conformal field theories}} (\bibinfo {year} {2023})\BibitemShut {NoStop}%
\bibitem [{\citenamefont {Cuomo}\ \emph {et~al.}(2022{\natexlab{b}})\citenamefont {Cuomo}, \citenamefont {Komargodski},\ and\ \citenamefont {Mezei}}]{Cuomo_2022}%
  \BibitemOpen
  \bibfield  {author} {\bibinfo {author} {\bibfnamefont {G.}~\bibnamefont {Cuomo}}, \bibinfo {author} {\bibfnamefont {Z.}~\bibnamefont {Komargodski}},\ and\ \bibinfo {author} {\bibfnamefont {M.}~\bibnamefont {Mezei}},\ }\bibfield  {title} {\bibinfo {title} {Localized magnetic field in the o(n) model},\ }\bibfield  {journal} {\bibinfo  {journal} {Journal of High Energy Physics}\ }\textbf {\bibinfo {volume} {2022}},\ \href {https://doi.org/10.1007/jhep02(2022)134} {10.1007/jhep02(2022)134} (\bibinfo {year} {2022}{\natexlab{b}})\BibitemShut {NoStop}%
\bibitem [{\citenamefont {He}(2025)}]{he2025freerealscalarcft}%
  \BibitemOpen
  \bibfield  {author} {\bibinfo {author} {\bibfnamefont {Y.-C.}\ \bibnamefont {He}},\ }\href {https://arxiv.org/abs/2506.14904} {\bibinfo {title} {Free real scalar cft on fuzzy sphere: spectrum, algebra and wavefunction ansatz}} (\bibinfo {year} {2025}),\ \Eprint {https://arxiv.org/abs/2506.14904} {arXiv:2506.14904 [hep-th]} \BibitemShut {NoStop}%
\bibitem [{\citenamefont {Guo}\ \emph {et~al.}(2025)\citenamefont {Guo}, \citenamefont {Zhou}, \citenamefont {Wei},\ and\ \citenamefont {He}}]{guo2025onfreescalarwilsonfisherconformal}%
  \BibitemOpen
  \bibfield  {author} {\bibinfo {author} {\bibfnamefont {W.}~\bibnamefont {Guo}}, \bibinfo {author} {\bibfnamefont {Z.}~\bibnamefont {Zhou}}, \bibinfo {author} {\bibfnamefont {T.-C.}\ \bibnamefont {Wei}},\ and\ \bibinfo {author} {\bibfnamefont {Y.-C.}\ \bibnamefont {He}},\ }\href {https://arxiv.org/abs/2512.02234} {\bibinfo {title} {The $o(n)$ free-scalar and wilson-fisher conformal field theories on the fuzzy sphere}} (\bibinfo {year} {2025}),\ \Eprint {https://arxiv.org/abs/2512.02234} {arXiv:2512.02234 [cond-mat.str-el]} \BibitemShut {NoStop}%
\bibitem [{sm()}]{sm}%
  \BibitemOpen
  \href@noop {} {\bibinfo  {journal} {Supplementary material}\ }\BibitemShut {NoStop}%
\bibitem [{\citenamefont {Casini}\ \emph {et~al.}(2016)\citenamefont {Casini}, \citenamefont {Landea},\ and\ \citenamefont {Torroba}}]{Casini_2016_gTheorem}%
  \BibitemOpen
\bibfield  {journal} {  }\bibfield  {author} {\bibinfo {author} {\bibfnamefont {H.}~\bibnamefont {Casini}}, \bibinfo {author} {\bibfnamefont {I.~S.}\ \bibnamefont {Landea}},\ and\ \bibinfo {author} {\bibfnamefont {G.}~\bibnamefont {Torroba}},\ }\bibfield  {title} {\bibinfo {title} {The g-theorem and quantum information theory},\ }\bibfield  {journal} {\bibinfo  {journal} {Journal of High Energy Physics}\ }\textbf {\bibinfo {volume} {2016}},\ \href {https://doi.org/10.1007/jhep10(2016)140} {10.1007/jhep10(2016)140} (\bibinfo {year} {2016})\BibitemShut {NoStop}%
\bibitem [{Note1()}]{Note1}%
  \BibitemOpen
  \bibinfo {note} {In this work, we only consider this particular scenario of spontaneous symmetry breaking. The condition $2g_{\protect \mathcal {D}}<1$ indicates that the corresponding symmetry-broken fixed point $\protect \mathcal {D}^+\oplus \protect \mathcal {D}^-$ is stable. Other possible spontaneous symmetry-breaking patterns may lead to different bounds on the value of $g-$function.}\BibitemShut {Stop}%
\bibitem [{\citenamefont {Metlitski}(2022)}]{Max_boundary_10.21468/SciPostPhys.12.4.131}%
  \BibitemOpen
  \bibfield  {author} {\bibinfo {author} {\bibfnamefont {M.~A.}\ \bibnamefont {Metlitski}},\ }\bibfield  {title} {\bibinfo {title} {{Boundary criticality of the O(N) model in d = 3 critically revisited}},\ }\href {https://doi.org/10.21468/SciPostPhys.12.4.131} {\bibfield  {journal} {\bibinfo  {journal} {SciPost Phys.}\ }\textbf {\bibinfo {volume} {12}},\ \bibinfo {pages} {131} (\bibinfo {year} {2022})}\BibitemShut {NoStop}%
\bibitem [{\citenamefont {Krishnan}\ and\ \citenamefont {Metlitski}(2023)}]{Krishnan_2023}%
  \BibitemOpen
  \bibfield  {author} {\bibinfo {author} {\bibfnamefont {A.}~\bibnamefont {Krishnan}}\ and\ \bibinfo {author} {\bibfnamefont {M.~A.}\ \bibnamefont {Metlitski}},\ }\bibfield  {title} {\bibinfo {title} {A plane defect in the 3d o(n) model},\ }\bibfield  {journal} {\bibinfo  {journal} {SciPost Physics}\ }\textbf {\bibinfo {volume} {15}},\ \href {https://doi.org/10.21468/scipostphys.15.3.090} {10.21468/scipostphys.15.3.090} (\bibinfo {year} {2023})\BibitemShut {NoStop}%
\bibitem [{\citenamefont {Cuomo}\ and\ \citenamefont {Zhang}(2024)}]{cuomo2024spontaneoussymmetrybreakingsurface}%
  \BibitemOpen
  \bibfield  {author} {\bibinfo {author} {\bibfnamefont {G.}~\bibnamefont {Cuomo}}\ and\ \bibinfo {author} {\bibfnamefont {S.}~\bibnamefont {Zhang}},\ }\href {https://arxiv.org/abs/2306.00085} {\bibinfo {title} {Spontaneous symmetry breaking on surface defects}} (\bibinfo {year} {2024}),\ \Eprint {https://arxiv.org/abs/2306.00085} {arXiv:2306.00085 [hep-th]} \BibitemShut {NoStop}%
\bibitem [{\citenamefont {Hu}\ \emph {et~al.}(2021)\citenamefont {Hu}, \citenamefont {Deng},\ and\ \citenamefont {Lv}}]{Hu_2021}%
  \BibitemOpen
  \bibfield  {author} {\bibinfo {author} {\bibfnamefont {M.}~\bibnamefont {Hu}}, \bibinfo {author} {\bibfnamefont {Y.}~\bibnamefont {Deng}},\ and\ \bibinfo {author} {\bibfnamefont {J.-P.}\ \bibnamefont {Lv}},\ }\bibfield  {title} {\bibinfo {title} {Extraordinary-log surface phase transition in the three-dimensional xy model},\ }\bibfield  {journal} {\bibinfo  {journal} {Physical Review Letters}\ }\textbf {\bibinfo {volume} {127}},\ \href {https://doi.org/10.1103/physrevlett.127.120603} {10.1103/physrevlett.127.120603} (\bibinfo {year} {2021})\BibitemShut {NoStop}%
\bibitem [{\citenamefont {Parisen~Toldin}\ and\ \citenamefont {Metlitski}(2022)}]{Parisen_Toldin_2022}%
  \BibitemOpen
  \bibfield  {author} {\bibinfo {author} {\bibfnamefont {F.}~\bibnamefont {Parisen~Toldin}}\ and\ \bibinfo {author} {\bibfnamefont {M.~A.}\ \bibnamefont {Metlitski}},\ }\bibfield  {title} {\bibinfo {title} {Boundary criticality of the 3d o(3) model: From normal to extraordinary},\ }\bibfield  {journal} {\bibinfo  {journal} {Physical Review Letters}\ }\textbf {\bibinfo {volume} {128}},\ \href {https://doi.org/10.1103/physrevlett.128.215701} {10.1103/physrevlett.128.215701} (\bibinfo {year} {2022})\BibitemShut {NoStop}%
\bibitem [{\citenamefont {Sun}\ \emph {et~al.}(2023)\citenamefont {Sun}, \citenamefont {Hu}, \citenamefont {Deng},\ and\ \citenamefont {Lv}}]{Sun_2023}%
  \BibitemOpen
  \bibfield  {author} {\bibinfo {author} {\bibfnamefont {Y.}~\bibnamefont {Sun}}, \bibinfo {author} {\bibfnamefont {M.}~\bibnamefont {Hu}}, \bibinfo {author} {\bibfnamefont {Y.}~\bibnamefont {Deng}},\ and\ \bibinfo {author} {\bibfnamefont {J.-P.}\ \bibnamefont {Lv}},\ }\bibfield  {title} {\bibinfo {title} {Extraordinary-log universality of critical phenomena in plane defects},\ }\bibfield  {journal} {\bibinfo  {journal} {Physical Review Letters}\ }\textbf {\bibinfo {volume} {131}},\ \href {https://doi.org/10.1103/physrevlett.131.207101} {10.1103/physrevlett.131.207101} (\bibinfo {year} {2023})\BibitemShut {NoStop}%
\bibitem [{\citenamefont {Parisen~Toldin}\ \emph {et~al.}(2025)\citenamefont {Parisen~Toldin}, \citenamefont {Krishnan},\ and\ \citenamefont {Metlitski}}]{Parisen_Toldin_2025}%
  \BibitemOpen
  \bibfield  {author} {\bibinfo {author} {\bibfnamefont {F.}~\bibnamefont {Parisen~Toldin}}, \bibinfo {author} {\bibfnamefont {A.}~\bibnamefont {Krishnan}},\ and\ \bibinfo {author} {\bibfnamefont {M.~A.}\ \bibnamefont {Metlitski}},\ }\bibfield  {title} {\bibinfo {title} {Universal finite-size scaling in the extraordinary-log boundary phase of three-dimensional o(n) model},\ }\bibfield  {journal} {\bibinfo  {journal} {Physical Review Research}\ }\textbf {\bibinfo {volume} {7}},\ \href {https://doi.org/10.1103/physrevresearch.7.023052} {10.1103/physrevresearch.7.023052} (\bibinfo {year} {2025})\BibitemShut {NoStop}%
\bibitem [{\citenamefont {Han}\ \emph {et~al.}(2024)\citenamefont {Han}, \citenamefont {Hu},\ and\ \citenamefont {Zhu}}]{chaohanPhysRevB.110.115113}%
  \BibitemOpen
  \bibfield  {author} {\bibinfo {author} {\bibfnamefont {C.}~\bibnamefont {Han}}, \bibinfo {author} {\bibfnamefont {L.}~\bibnamefont {Hu}},\ and\ \bibinfo {author} {\bibfnamefont {W.}~\bibnamefont {Zhu}},\ }\bibfield  {title} {\bibinfo {title} {Conformal operator content of the wilson-fisher transition on fuzzy sphere bilayers},\ }\href {https://doi.org/10.1103/PhysRevB.110.115113} {\bibfield  {journal} {\bibinfo  {journal} {Phys. Rev. B}\ }\textbf {\bibinfo {volume} {110}},\ \bibinfo {pages} {115113} (\bibinfo {year} {2024})}\BibitemShut {NoStop}%
\bibitem [{\citenamefont {Feng}\ and\ \citenamefont {Wang}(2026)}]{feng2026studying3donsurface}%
  \BibitemOpen
  \bibfield  {author} {\bibinfo {author} {\bibfnamefont {J.}~\bibnamefont {Feng}}\ and\ \bibinfo {author} {\bibfnamefont {T.}~\bibnamefont {Wang}},\ }\href {https://arxiv.org/abs/2604.21091} {\bibinfo {title} {Studying 3d o(n) surface cft on the fuzzy sphere}} (\bibinfo {year} {2026}),\ \Eprint {https://arxiv.org/abs/2604.21091} {arXiv:2604.21091 [cond-mat.str-el]} \BibitemShut {NoStop}%
\bibitem [{\citenamefont {Padayasi}\ \emph {et~al.}(2022)\citenamefont {Padayasi}, \citenamefont {Krishnan}, \citenamefont {Metlitski}, \citenamefont {Gruzberg},\ and\ \citenamefont {Meineri}}]{Padayasi_2022}%
  \BibitemOpen
  \bibfield  {author} {\bibinfo {author} {\bibfnamefont {J.}~\bibnamefont {Padayasi}}, \bibinfo {author} {\bibfnamefont {A.}~\bibnamefont {Krishnan}}, \bibinfo {author} {\bibfnamefont {M.}~\bibnamefont {Metlitski}}, \bibinfo {author} {\bibfnamefont {I.}~\bibnamefont {Gruzberg}},\ and\ \bibinfo {author} {\bibfnamefont {M.}~\bibnamefont {Meineri}},\ }\bibfield  {title} {\bibinfo {title} {The extraordinary boundary transition in the 3d o(n) model via conformal bootstrap},\ }\bibfield  {journal} {\bibinfo  {journal} {SciPost Physics}\ }\textbf {\bibinfo {volume} {12}},\ \href {https://doi.org/10.21468/scipostphys.12.6.190} {10.21468/scipostphys.12.6.190} (\bibinfo {year} {2022})\BibitemShut {NoStop}%
\bibitem [{\citenamefont {Sun}\ \emph {et~al.}(2026)\citenamefont {Sun}, \citenamefont {Jian},\ and\ \citenamefont {Yao}}]{sun2026analyticbootstraponboundary}%
  \BibitemOpen
  \bibfield  {author} {\bibinfo {author} {\bibfnamefont {X.}~\bibnamefont {Sun}}, \bibinfo {author} {\bibfnamefont {S.-K.}\ \bibnamefont {Jian}},\ and\ \bibinfo {author} {\bibfnamefont {H.}~\bibnamefont {Yao}},\ }\href {https://arxiv.org/abs/2605.28933} {\bibinfo {title} {Analytic bootstrap for $o(n)$ boundary conformal field theories with interacting boundaries}} (\bibinfo {year} {2026}),\ \Eprint {https://arxiv.org/abs/2605.28933} {arXiv:2605.28933 [hep-th]} \BibitemShut {NoStop}%
\bibitem [{\citenamefont {Zhang}\ and\ \citenamefont {Wang}(2017)}]{long_zhang_Fa_wang_PhysRevLett.118.087201}%
  \BibitemOpen
  \bibfield  {author} {\bibinfo {author} {\bibfnamefont {L.}~\bibnamefont {Zhang}}\ and\ \bibinfo {author} {\bibfnamefont {F.}~\bibnamefont {Wang}},\ }\bibfield  {title} {\bibinfo {title} {Unconventional surface critical behavior induced by a quantum phase transition from the two-dimensional affleck-kennedy-lieb-tasaki phase to a n\'eel-ordered phase},\ }\href {https://doi.org/10.1103/PhysRevLett.118.087201} {\bibfield  {journal} {\bibinfo  {journal} {Phys. Rev. Lett.}\ }\textbf {\bibinfo {volume} {118}},\ \bibinfo {pages} {087201} (\bibinfo {year} {2017})}\BibitemShut {NoStop}%
\bibitem [{\citenamefont {Ding}\ \emph {et~al.}(2018)\citenamefont {Ding}, \citenamefont {Zhang},\ and\ \citenamefont {Guo}}]{ding_chengxiang_PhysRevLett.120.235701}%
  \BibitemOpen
  \bibfield  {author} {\bibinfo {author} {\bibfnamefont {C.}~\bibnamefont {Ding}}, \bibinfo {author} {\bibfnamefont {L.}~\bibnamefont {Zhang}},\ and\ \bibinfo {author} {\bibfnamefont {W.}~\bibnamefont {Guo}},\ }\bibfield  {title} {\bibinfo {title} {Engineering surface critical behavior of ($2+1$)-dimensional o(3) quantum critical points},\ }\href {https://doi.org/10.1103/PhysRevLett.120.235701} {\bibfield  {journal} {\bibinfo  {journal} {Phys. Rev. Lett.}\ }\textbf {\bibinfo {volume} {120}},\ \bibinfo {pages} {235701} (\bibinfo {year} {2018})}\BibitemShut {NoStop}%
\bibitem [{\citenamefont {Weber}\ \emph {et~al.}(2018)\citenamefont {Weber}, \citenamefont {Parisen~Toldin},\ and\ \citenamefont {Wessel}}]{StefanWessel__PhysRevB.98.140403}%
  \BibitemOpen
  \bibfield  {author} {\bibinfo {author} {\bibfnamefont {L.}~\bibnamefont {Weber}}, \bibinfo {author} {\bibfnamefont {F.}~\bibnamefont {Parisen~Toldin}},\ and\ \bibinfo {author} {\bibfnamefont {S.}~\bibnamefont {Wessel}},\ }\bibfield  {title} {\bibinfo {title} {Nonordinary edge criticality of two-dimensional quantum critical magnets},\ }\href {https://doi.org/10.1103/PhysRevB.98.140403} {\bibfield  {journal} {\bibinfo  {journal} {Phys. Rev. B}\ }\textbf {\bibinfo {volume} {98}},\ \bibinfo {pages} {140403(R)} (\bibinfo {year} {2018})}\BibitemShut {NoStop}%
\bibitem [{\citenamefont {Liu}\ \emph {et~al.}(2021)\citenamefont {Liu}, \citenamefont {Shapourian}, \citenamefont {Vishwanath},\ and\ \citenamefont {Metlitski}}]{shangliu_PhysRevB.104.104201}%
  \BibitemOpen
  \bibfield  {author} {\bibinfo {author} {\bibfnamefont {S.}~\bibnamefont {Liu}}, \bibinfo {author} {\bibfnamefont {H.}~\bibnamefont {Shapourian}}, \bibinfo {author} {\bibfnamefont {A.}~\bibnamefont {Vishwanath}},\ and\ \bibinfo {author} {\bibfnamefont {M.~A.}\ \bibnamefont {Metlitski}},\ }\bibfield  {title} {\bibinfo {title} {Magnetic impurities at quantum critical points: Large-$n$ expansion and connections to symmetry-protected topological states},\ }\href {https://doi.org/10.1103/PhysRevB.104.104201} {\bibfield  {journal} {\bibinfo  {journal} {Phys. Rev. B}\ }\textbf {\bibinfo {volume} {104}},\ \bibinfo {pages} {104201} (\bibinfo {year} {2021})}\BibitemShut {NoStop}%
\bibitem [{\citenamefont {Zhu}\ \emph {et~al.}(2026)\citenamefont {Zhu}, \citenamefont {Liu}, \citenamefont {Wang}, \citenamefont {Wang},\ and\ \citenamefont {Yan}}]{zhu_yanzhang_tkx5-kzhh}%
  \BibitemOpen
  \bibfield  {author} {\bibinfo {author} {\bibfnamefont {Y.}~\bibnamefont {Zhu}}, \bibinfo {author} {\bibfnamefont {Z.}~\bibnamefont {Liu}}, \bibinfo {author} {\bibfnamefont {Z.}~\bibnamefont {Wang}}, \bibinfo {author} {\bibfnamefont {Y.-C.}\ \bibnamefont {Wang}},\ and\ \bibinfo {author} {\bibfnamefont {Z.}~\bibnamefont {Yan}},\ }\bibfield  {title} {\bibinfo {title} {Bipartite entanglement and surface criticality: The extra contribution of the nonordinary edge in entanglement},\ }\href {https://doi.org/10.1103/tkx5-kzhh} {\bibfield  {journal} {\bibinfo  {journal} {Phys. Rev. Lett.}\ }\textbf {\bibinfo {volume} {136}},\ \bibinfo {pages} {046501} (\bibinfo {year} {2026})}\BibitemShut {NoStop}%
\bibitem [{\citenamefont {Toldin}\ \emph {et~al.}(2025)\citenamefont {Toldin}, \citenamefont {Assaad},\ and\ \citenamefont {Metlitski}}]{toldin2025extraordinarytransitionedgecorrelated}%
  \BibitemOpen
  \bibfield  {author} {\bibinfo {author} {\bibfnamefont {F.~P.}\ \bibnamefont {Toldin}}, \bibinfo {author} {\bibfnamefont {F.~F.}\ \bibnamefont {Assaad}},\ and\ \bibinfo {author} {\bibfnamefont {M.~A.}\ \bibnamefont {Metlitski}},\ }\href {https://arxiv.org/abs/2508.00999} {\bibinfo {title} {Extraordinary transition at the edge of a correlated topological insulator}} (\bibinfo {year} {2025}),\ \Eprint {https://arxiv.org/abs/2508.00999} {arXiv:2508.00999 [cond-mat.str-el]} \BibitemShut {NoStop}%
\bibitem [{\citenamefont {Liu}\ \emph {et~al.}(2025)\citenamefont {Liu}, \citenamefont {Sato}, \citenamefont {Hou}, \citenamefont {Wang}, \citenamefont {Guo},\ and\ \citenamefont {Assaad}}]{liu2025edgemodestopologicalmott}%
  \BibitemOpen
  \bibfield  {author} {\bibinfo {author} {\bibfnamefont {Y.}~\bibnamefont {Liu}}, \bibinfo {author} {\bibfnamefont {T.}~\bibnamefont {Sato}}, \bibinfo {author} {\bibfnamefont {D.}~\bibnamefont {Hou}}, \bibinfo {author} {\bibfnamefont {Z.}~\bibnamefont {Wang}}, \bibinfo {author} {\bibfnamefont {W.}~\bibnamefont {Guo}},\ and\ \bibinfo {author} {\bibfnamefont {F.~F.}\ \bibnamefont {Assaad}},\ }\href {https://arxiv.org/abs/2508.04455} {\bibinfo {title} {Edge modes of topological mott insulators and deconfined quantum critical points}} (\bibinfo {year} {2025}),\ \Eprint {https://arxiv.org/abs/2508.04455} {arXiv:2508.04455 [cond-mat.str-el]} \BibitemShut {NoStop}%
\bibitem [{\citenamefont {Ma}\ \emph {et~al.}(2022)\citenamefont {Ma}, \citenamefont {Zou},\ and\ \citenamefont {Wang}}]{Ma_2022}%
  \BibitemOpen
  \bibfield  {author} {\bibinfo {author} {\bibfnamefont {R.}~\bibnamefont {Ma}}, \bibinfo {author} {\bibfnamefont {L.}~\bibnamefont {Zou}},\ and\ \bibinfo {author} {\bibfnamefont {C.}~\bibnamefont {Wang}},\ }\bibfield  {title} {\bibinfo {title} {Edge physics at the deconfined transition between a quantum spin hall insulator and a superconductor},\ }\bibfield  {journal} {\bibinfo  {journal} {SciPost Physics}\ }\textbf {\bibinfo {volume} {12}},\ \href {https://doi.org/10.21468/scipostphys.12.6.196} {10.21468/scipostphys.12.6.196} (\bibinfo {year} {2022})\BibitemShut {NoStop}%
\bibitem [{\citenamefont {Voinea}\ \emph {et~al.}(2025)\citenamefont {Voinea}, \citenamefont {Fan}, \citenamefont {Regnault},\ and\ \citenamefont {Papić}}]{Voinea_2025}%
  \BibitemOpen
  \bibfield  {author} {\bibinfo {author} {\bibfnamefont {C.}~\bibnamefont {Voinea}}, \bibinfo {author} {\bibfnamefont {R.}~\bibnamefont {Fan}}, \bibinfo {author} {\bibfnamefont {N.}~\bibnamefont {Regnault}},\ and\ \bibinfo {author} {\bibfnamefont {Z.}~\bibnamefont {Papić}},\ }\bibfield  {title} {\bibinfo {title} {Regularizing 3d conformal field theories via anyons on the fuzzy sphere},\ }\bibfield  {journal} {\bibinfo  {journal} {Physical Review X}\ }\textbf {\bibinfo {volume} {15}},\ \href {https://doi.org/10.1103/bf4k-phl9} {10.1103/bf4k-phl9} (\bibinfo {year} {2025})\BibitemShut {NoStop}%
\bibitem [{\citenamefont {Zhou}\ and\ \citenamefont {He}(2025)}]{Zhou_2025}%
  \BibitemOpen
  \bibfield  {author} {\bibinfo {author} {\bibfnamefont {Z.}~\bibnamefont {Zhou}}\ and\ \bibinfo {author} {\bibfnamefont {Y.-C.}\ \bibnamefont {He}},\ }\bibfield  {title} {\bibinfo {title} {3d conformal field theories with sp(n) global symmetry on a fuzzy sphere},\ }\bibfield  {journal} {\bibinfo  {journal} {Physical Review Letters}\ }\textbf {\bibinfo {volume} {135}},\ \href {https://doi.org/10.1103/xstj-xvcy} {10.1103/xstj-xvcy} (\bibinfo {year} {2025})\BibitemShut {NoStop}%
\bibitem [{\citenamefont {Zhou}\ \emph {et~al.}(2025)\citenamefont {Zhou}, \citenamefont {Wang},\ and\ \citenamefont {He}}]{zhou2025chernsimonsmatterconformalfieldtheory}%
  \BibitemOpen
  \bibfield  {author} {\bibinfo {author} {\bibfnamefont {Z.}~\bibnamefont {Zhou}}, \bibinfo {author} {\bibfnamefont {C.}~\bibnamefont {Wang}},\ and\ \bibinfo {author} {\bibfnamefont {Y.-C.}\ \bibnamefont {He}},\ }\href {https://arxiv.org/abs/2507.19580} {\bibinfo {title} {Chern-simons-matter conformal field theory on fuzzy sphere: Confinement transition of kalmeyer-laughlin chiral spin liquid}} (\bibinfo {year} {2025}),\ \Eprint {https://arxiv.org/abs/2507.19580} {arXiv:2507.19580 [cond-mat.str-el]} \BibitemShut {NoStop}%
\bibitem [{\citenamefont {Zhou}\ \emph {et~al.}(2026)\citenamefont {Zhou}, \citenamefont {Gaiotto},\ and\ \citenamefont {He}}]{Zhou_2026}%
  \BibitemOpen
  \bibfield  {author} {\bibinfo {author} {\bibfnamefont {Z.}~\bibnamefont {Zhou}}, \bibinfo {author} {\bibfnamefont {D.}~\bibnamefont {Gaiotto}},\ and\ \bibinfo {author} {\bibfnamefont {Y.-C.}\ \bibnamefont {He}},\ }\bibfield  {title} {\bibinfo {title} {Free and interacting fermionic conformal field theories on the fuzzy sphere},\ }\bibfield  {journal} {\bibinfo  {journal} {Physical Review X}\ }\textbf {\bibinfo {volume} {16}},\ \href {https://doi.org/10.1103/fkkl-9tw2} {10.1103/fkkl-9tw2} (\bibinfo {year} {2026})\BibitemShut {NoStop}%
\bibitem [{\citenamefont {Voinea}\ \emph {et~al.}(2026)\citenamefont {Voinea}, \citenamefont {Zhu}, \citenamefont {Regnault},\ and\ \citenamefont {Papić}}]{Voinea_2026}%
  \BibitemOpen
  \bibfield  {author} {\bibinfo {author} {\bibfnamefont {C.}~\bibnamefont {Voinea}}, \bibinfo {author} {\bibfnamefont {W.}~\bibnamefont {Zhu}}, \bibinfo {author} {\bibfnamefont {N.}~\bibnamefont {Regnault}},\ and\ \bibinfo {author} {\bibfnamefont {Z.}~\bibnamefont {Papić}},\ }\bibfield  {title} {\bibinfo {title} {Critical majorana fermion at a topological quantum hall bilayer transition},\ }\bibfield  {journal} {\bibinfo  {journal} {Physical Review Letters}\ }\textbf {\bibinfo {volume} {136}},\ \href {https://doi.org/10.1103/mztz-fyk3} {10.1103/mztz-fyk3} (\bibinfo {year} {2026})\BibitemShut {NoStop}%
\bibitem [{\citenamefont {Tang}\ \emph {et~al.}(2025)\citenamefont {Tang}, \citenamefont {Voinea}, \citenamefont {Hu}, \citenamefont {Papić},\ and\ \citenamefont {Zhu}}]{tang2025emergence3dsuperconformalising}%
  \BibitemOpen
  \bibfield  {author} {\bibinfo {author} {\bibfnamefont {Y.}~\bibnamefont {Tang}}, \bibinfo {author} {\bibfnamefont {C.}~\bibnamefont {Voinea}}, \bibinfo {author} {\bibfnamefont {L.}~\bibnamefont {Hu}}, \bibinfo {author} {\bibfnamefont {Z.}~\bibnamefont {Papić}},\ and\ \bibinfo {author} {\bibfnamefont {W.}~\bibnamefont {Zhu}},\ }\href {https://arxiv.org/abs/2512.25054} {\bibinfo {title} {Emergence of 3d superconformal ising criticality on the fuzzy sphere}} (\bibinfo {year} {2025}),\ \Eprint {https://arxiv.org/abs/2512.25054} {arXiv:2512.25054 [cond-mat.str-el]} \BibitemShut {NoStop}%
\bibitem [{\citenamefont {Jiang}\ \emph {et~al.}(2025)\citenamefont {Jiang}, \citenamefont {Ge},\ and\ \citenamefont {Jian}}]{BCFT_GNY_MC}%
  \BibitemOpen
  \bibfield  {author} {\bibinfo {author} {\bibfnamefont {H.}~\bibnamefont {Jiang}}, \bibinfo {author} {\bibfnamefont {Y.}~\bibnamefont {Ge}},\ and\ \bibinfo {author} {\bibfnamefont {S.-K.}\ \bibnamefont {Jian}},\ }\bibfield  {title} {\bibinfo {title} {Boundary criticality for the gross-neveu-yukawa models},\ }\href {https://doi.org/10.1103/tjfk-84f8} {\bibfield  {journal} {\bibinfo  {journal} {Phys. Rev. Lett.}\ }\textbf {\bibinfo {volume} {135}},\ \bibinfo {pages} {141602} (\bibinfo {year} {2025})}\BibitemShut {NoStop}%
\bibitem [{\citenamefont {Shen}\ \emph {et~al.}(2025)\citenamefont {Shen}, \citenamefont {Wu},\ and\ \citenamefont {Jian}}]{Shen_2025_BCFT_TI_TSC}%
  \BibitemOpen
  \bibfield  {author} {\bibinfo {author} {\bibfnamefont {X.}~\bibnamefont {Shen}}, \bibinfo {author} {\bibfnamefont {Z.}~\bibnamefont {Wu}},\ and\ \bibinfo {author} {\bibfnamefont {S.-K.}\ \bibnamefont {Jian}},\ }\bibfield  {title} {\bibinfo {title} {Boundary and defect criticality in topological insulators and superconductors},\ }\bibfield  {journal} {\bibinfo  {journal} {Physical Review B}\ }\textbf {\bibinfo {volume} {112}},\ \href {https://doi.org/10.1103/4lv4-mc81} {10.1103/4lv4-mc81} (\bibinfo {year} {2025})\BibitemShut {NoStop}%
\bibitem [{\citenamefont {Ge}\ \emph {et~al.}(2025)\citenamefont {Ge}, \citenamefont {Jiang}, \citenamefont {Yao},\ and\ \citenamefont {Jian}}]{ge2025boundarycriticalitytwodimensionalcorrelated}%
  \BibitemOpen
  \bibfield  {author} {\bibinfo {author} {\bibfnamefont {Y.}~\bibnamefont {Ge}}, \bibinfo {author} {\bibfnamefont {H.}~\bibnamefont {Jiang}}, \bibinfo {author} {\bibfnamefont {H.}~\bibnamefont {Yao}},\ and\ \bibinfo {author} {\bibfnamefont {S.-K.}\ \bibnamefont {Jian}},\ }\href {https://arxiv.org/abs/2510.05230} {\bibinfo {title} {Boundary criticality in two-dimensional correlated topological superconductors}} (\bibinfo {year} {2025}),\ \Eprint {https://arxiv.org/abs/2510.05230} {arXiv:2510.05230 [cond-mat.str-el]} \BibitemShut {NoStop}%
\bibitem [{\citenamefont {Zhou}(2025)}]{zhou2025fuzzifiedjuliapackage}%
  \BibitemOpen
  \bibfield  {author} {\bibinfo {author} {\bibfnamefont {Z.}~\bibnamefont {Zhou}},\ }\href {https://arxiv.org/abs/2503.00100} {\bibinfo {title} {Fuzzified : Julia package for numerics on the fuzzy sphere}} (\bibinfo {year} {2025}),\ \Eprint {https://arxiv.org/abs/2503.00100} {arXiv:2503.00100 [cond-mat.str-el]} \BibitemShut {NoStop}%
\bibitem [{\citenamefont {Fishman}\ \emph {et~al.}(2022)\citenamefont {Fishman}, \citenamefont {White},\ and\ \citenamefont {Stoudenmire}}]{Fishman_2022}%
  \BibitemOpen
  \bibfield  {author} {\bibinfo {author} {\bibfnamefont {M.}~\bibnamefont {Fishman}}, \bibinfo {author} {\bibfnamefont {S.}~\bibnamefont {White}},\ and\ \bibinfo {author} {\bibfnamefont {E.}~\bibnamefont {Stoudenmire}},\ }\bibfield  {title} {\bibinfo {title} {The itensor software library for tensor network calculations},\ }\bibfield  {journal} {\bibinfo  {journal} {SciPost Physics Codebases}\ }\href {https://doi.org/10.21468/scipostphyscodeb.4} {10.21468/scipostphyscodeb.4} (\bibinfo {year} {2022})\BibitemShut {NoStop}%
\bibitem [{\citenamefont {Kos}\ \emph {et~al.}(2014)\citenamefont {Kos}, \citenamefont {Poland},\ and\ \citenamefont {Simmons-Duffin}}]{Kos_2014}%
  \BibitemOpen
  \bibfield  {author} {\bibinfo {author} {\bibfnamefont {F.}~\bibnamefont {Kos}}, \bibinfo {author} {\bibfnamefont {D.}~\bibnamefont {Poland}},\ and\ \bibinfo {author} {\bibfnamefont {D.}~\bibnamefont {Simmons-Duffin}},\ }\bibfield  {title} {\bibinfo {title} {Bootstrapping the o(n ) vector models},\ }\bibfield  {journal} {\bibinfo  {journal} {Journal of High Energy Physics}\ }\textbf {\bibinfo {volume} {2014}},\ \href {https://doi.org/10.1007/jhep06(2014)091} {10.1007/jhep06(2014)091} (\bibinfo {year} {2014})\BibitemShut {NoStop}%
\end{thebibliography}%

\clearpage
\onecolumngrid
 \appendix
 \section{The analysis of $SO(N)$ group}
 In this section, we discuss the group structure of the $O(N)$ critical spectrum. The irreducible representations of the $SO(N)$ group are specified by a set of highest weights $\lambda_i, i = 1, \dots, r = [N/2]$. 
 
 For the case of odd $N = 2r + 1$, the dimension of a representation and the eigenvalue of its quadratic Casimir operator are given by the formulas:

\[
\text{dim}(\lambda) = \prod_{1 \le i < j \le r} \frac{\lambda_i - \lambda_j + j - i}{j - i} \cdot \frac{\lambda_i + \lambda_j + 2r + 1 - i - j}{2r + 1 - i - j} \cdot \prod_{i=1}^{r} \frac{\lambda_i + r + 1/2 - i}{r + 1/2 - i},
\]

\[
C_2(\lambda) = \sum_{i=1}^{r} \lambda_i (\lambda_i + 2r + 1 - 2i).
\]

For the even $N = 2r$ case, the corresponding formulas are:

\[
\text{dim}(\lambda) = \prod_{1 \le i < j \le r} \frac{\lambda_i - \lambda_j + j - i}{j - i} \cdot \frac{\lambda_i + \lambda_j + 2r - i - j}{2r - i - j},
\]

\[
C_2(\lambda) = \sum_{i=1}^{r} \lambda_i (\lambda_i + 2r - 2i).
\]

When a line defect is introduced, the flavor symmetry is reduced from $O(N)$ to $O(N-1)$. The branching of a higher-dimensional $SO(N)$ representation into a direct sum of lower-dimensional $SO(N-1)$ representations under this symmetry reduction can be systematically described using the Gelfand–Tsetlin pattern. Given an $SO(N)$ irreducible representation $[\lambda]$, its decomposition into $SO(N-1)$ irreducible representations $[\mu]$ follows the rule:

\[
(\lambda_1, \dots, \lambda_r) \rightarrow \bigoplus_{\mu} (\mu_1, \dots, \mu_{r'}),
\]

where the interleaving condition must be satisfied:

\[
\lambda_1 \ge \mu_1 \ge \lambda_2 \ge \mu_2 \ge \cdots.
\]

Based on this, we present some concrete examples:
\begin{itemize}
    \item The $O(N)$ vector $\phi_i$ decomposes into $\phi_a$ and $\phi_N$, with the dimension decomposed as: 
    $$
\begin{array}{c}
	\mathbf{N}\\
	\text{vector}\\
\end{array}\rightarrow \begin{array}{c}
	\mathbf{N}-\mathbf{1}\\
	\text{vector}\\
\end{array}\oplus \begin{array}{c}
	\mathbf{1}\\
	\text{scalar}\\
\end{array}
$$
    \item The rank-2 symmetric traceless tensor decomposes as follows, corresponding to the dimensional splitting:
 $$
\begin{array}{c}
	\mathbf{\left( N+2 \right) \left( N-1 \right) /2}\\
	\text{tensor}\\
\end{array}\rightarrow \begin{array}{c}
	\mathbf{{N}\left( \text{N}-\mathbf{1} \right) /\mathbf{2}-\mathbf{1}}\\
	\text{tensor}\\
\end{array}\oplus \begin{array}{c}
	\mathbf{N}-\mathbf{1}\\
	\text{vector}\\
\end{array}\oplus \begin{array}{c}
	\mathbf{1}\\
	\text{scalar}\\
\end{array}
$$

\end{itemize}

  \section{locating the Wilson-Fisher conformal fixed point up to $O(N=5,6)$ case}
  To fully illustrate the possibility of spontaneous symmetry breaking for the one-dimensional line defect, we extend the optimal conformal fixed point defined for the $O(N)$ Wilson-Fisher model in Ref. \cite{guo2025onfreescalarwilsonfisherconformal} to the cases of $N=5$ and $6$. Specifically, we use the following cost function, with tunable parameters ${U, h}$, to search for the parameter point that best manifests the conformal tower structure in the energy spectrum, within the maximum system size accessible by exact diagonalization ($N_o=8$ for $N=5,6$ cases).

\[
Q(U, h) = \{ \Delta_\phi - \Delta_\phi^{\text{BS}}, \Delta_S - \Delta_S^{\text{BS}},\Delta_T-\Delta_T^{\text{BS}},\Delta_{\partial_\mu\phi}-\Delta_\phi,\Delta_{\partial_\mu S}-\Delta_S,\Delta_{\partial_\mu T}-\Delta_T,\Delta_{T^{\mu\nu}}-3 \}
\]

For the $O(N=5)$ case, we use the following conformal data determined by the bootstrap method\cite{Kos_2014}:

\[
\Delta_\phi^{\text{BS}} = 0.5155,\Delta_T^{\text{BS}}=1.1568,\Delta_S^{\text{BS}}=1.682
\]

For the $O(N=6)$ case, we use the following conformal data determined by the bootstrap method\cite{Kos_2014}:

\[
\Delta_\phi^{\text{BS}} = 0.5145,\Delta_T^{\text{BS}}=1.1401,\Delta_S^{\text{BS}}=1.725
\]

The optimization of the cost function yields the following optimal points:
\begin{equation}
    \begin{aligned}
        O\left( N=5 \right) \ \text{WF\ CFT:\ }\left\{ U=0.20942,\ h=0.10246 \right\} \\
\text{O}\left( N=6 \right) \ \text{WF\ CFT:\ }\left\{ U=0.19759,\ h=0.10834 \right\} 
    \end{aligned}
\end{equation}



The defect conformal data for the bulk $O(N=5,6)$ WF CFTs in the main text are based on the bulk critical points determined in this manner.

\section{another realization of line defect CFT in bilayer fuzzy sphere model}
In this section, we present an alternative implementation of the $O(3)$ line defect. We realise the \(h_d \to \infty\) limit by pinning the spins on the orbitals with \(m = \pm s\) in the orbital space. The bulk Hamiltonian we consider is the bilayer model proposed in the literature\cite{chaohanPhysRevB.110.115113}. The pinning field we apply takes the following form:
\begin{equation}
    H_d=\lim_{h_d\rightarrow\infty}\sum_{m=\pm s}\boldsymbol{c}_m^\dagger\tau^z\sigma^z \boldsymbol{c}_m
\end{equation}
By rescaling the defect spectrum using the energy gap of the stress-energy tensor in the bulk spectrum, we again observe a clear conformal multiplet structure, as shown in the Fig.\ref{fig:O3_hsb_wf_multi} below.
\begin{figure}[!htbp] 
    \centering
\includegraphics[width=0.16\textwidth]{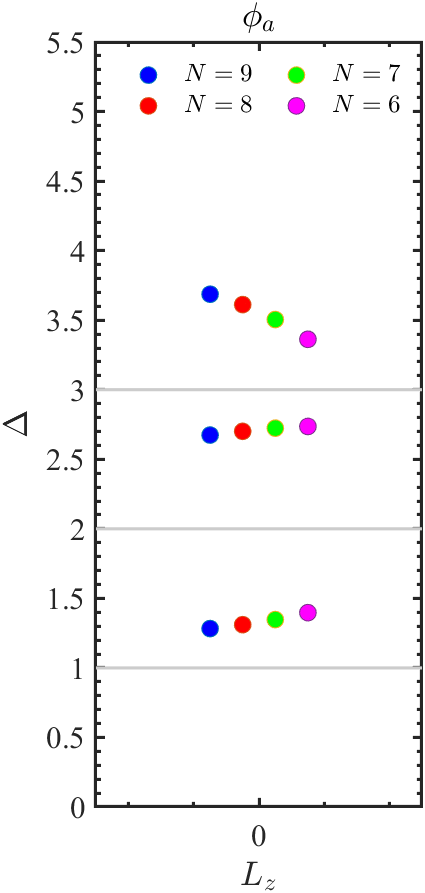} 
\includegraphics[width=0.16\textwidth]{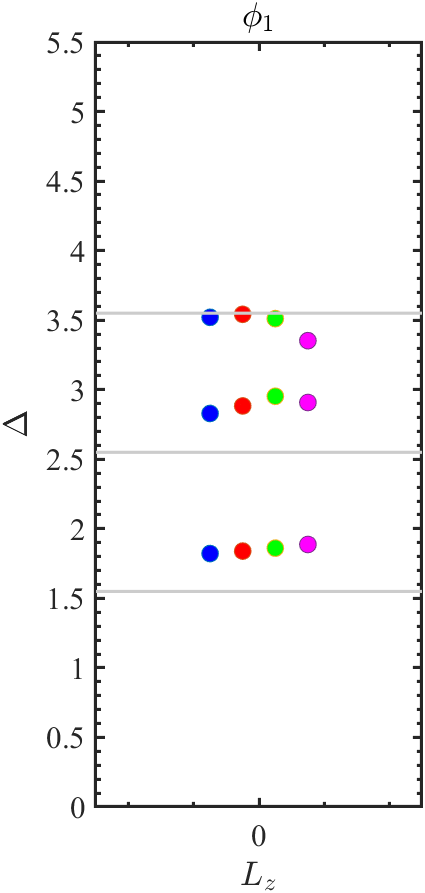}
\includegraphics[width=0.16\textwidth]{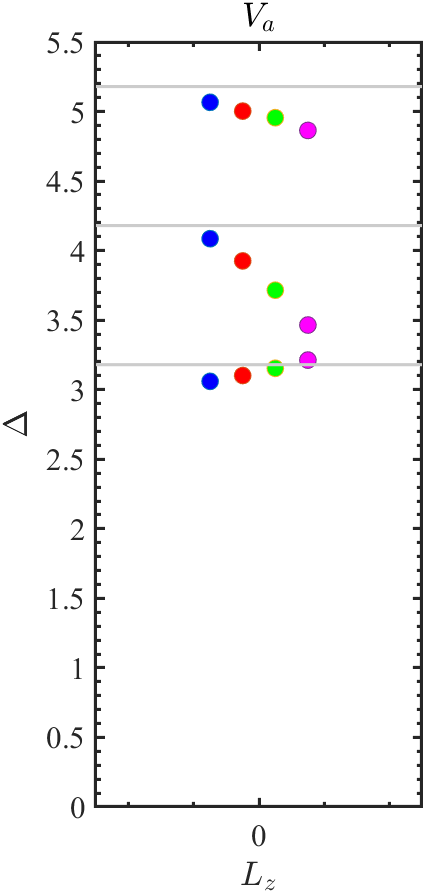}
\includegraphics[width=0.16\textwidth]{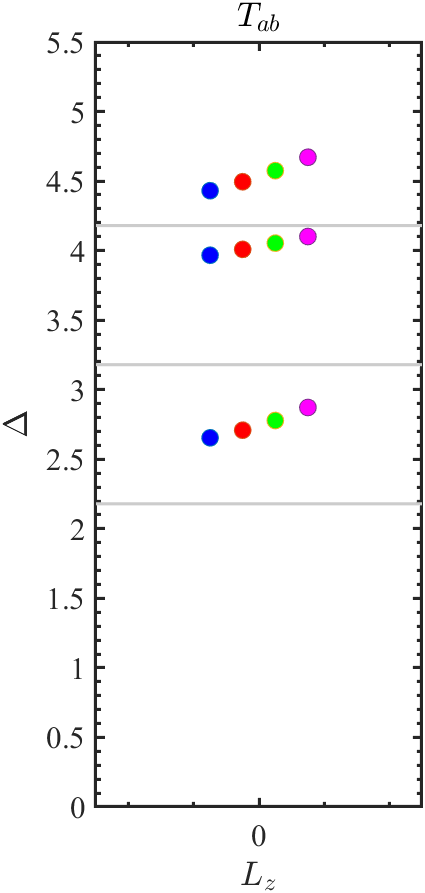}
\includegraphics[width=0.16\textwidth]{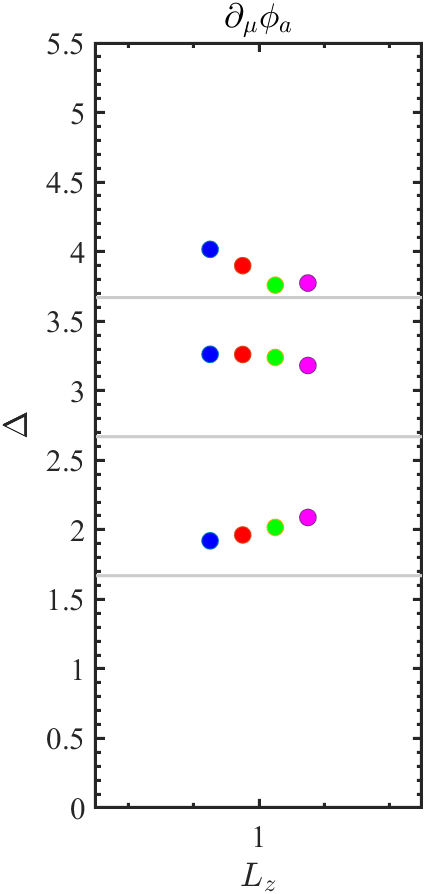}\includegraphics[width=0.16\textwidth]{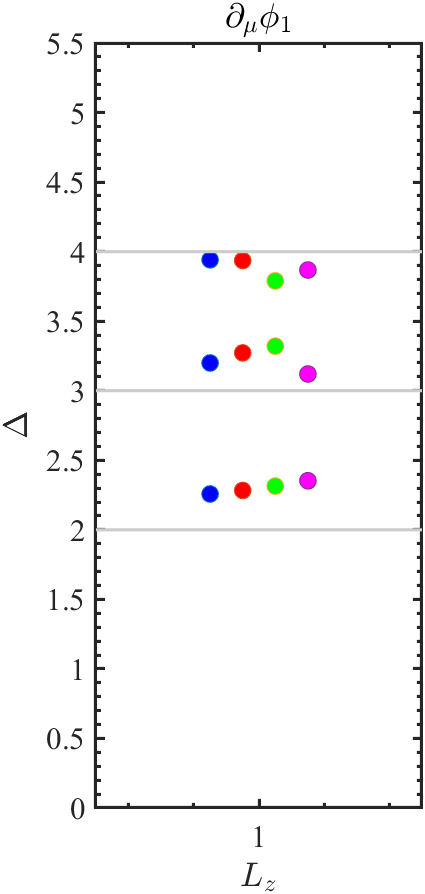}
    
    \caption{Line defect conformal multiplets. The gray horizontal line is taken from the result of the epsilon expansion. These results are produced at $h_d=\infty$ with subject to critical $O(3)$ bulk theory. The bulk parameters are set to be identical with Ref.\cite{chaohanPhysRevB.110.115113}}
    \label{fig:O3_hsb_wf_multi}
\end{figure}

\section{More data}
\subsection{The primary operators for $O(N)$ line defects}
In this section, for different $O(N=2,\cdots,6)$ Wilson–Fisher bulk CFTs, we present in Tab.\ref{tab:o2_wf_operator_dimensions},\ref{tab:o3_wf_operator_dimensions},\ref{tab:o4_wf_operator_dimensions},\ref{tab:o5_wf_operator_dimensions},\ref{tab:o6_wf_operator_dimensions} the scaling dimensions of several low-lying primary operators obtained from exact diagonalization calculations on finite-size systems. The operators considered include the lowest-energy $O(N-1)$ vector operator, scalar operator, and tensor operator, as well as the defect-changing operator and the defect-creation operator.

\begin{table}[htbp!]
  \centering
  
  \caption{Scaling dimensions of several primary operators in the defect CFT obtained from finite-size calculations. The bulk Hamiltonian is tuned to the critical point of the $O(N=2)$ Wilson–Fisher CFT. Here, $\text{Dim}$ and $\mathbb{Z}_2$ denote the dimension and the charge under the residual $O(1)\sim Z_2$ group.}
  
  \begin{tabular}{c|c|ccc|cccc|c}  
    \hline\hline
    Operator & $\epsilon$ expansion & $L_z$ & $\text{Dim}_{\mathrm{O(1)}}$ & $\mathbb{Z}_2$ & $N_o=11$ & $N_o=10$& $N_o=9$ & $N_o=8$ & Notes \\
    \hline
    $\phi_1$ & 2(1.55) & 0 & 1 & $+$ & 1.5607 & 1.5610 &1.5614 &1.5623 &   \\
    $\phi_a$ & 1 & 0 & 1 & $-$ & 1.2407 & 1.2616 &1.2869 &1.3181  & Only $N \geq 2$ \\
    $S_+$ & 4.3050 & 0 & 1 & $+$ & 3.5186 & 3.4887 &3.4465 &3.3869  &  \\
    $S_-$ & 2.2950 & 0 & 1 & $+$ & 2.6232 & 2.6602 &2.7051 &2.7608  & Only $N \geq 2$ \\
    $V_a$ & 3.2000 & 0 & 1 & $-$ & 2.7685 & 2.7897 &2.8151 &2.8465  & Only $N \geq 2$ \\
    $\nabla \phi_1$ & 2 & 1 & 1 & $+$ & 2.0286 & 2.0361 &2.0458 &2.0586 &  \\
    $\nabla \phi_a$ & 1.6667 & 1 & 1 & $-$ & 1.7447 & 1.7568 &1.7710 &1.7878  & Only $N \geq 2$ \\
    \hline
    $\phi^{+-}$ & $-$ & 0 & 1 & $+$ & 0.8528 & 0.8510 &0.8491 &0.8471 & defect changing op.\\
    $\phi^{+0}$ & $-$ & 0 & 1 & $+$ & 0.1175 & 0.1175 &0.1176 &0.1178 & defect creation op. \\
    \hline
    \hline
    \label{tab:o2_wf_operator_dimensions}
  \end{tabular}
\end{table}

\begin{table}[htbp!]
  \centering
  \caption{Scaling dimensions of several primary operators in the defect CFT obtained from finite-size calculations. The bulk Hamiltonian is tuned to the critical point of the $O(N=3)$ Wilson–Fisher CFT. Here, $\text{Dim}$ and $C_2$ denote the dimension and the quadratic Casimir eigenvalue of the representations under the residual $SO(N-1)$ group.}

  \begin{tabular}{c|c|ccc|cccc|c}  
    \hline\hline
    Operator & $\epsilon$ expansion & $L_z$ & $\text{Dim}_{SO(2)}$  & $C_2^{\mathrm{SO(2)}}$ & $N_o=10$ & $N_o=9$ &$N_o=8$ &$N_o=7$ & Notes \\
    \hline
    $\phi_1$ & 2(1.55) & 0 & 1 & 0 & 1.5372 & 1.5375 &1.5383 &1.5399 &  \\
    $\phi_a$ & 1 & 0 & 2 & 1 & 1.2323 & 1.2576 &1.2891 &1.3295 & Only $N \geq 2$ \\
    $S_+$ & 4.2813 & 0 & 1 & 0 & 3.4619 & 3.4536 &3.3915 &3.3034 &  \\
    $S_-$ & 2.3550 & 0 & 1 & 0 & 2.6314 & 2.6714 &2.7220 &2.7873 & Only $N \geq 2$ \\
    $V_a$ & 3.1818 & 0 & 2 & 1 & 2.7313 & 2.7564 &2.7877 &2.8283 & Only $N \geq 2$ \\
    $T_{ab}$ & 2.1818 & 0 & 2 & 4 & 2.5563 & 2.6061 &2.6678 &2.7464 & Only $N \geq 3$ \\
    $\nabla \phi_1$ & 2 & 1 & 1 & 0 & 2.0184 & 2.0273 &2.0394 &2.0566 &  \\
    $\nabla \phi_a$ & 1.6667 & 1 & 2 & 1 & 1.7307 & 1.7445 &1.7611 &1.7814 & Only $N \geq 2$ \\
    \hline
    $\phi^{+-}$ & $-$ & 0 & 1 & 0 & 0.9041 & 0.8967 &0.8887 &0.8799 & defect changing op.\\
    $\phi^{+0}$ & $-$ & 0 & 1 & 0 & 0.1330 & 0.1330 &0.1330 &0.1333 & defect creation op. \\
    \hline\hline
    \label{tab:o3_wf_operator_dimensions}
  \end{tabular}
\end{table}

\begin{table}[htbp!]
  \centering
  \caption{Scaling dimensions of several primary operators in the defect CFT obtained from finite-size calculations. The bulk Hamiltonian is tuned to the critical point of the $O(N=4)$ Wilson–Fisher CFT. Here, $\text{Dim}$ and $C_2$ denote the dimension and the quadratic Casimir eigenvalue of the representations under the residual $SO(N-1)$ group.}
  
  \begin{tabular}{c|c|ccc|cccc|c}  
    \hline\hline
    Operator & $\epsilon$ expansion & $L_z$ & $\text{Dim}_{SO(3)}$  & $C_2^{\mathrm{SO(3)}}$ & $N_o=9$ & $N_o=8$ &$N_o=7$ &$N_o=6$ & Notes \\
    \hline
    $\phi_1$ & 2(1.55) & 0 & 1 & 0 & 1.5210 & 1.5216 &1.5232 &1.5270 &  \\
    $\phi_a$ & 1 & 0 & 3 & 2 & 1.2298 & 1.2619 &1.3034 &1.3590 & Only $N \geq 2$ \\
    $S_+$ & 4.2613 & 0 & 1 & 0 & 3.4944 & 3.4289 &3.3358 &3.2038 &  \\
    $S_-$ & 2.4054 & 0 & 1 & 0 & 2.6419 & 2.6872 &2.7471 &2.8194 & Only $N \geq 2$ \\
    $V_a$ & 3.1667 & 0 & 3 & 2 & 2.7044 & 2.7360 &2.7778 &2.8354 & Only $N \geq 2$ \\
    $T_{ab}$ & 2.1667 & 0 & 5 & 6 & 2.5415 & 2.6045 &2.6855 &2.7934 & Only $N \geq 3$ \\
    $\nabla \phi_1$ & 2 & 1 & 1 & 0 & 2.0202 & 2.0321 &2.0494 &2.0757 &  \\
    $\nabla \phi_a$ & 1.6667 & 1 & 3 & 2 & 1.7241 & 1.7408 &1.7617 &1.7884 & Only $N \geq 2$ \\
    \hline
    $\phi^{+-}$ & $-$ & 0 & 1 & 0 & 0.9705 & 0.9543 &0.9365 &0.9167 & defect changing Op.\\
    $\phi^{+0}$ & $-$ & 0 & 1 & 0 & 0.1489 & 0.1486 &0.1486 &0.1490 & defect creation Op. \\
    \hline\hline
    \label{tab:o4_wf_operator_dimensions}
  \end{tabular}
\end{table}

\begin{table}[htbp!]
  \centering
  \caption{Scaling dimensions of several primary operators in the defect CFT obtained from finite-size calculations. The bulk Hamiltonian is tuned to the critical point of the $O(N=5)$ Wilson–Fisher CFT. Here, $\text{Dim}$ and $C_2$ denote the dimension and the quadratic Casimir eigenvalue of the representations under the residual $SO(N-1)$ group.}
  
  \begin{tabular}{c|c|ccc|cccc|c}  
    \hline\hline
    Operator & $\epsilon$ expansion & $L_z$ & $\text{Dim}_{SO(4)}$  & $C_2^{\mathrm{SO(4)}}$ & $N_o=9$ & $N_o=8$ &$N_o=7$ &$N_o=6$ & Notes \\
    \hline
    $\phi_1$ & 2(1.55) & 0 & 1 & 0 & 1.4531 & 1.4556 &1.4603 &1.4695 &  \\
    $\phi_a$ & 1 & 0 & 4 & 3 & 1.2006 & 1.2326 &1.2748 &1.3326 & Only $N \geq 2$ \\
    $S_+$ & 4.2440 & 0 & 1 & 0 & 3.6069 & 3.5394 &3.4404 &3.2970 &  \\
    $S_-$ & 2.4483 & 0 & 1 & 0 & 2.6383 & 2.6510 &2.6787 &2.7370 & Only $N \geq 2$ \\
    $V_a$ & 3.1538 & 0 & 4 & 3 & 2.6611 & 2.6996 &2.7340 &2.7557 & Only $N \geq 2$ \\
    $T_{ab}$ & 2.1538 & 0 & 9 & 8 & 2.4719 & 2.5345 &2.6166 &2.7285 & Only $N \geq 3$ \\
    $\nabla \phi_1$ & 2 & 1 & 1 & 0 & 1.9879 & 2.0046 &2.0286 &2.0644 &  \\
    $\nabla \phi_a$ & 1.6667 & 1 & 4 & 3 & 1.7043 & 1.7221 &1.7449 &1.7747 & Only $N \geq 2$ \\
    \hline
    $\phi^{+-}$ & $-$ & 0 & 1 & 0 & 0.9895 & 0.9734 &0.9552 &0.9343 & defect changing Op.\\
    $\phi^{+0}$ & $-$ & 0 & 1 & 0 & 0.1601 & 0.1597 &0.1595 &0.1598 & defect creation Op. \\
    \hline\hline
    \label{tab:o5_wf_operator_dimensions}
  \end{tabular}
\end{table}

\begin{table}[htbp!]
  \centering
  \caption{Scaling dimensions of several primary operators in the defect CFT obtained from finite-size calculations. The bulk Hamiltonian is tuned to the critical point of the $O(N=6)$ Wilson–Fisher CFT. Here, $\text{Dim}$ and $C_2$ denote the dimension and the quadratic Casimir eigenvalue of the representations under the residual $SO(N-1)$ group.}
  
  \begin{tabular}{c|c|ccc|cccc|c}  
    \hline\hline
    Operator & $\epsilon$ expansion & $L_z$ & $\text{Dim}_{SO(5)}$  & $C_2^{\mathrm{SO(5)}}$ & $N_o=9$ & $N_o=8$ &$N_o=7$ &$N_o=6$ & Notes \\
    \hline
    $\phi_1$ & 2(1.55) & 0 & 1 & 0 & 1.4327 & 1.4388 &1.4499 &1.4717 &  \\
    $\phi_a$ & 1 & 0 & 5 & 4 & 1.2129 & 1.2562 &1.3158 &1.4016 & Only $N \geq 2$ \\
    $S_+$ & 4.2290 & 0 & 1 & 0 & 3.5872 & 3.4838 &3.3334 &3.1174 &  \\
    $S_-$ & 2.4852 & 0 & 1 & 0 & 2.6703 & 2.6808 &2.6984 &2.7545 & Only $N \geq 2$ \\
    $V_a$ & 3.1429 & 0 & 5 & 4 & 2.7155 & 2.7512 & 2.7734 & 2.8206 & Only $N \geq 2$ \\
    $T_{ab}$ & 2.1429 & 0 & 14 & 10 & 2.6703 & 2.6808 & 2.6887 &2.8554 & Only $N \geq 3$ \\
    $\nabla \phi_1$ & 2 & 1 & 1 & 0 & 1.9958 & 2.0220 &2.0609 &2.1212 &  \\
    $\nabla \phi_a$ & 1.6667 & 1 & 5 & 4 & 1.7090 & 1.7327 &1.7641 &1.8072 & Only $N \geq 2$ \\
    \hline
    $\phi^{+-}$ & $-$ & 0 & 1 & 0 & 1.0178 & 0.9933 &0.9652 &0.9328 & defect changing Op.\\
    $\phi^{+0}$ & $-$ & 0 & 1 & 0 & 0.1705 & 0.1709 &0.1708 & 0.1714 & defect creation Op. \\
    \hline\hline
    \label{tab:o6_wf_operator_dimensions}
  \end{tabular}
\end{table}

\end{document}